\documentclass{aa}
\usepackage{natbib,twoopt}
\usepackage{graphicx}
\usepackage{float}
\usepackage{caption}
\usepackage{txfonts}
\usepackage{multirow}
\usepackage{tablefootnote}
\usepackage{amsmath}
\usepackage{siunitx}
\usepackage{subcaption}
\usepackage{multirow}
\usepackage[dvipsnames]{xcolor}
\usepackage{xcolor}
\usepackage{tikz}
\usepackage{pdflscape}
\usetikzlibrary{shapes, arrows.meta, positioning, matrix, calc}
\definecolor{darkgoldenrod}{rgb}{0.72, 0.53, 0.04}

\usepackage[colorlinks = true,
            linkcolor = RoyalBlue,
            urlcolor = darkgoldenrod,
            citecolor = RoyalBlue,
            anchorcolor = RoyalBlue, 
            draft = false]{hyperref}

\newcommand{\comment}[1]{}

\definecolor{diagramdarkblue}{rgb}{0,0,0.55}
\definecolor{diagramdarkred}{rgb}{0.55,0,0}

\begin{document} 
   \title{Machine learning technique for morphological classification of galaxies from SDSS. IV. CNN-Assisted Expert Validation of Morphological Catalogues for Merging, Irregular, Edge-on, Barred, Ringed, and Dust-lane Galaxies at 0.003 < z < 0.1} 

  \author{Dobrycheva, D.V.$^{1}$, Vavilova, I.B.$^{1}$, Khramtsov, V.$^{2}$, Kompaniiets, O.V.$^{1}$, \\ Vasylenko, M.Yu.$^{1}$, Hetmantsev, O.O.$^{1}$, Melnyk, O.V.$^{1}$, Karachentseva, V.E.$^{1}$}

   \institute{
$^{1}$ Main Astronomical Observatory of the NAS of Ukraine, 27, Akademik Zabolotnyi St., Kyiv, 03143, Ukraine\\
$^{2}$ Institute of Astronomy, V.N. Karazin Kharkiv National University, 35, Sumska St., Kharkiv, 61022, Ukraine\\
}

\titlerunning{ML-IV}
\authorrunning{Dobrycheva, D.V. et al.}

   \date{Received 2025; accepted 04 November 2026}
\abstract
{Convolutional neural networks (CNNs) have become an efficient tool for automated galaxy morphological classification in large astronomical surveys. However, identifying complex morphological features remains limited by projection effects, image artefacts, and intrinsic ambiguities, making expert visual validation essential to building reliable morphological catalogues.}
{This paper introduces a CNN-assisted expert validation framework for constructing reliable galaxy image-based morphological catalogues. The aim is to perform a comprehensive visual validation of SDSS galaxies at $0.003<z<0.1$ previously classified by a CNN as merging, irregular, edge-on, barred, ringed, and dust-lane galaxies; to quantify the reliability and failure modes of CNN classifications; to construct homogeneous visually verified catalogues of these morphological classes; and to characterise the central ionisation of the verified galaxies with emission-line diagnostics.}
{We visually inspected all CNN-selected candidates in four morphological categories: merging (2\,574), irregular (9\,432), ringed (13\,882), and dust-lane (588) galaxies, covering the full range of their catalogue feature scores. For edge-on and barred galaxies, we inspected the 17\,000 and 4\,323 highest-scoring candidates, respectively, selected from larger CNN catalogues. We cross-matched the refined samples with the GZ2 catalogue, and we visually classified galaxies outside its selection domain, which carry no GZ2 morphological label. For the subset with SDSS spectroscopy, we applied the BPT and WHAN diagnostics to MPA--JHU emission-line measurements. 
}
{We present homogeneous visually verified catalogues containing 612 merging galaxies, 9\,372 irregular galaxies, 16\,822 edge-on galaxies, 811 barred galaxies, 2\,145 ringed galaxies, and 575 dust-lane galaxies. Comparison with GZ2 demonstrates that CNN-assisted expert validation both refines existing classifications and extends reliable morphological catalogues beyond the effective Galaxy Zoo selection domain. The principal causes of CNN misclassification are projection effects, foreground stars, image artefacts, faint tidal structures, and intrinsically ambiguous morphologies. Emission-line diagnostics show that 87.8\% of the ringed galaxies with LINER-like line ratios have H$\alpha$ equivalent widths consistent with ionisation by evolved stars rather than by an active nucleus.
We also identified five strong new candidates for polar-ring galaxies.}
{Our results demonstrate that CNN-assisted expert validation remains essential for constructing reliable galaxy morphological catalogues. The resulting homogeneous datasets provide valuable resources for studies of galaxy morphology, secular evolution, and nuclear activity, and constitute high-quality benchmark samples for future machine-learning applications and forthcoming wide-field surveys.}
{}
   \maketitle

\section{Introduction}

The automated galaxy morphological classification using machine learning and other artificial intelligence methods is a key component of processing the present and forthcoming large-scale sky surveys. It allows us to rapidly determine galaxies' physical properties, given the well-established strong correlations between morphology, luminosity, colour indices, density environment, star formation rate, evolutionary stage, etc. The morphological diversity of galaxies revealed by large sky surveys is fundamental to understanding the formation history and evolutionary pathways of the large-scale structure of the Universe. Since galaxy morphology is intricately linked to dynamical processes such as past and ongoing mergers and substructural modifications during the internal secular evolution, its classification remains a cornerstone of extragalactic research (see reviews by \cite{Buta2015} and \cite{Masters2025}). 

In this context, the photometry-based morphological classification is more developed, with machine learning methods such as Random Forest, Support Vector Machine, k-NN, and others already widely used. We note recent works based on the optical observations by \cite{Clarke2020} and \cite{Zeraatgari2024b} for galaxies, quasars, and stars; \cite{Vavilova2021} for SDSS galaxies at $z$ < 0.1; \cite{Vehlova2022} on low surface brightness galaxies, \cite{Ren2023} for merging galaxies; \cite{Argudo2024} for galaxies within voids; \cite{Martin2020} on the LSST morphological classificator, which was verified and implemented for future observations\footnote{\href{https://sites.google.com/view/lsstgsc/working-groups/morphology}{LSST Galaxy Science Collaboration Morphology Working Group}.}

Recent innovations in machine learning, particularly convolutional neural networks (CNNs), have revolutionised this field. They enable efficient image-based classification of galaxies by both their shape and inner structural morphological features. We note, for example, research by \cite{Cheng2020, Walmsley2022, Dominguez2022, Wei2022, Mao2025, Goh2025}, as well as \cite{Vavilova2022, Khramtsov2022} for SDSS galaxies at $z$ < 0.1 and \cite{Xu2023} on detailed galaxy morphology, \cite{Chrobakova2025} on edge-on and highly inclined galaxies, \cite{Heestars2025} and \cite{Thuruthipilly2025} on low surface brightness galaxies, \cite{Abraham2025} on galactic rings, and \cite{Mukundan2024} and \cite{Aguilar2025} on morphological segregation by T-types. An overview of the applications and impact of deep learning on big data galaxy surveys can be found in the work by \cite{Huertas2023}. 

Does manual classification become infeasible in the era of sky surveys such as SDSS, Pan-STARRS, DESI Legacy, Euclid, and the upcoming LSST? We consider that a visual inspection remains critical as it underpins supervised machine learning approaches, which in turn form the basis for most unsupervised and AI-based classification systems (see \cite{Nair2010, Baillard2011, Buta2015}, and references in \cite{Masters2025}). Most image-based morphological classifications are based on the volunteer Galaxy Zoo project \citep{Lintott2008}. The labelled data of this project is usefully exploited as training samples for many supervised machine learning methods (see, for example, \cite{Banerji2010, Dieleman2015}, Zoobot \cite{Aussel2024, Euclid2025b}, and us \cite{Vavilova2022, Dobrycheva2025} for verifying the results of deep learning implementation). The Euclid Quick Data Release (Q1) is also based on a visual Zoobot classification \citep{Walmsley2025} and allows the catalogue of about 380,000 bright galaxies with morphological features such as bars, spiral arms, and ongoing mergers.

Despite the efficiency of CNN-based classifications, an important methodological gap remains. Although CNN classifiers often achieve high global performance metrics on benchmark datasets, these metrics do not necessarily reflect the reliability of the resulting morphological catalogues.  Projection effects, low angular resolution, limited signal-to-noise ratio, overlapping structural signatures, and imbalances in the training data may cause confusion between merging and irregular galaxies with bars, rings, and dust lanes, as well as edge-on systems. Moreover, a high CNN confidence does not by itself guarantee that a predicted feature is physically present in an individual galaxy. A systematic visual audit is therefore required to determine which classes are robustly recovered, identify characteristic failure modes, and establish the degree of contamination in the resulting catalogues. The present study addresses this problem by validating the final CNN-derived morphological catalogue \citep{Vavilova2023Cat} rather than the CNN training procedure itself.

The purpose is to provide a visual inspection of SDSS galaxies at $z$<0.1, which were classified with CNN by their morphological features as merging, irregular, edge-on, with bar, rings and dust lane; to define catalogue purities and the principal causes of CNN misclassification, analysis of disagreement cases of CNN predictions; to create the six catalogues of verified galaxies with these structural characteristics; to identify astrophysically interesting systems. 

The data sets and CNN-based methodology for classifying galaxies by their morphological features are described in Section 2; results of CNN-assisted expert validation are discussed in Sections 3--4 for galaxies with the aforementioned morphological features. Additionally, in Section 5, we analyse the optical emission-line excitation properties of seven visually confirmed morphological samples using BPT and WHAN diagnostics; the Milky Way galaxy analogues are discussed in Section 6, and the general conclusion with references to the new relevant catalogues is summarised in Section 7.  

\section{Data and Methodology.}

The present work is the fourth paper in our series devoted to the machine-learning-based morphological classification of SDSS galaxies at $z<0.1$. The previous studies established the parent galaxy sample and investigated photometry-based morphological classification using classical machine-learning techniques \citep{Vavilova2021, Vavilova2021Cat}; developed an image-based CNN classifier trained and evaluated using labelled galaxy samples \citep{Khramtsov2022, Vavilova2022, Vavilova2024}; and applied the trained classifier to the larger SDSS galaxy sample to construct catalogues of galaxies with different morphological types and structural features \citep{Vavilova2023Cat}. 

\subsection{Previous studies and the CNN-based morphological catalogue}
\label{subsec:previous_studies}
In the first paper of the series, \citet{Vavilova2021} constructed a parent sample of low-redshift SDSS galaxies and applied classical supervised machine learning methods to their binary morphological classification. The preliminary SDSS sample contained $\sim$ 724,000 galaxies at $z<0.1$. After applying the adopted redshift and magnitude limits, photometric-quality criteria, and removing duplicate and artificial objects, the final photometric catalogue included 316,031 galaxies with $0.003<z<0.1$ and $-24<M_r<-13$ \citep{Vavilova2021Cat}. This catalogue subsequently served as the parent SDSS galaxy sample for the development of image-based morphological classifiers.

The second paper \citep{Vavilova2022} extended the morphological analysis from photometric parameters to SDSS galaxy images. We developed a CNN-based approach to classify galaxies into five broad visual morphological types---completely rounded, rounded in-between, cigar-shaped, edge-on, and spiral---and to recognise detailed morphological and structural features. Galaxy Zoo 2 (GZ2; \citealt{Willett2013}) classifications were used to construct the labelled sample required for supervised CNN training and evaluation. The CNN training labels were constructed from the original Galaxy Zoo 2 vote counts (\texttt{*\_count}) and the corresponding redshift-debiased vote fractions (\texttt{*\_debiased}). We retained this original GZ2 labelling throughout the series for methodological consistency. The revised debiasing of \cite{Hart2016} was therefore not adopted. Hereafter, $N_{\rm GZ2}$ denotes objects from the GZ2-matched training subset, whereas $N_{\rm Inf}$ denotes galaxies outside GZ2 for which the morphological features were inferred by the CNN. The study demonstrated the applicability of deep-learning methods to the morphological classification of the larger SDSS galaxy sample established in the first paper and identified systematic differences in magnitude and angular-size distributions between the labelled GZ2 sample and the remaining SDSS galaxies to which the trained classifier was subsequently applied (see Section \ref{subsec:CNNmodel}).

The third paper \citep{Khramtsov2022} explicitly addressed this difference by analysing the domain shift between the labelled GZ2 sample and the morphological catalogue of \cite{Vavilova2021Cat}. The trained CNN provided probabilistic predictions for broad morphological classes and detailed structural features of galaxies (bars, rings, dust lanes, etc.). The resulting  CNN-based morphological catalogue contains 315,782 galaxies with $m_r<17.7$ at $0.003<z<0.1$ \citep{Vavilova2023Cat} and constitutes the starting point for the present work.

\subsection{Construction of the CNN-based morphological catalogue}
\label{subsec:CNNmodel}
To help readers understand the present paper without requiring the previous publications, we briefly describe the complete workflow leading to the CNN-based morphological catalogue and the visual verification presented here. \autoref{fig:pipeline/diagram} summarises the complete workflow adopted throughout Papers II-IV and provides the methodological framework for the present study. It illustrates the construction of the labelled and inference samples, CNN training and application, and the subsequent visual verification performed in this work. The CNN model was trained using the Galaxy Zoo 2 (GZ2) labelled sample and subsequently applied to the entire SDSS parent sample. 

\begin{figure*}[htbp]
    \centering
    \begin{tikzpicture}[
    scale=0.88,
    transform shape,
    node distance=1.0cm,
    every node/.style={font=\small},
    box/.style={rectangle, rounded corners, draw, align=center, minimum width=5cm, minimum height=0.85cm},
    databox/.style={box, fill=white},
    modelbox/.style={box, fill=gray!20},
    gzbox/.style={box, fill=diagramdarkblue, text=white, minimum width=4.2cm},
    inferbox/.style={box, fill=diagramdarkred, text=white, minimum width=4.2cm},
    subbox/.style={rectangle, rounded corners, draw, align=center, fill=white, font=\scriptsize, minimum width=2.45cm, minimum height=0.75cm},
    arrow/.style={->, thick}
    ]

    \node[databox] (data) {SDSS Galaxy Dataset\\315\,782 galaxies};
    \node[gzbox, below left=of data, xshift=-0.8cm] (train) {Training (GZ2)\\172\,372 galaxies};
    \node[inferbox, below right=of data, xshift=0.8cm] (infer) {Inference\\143\,410 galaxies};
    \coordinate (advjoin) at ($(train.south)!0.5!(infer.south)+(0,-0.12cm)$);
    \node[databox, below=0.18cm of advjoin, anchor=north] (adv) {Adversarial Validation\\(train vs inference)};
    \node[modelbox, below=of adv] (aug) {Data Augmentation\\(scaling, noise, degradation)};
    \node[modelbox, below=of aug] (cnn) {CNN Model\\(DenseNet-201)};
    \node[modelbox, below=of cnn] (pred) {Multi-label Prediction\\(34 features)};
    \node[databox, below=of pred, minimum width=7.8cm] (cat) {Morphological Catalog\\34 subcatalogs};

    \matrix[
        matrix anchor=north,
        below=0.3cm of cat,
        column sep=0.08cm,
        row sep=0.08cm
    ] (subm) {
        \node[subbox] (sc1) {Merging\\2\,574 galaxies}; &
        \node[subbox] (sc2) {Irregular\\9\,432 galaxies}; &
        \node[subbox] (sc3) {Edge-on\\34\,420 galaxies}; &
        \node[subbox] (sc4) {Dust lane\\588 galaxies}; &
        \node[subbox] (sc5) {Barred galaxies\\29\,892 galaxies}; &
        \node[subbox] (sc6) {Ring\\13\,882 galaxies}; \\
    };

    \draw[arrow] (data) -- (train);
    \draw[arrow] (data) -- (infer);
    \draw[arrow] (train) -- (adv);
    \draw[arrow] (infer) -- (adv);
    \draw[arrow] (adv) -- (aug);
    \draw[arrow] (aug) -- (cnn);
    \draw[arrow] (cnn) -- (pred);
    \draw[arrow] (infer) |- (pred);
    \draw[arrow] (pred) -- (cat);
    \draw[arrow] (cat) -- (sc1.north);
    \draw[arrow] (cat) -- (sc2.north);
    \draw[arrow] (cat) -- (sc3.north);
    \draw[arrow] (cat) -- (sc4.north);
    \draw[arrow] (cat) -- (sc5.north);
    \draw[arrow] (cat) -- (sc6.north);

\end{tikzpicture}
    \caption{
    Schematic representation of the pipeline for multi-label morphological classification of galaxies. The initial SDSS dataset is split into a labelled training set (Galaxy Zoo 2) and an unlabeled inference set. Adversarial validation is applied to quantify the domain shift between the datasets. Data augmentation is then used to simulate observational conditions and improve model robustness. A convolutional neural network (DenseNet-201) predicts 34 morphological features (six of which are visually analysed in this paper), resulting in a homogeneous CNN-based morphological catalogue, as reported by \cite{Vavilova2023Cat}.
    }
    \label{fig:pipeline/diagram}
\end{figure*}

\begin{figure}[h]
    \centering
    {\includegraphics[width=8.6cm]{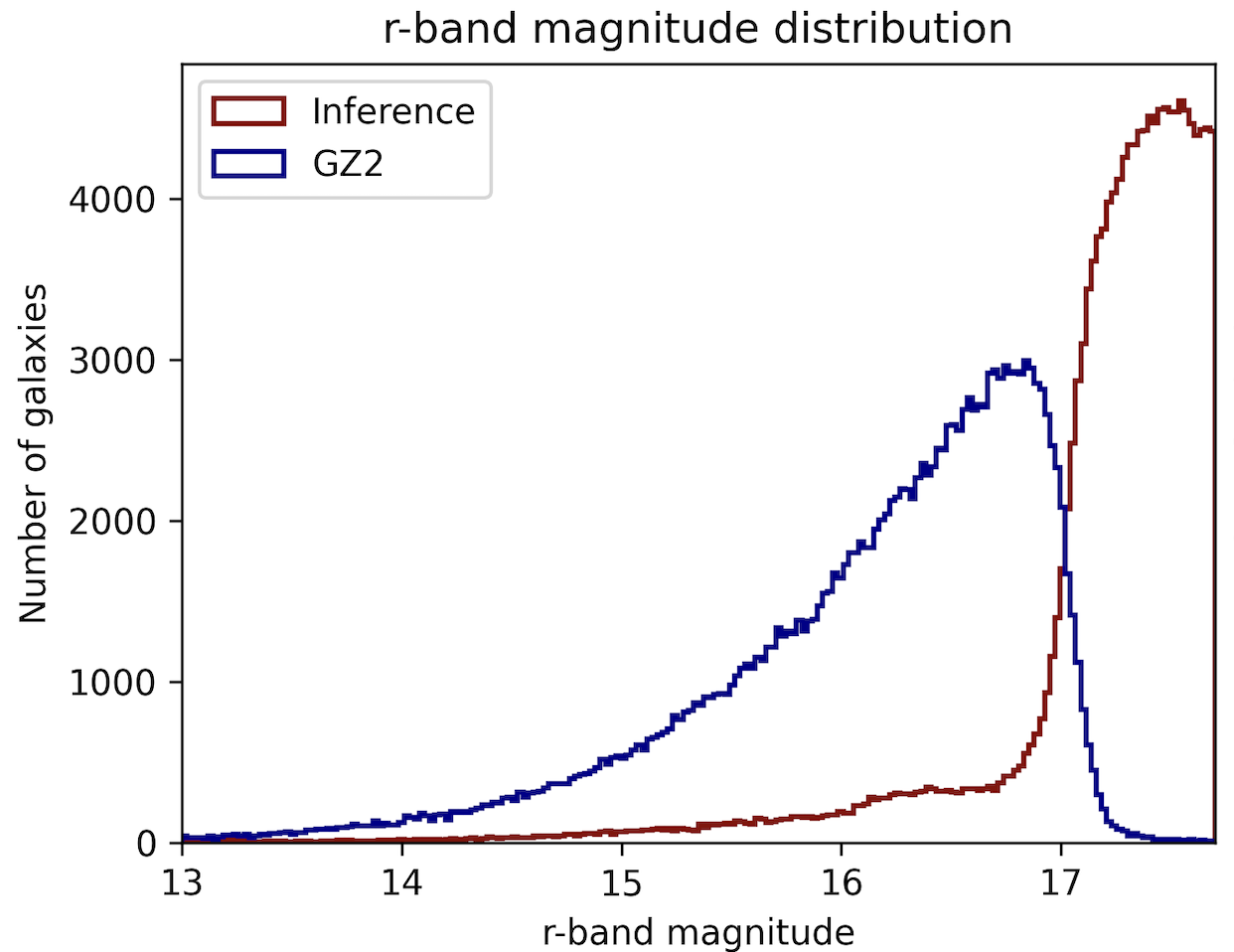}}
    \qquad
    {\includegraphics[width=9cm]{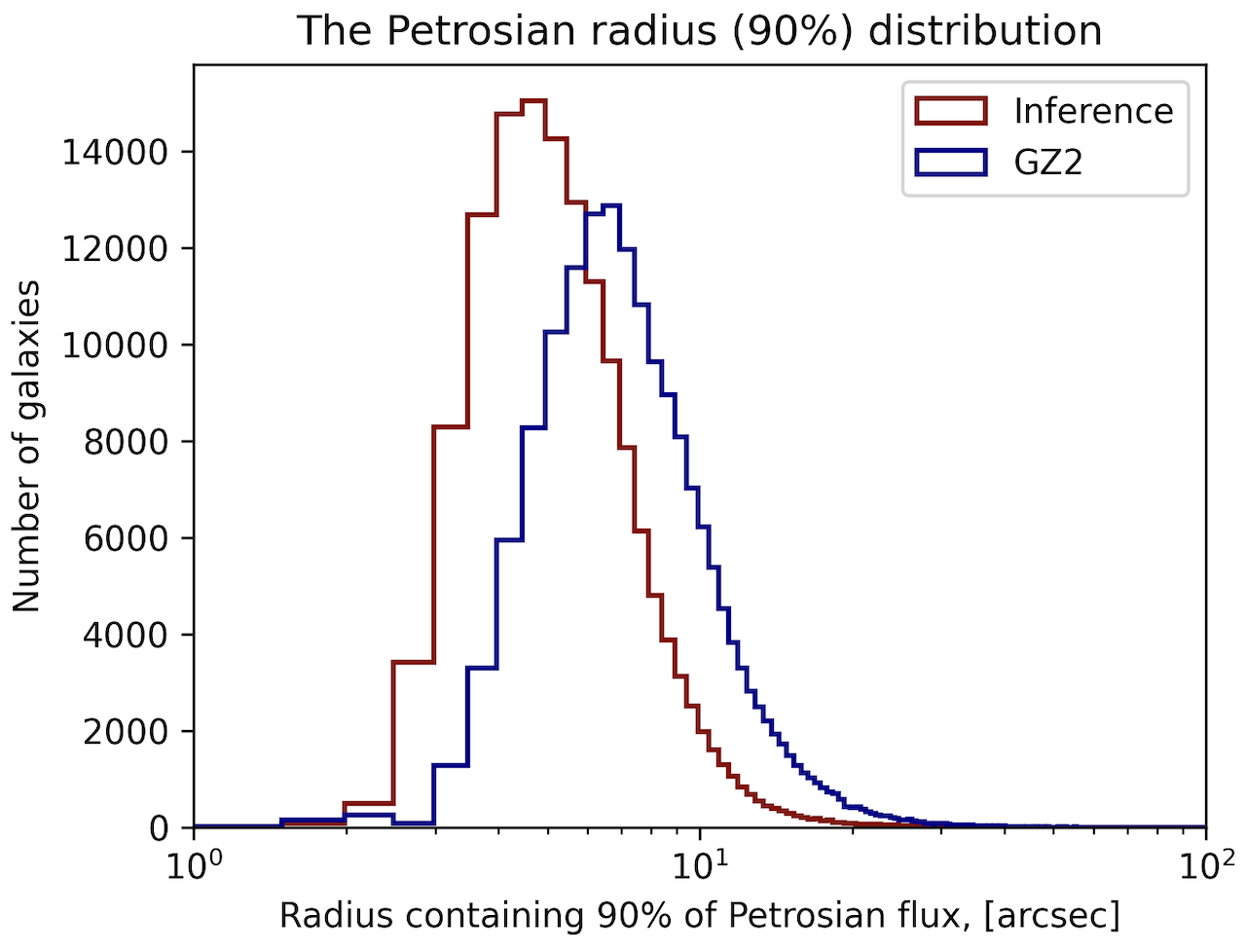}}
    \caption{Distribution of $r$-band magnitude and Petrosian radius for inference and GZ2 datasets. }
    \label{fig:GZ2_inference}
\end{figure}

The workflow shown in \autoref{fig:pipeline/diagram} consists of four main stages: (i) construction of the parent SDSS sample for the CNN-based catalogue, (ii) development and validation of the CNN classifier, (iii) application of the trained CNN to the complete SDSS sample, (iv) visual verification of selected morphological classes performed in the present paper. We validate the resulting CNN-based morphological catalogue through expert inspection of the original GZ2 galaxies and inference galaxies
outside the GZ2 selection domain~(\autoref{fig:GZ2_inference}).

The input SDSS photometric catalogue served as the parent galaxy sample for constructing the CNN-based morphological catalogue of galaxies with $0.003<z<0.1$ and $-24^{m}<M_{r}<-13^{m}$ at the first stage of the pipeline. The lower redshift bound follows from the restriction $V_{\rm LG}>1500$~km\,s$^{-1}$ on the Local-Group-corrected recession velocity adopted for the parent sample \citep{Vavilova2022}. RGB images of all galaxies in the $gri$ bands were downloaded from SDSS. After removing duplicate objects, stars, and images with artefacts, the parent sample contained 315,782  galaxies with $m_{r}<17.7$. The galaxies entering the catalogues presented here span $0.003 \le z \le 0.100$, $11.09 \le m_{r} \le 17.70$, and $-23.95 \le M_{r} \le -13.09$. Each galaxy image was prepared as a $100 \times 100$ pixel colour cutout, corresponding to $39.6 \times 39.6$ arcsec.

The second stage constructs the labelled GZ2 sample for further CNN training and evaluation. We considered only those galaxies for which GZ2's volunteers cast the most votes. Galaxies overlapping the Galaxy Zoo 2 catalogue formed the labelled sample used for CNN training and validation, while the remaining galaxies from the same parent SDSS sample, with no GZ2 morphological labels, formed the inference sample. The GZ2 labels include detailed answers for bars, rings, mergers, dust lanes, irregular structures, edge-on orientations, and other morphological features. These GZ2 labels served exclusively as the training labels for the CNN.

Before creating the inference sample, we applied adversarial validation to estimate how similar the GZ2 training images and the inference images are, because non-GZ2 galaxies are, on average, fainter and smaller. The labelled GZ2 sample was divided into training and test subsets for model development and evaluation. Image transformations and data augmentation were additionally used to simulate the lower apparent brightness and smaller angular sizes characteristic of galaxies in the inference sample. These transformations improved the CNN's robustness to observational differences between the labelled GZ2 sample and the inference sample by simulating lower signal-to-noise ratios, smaller angular sizes, and fainter apparent magnitudes. The differences in apparent magnitude and angular size between the GZ2-labelled and inference samples are illustrated in \autoref{fig:GZ2_inference}. The shift of the inference-sample galaxies toward fainter $r$-band magnitudes is an inevitable consequence of the different magnitude limits of the GZ2-labelled and parent SDSS samples and should not be interpreted as a preference of the CNN for faint or low-surface-brightness galaxies.

The CNN classifier was based on the DenseNet-201 architecture and had good robustness when applied beyond the labelled GZ2 sample. The six categories analysed here are detailed morphological features predicted by the multi-label model of \citet{Khramtsov2022}, not the five mutually exclusive broad morphological classes used in the preceding classification stage. For each galaxy, the final layer returns 34 independent sigmoid scores (one per feature), and the model was trained with binary cross-entropy, which means that a galaxy may satisfy several feature criteria simultaneously. A galaxy enters a feature catalogue when the corresponding sigmoid score exceeds the feature-specific threshold. These scores are not constrained to sum to unity and should therefore not be compared as a probability distribution across the 34 features. After training, the CNN was applied uniformly to every galaxy in the parent SDSS sample (third stage in \autoref{fig:pipeline/diagram}). The resulting score vector constituted the basis of the homogeneous CNN-based morphological catalogue \citet{Vavilova2023Cat}.

On the GZ2 test set, the feature-level ROC-AUC values reported by \citet{Khramtsov2022} range from 83.3\% to 99.4\% for 32 of the 34 features; the exceptions are ``disturbed'' (68.6\%) and ``arms winding medium'' (77.4\%). As a result, our CNN-based morphological catalogue contains detailed classifications for 315,782   galaxies with $m_r<17.7$ at $0.003<z<0.1$, including fainter galaxies beyond the GZ2-labelled sample of galaxies \citep{Vavilova2023Cat}. 

The resulting homogeneous CNN-based catalogue constituted the starting point of the present study (final stage in \autoref{fig:pipeline/diagram}). We visually inspected galaxies from both the labelled GZ2 subset and the inference subset. For the GZ2 galaxies, visual inspection let us assess the consistency between the original GZ2 classifications and the CNN predictions. For the inference galaxies, which had no prior GZ2 morphological labels, the visual inspection provided the first expert validation of the CNN-derived classifications.

\subsection{Selection of six CNN-classified galaxy samples}
\label{subsec:sixcats}
This fourth paper presents the visual verification of selected morphological samples from the previously published CNN-based catalogue. From the homogeneous CNN-based morphological catalogue, we selected galaxies assigned to six morphological categories: merging, irregular, edge-on, barred, ringed, and dust-lane galaxies. Each selected galaxy was visually inspected to determine whether the morphological feature assigned by the CNN was genuinely present. Objects were either confirmed, rejected, or reassigned to a more appropriate morphological category. For rejected or ambiguous classifications, we identified the main reasons for CNN disagreement, including projection effects, foreground stars, low-surface-brightness structures, image artefacts, and morphological similarities among features. The final products are six homogeneous catalogues of visually verified galaxies, each with corresponding morphological characteristics.

For this study, we selected candidates using six relevant feature scores: merger, irregular, edge-on, dust lane, bar, and ring. The 5.0\% catalogue threshold for each feature was retained so that the subsequent visual audit sampled a broad score range. Both real features and false positives occur across the inspected score ranges due to projection effects, foreground stars, spiral arms, central light concentrations, and image artefacts that can affect the network output. Therefore, the CNN-selected catalogues served as candidate lists for visual verification: all candidates were inspected for merging, irregular, ring, and dust-lane features, whereas the edge-on and bar samples were restricted to the highest-scoring candidates.

Our input data sets for visual inspection contained 2\,574 candidate merging galaxies, 9\,432 candidate irregular galaxies, 17\,000 candidate edge-on galaxies selected with the largest edge-on scores, 588 galaxies with dust lane candidates, 4\,323 candidate barred galaxies selected in decreasing scores, and 13\,882 candidate ring galaxies. For each object, we retained the SDSS coordinates, redshift, magnitude, feature scores, and whether the object matched GZ2 or belonged to the preliminary inference set. The expert visual inspection protocol is as follows: all authors of this paper acted as expert classifiers, and each candidate galaxy was independently inspected by at least two of them. The inspection was coordinated by D. Dobrycheva, who also discovered new polar-ring galaxies visually. Cases of disagreement were discussed jointly until a consensus was reached, and final label assignments were made according to the classification rubrics. The results are summarized in~\autoref{tab:VI}. 

Of the 20\,965 confirmed morphological entries in the merging, edge-on, barred, ringed, and dust-lane classes, 4\,716 (22.5\%) either have no previous visual classification for the corresponding feature or have only an automated classification, and therefore receive their first expert visual classification. These counts refer to morphological entries rather than necessarily unique galaxies, because an object may occur in more than one feature catalogue. Throughout this paper, galaxies outside the GZ2-matched subset are referred to as not present in GZ2.

\begin{figure*}[h]
    \centering    
    \includegraphics[width=\textwidth]{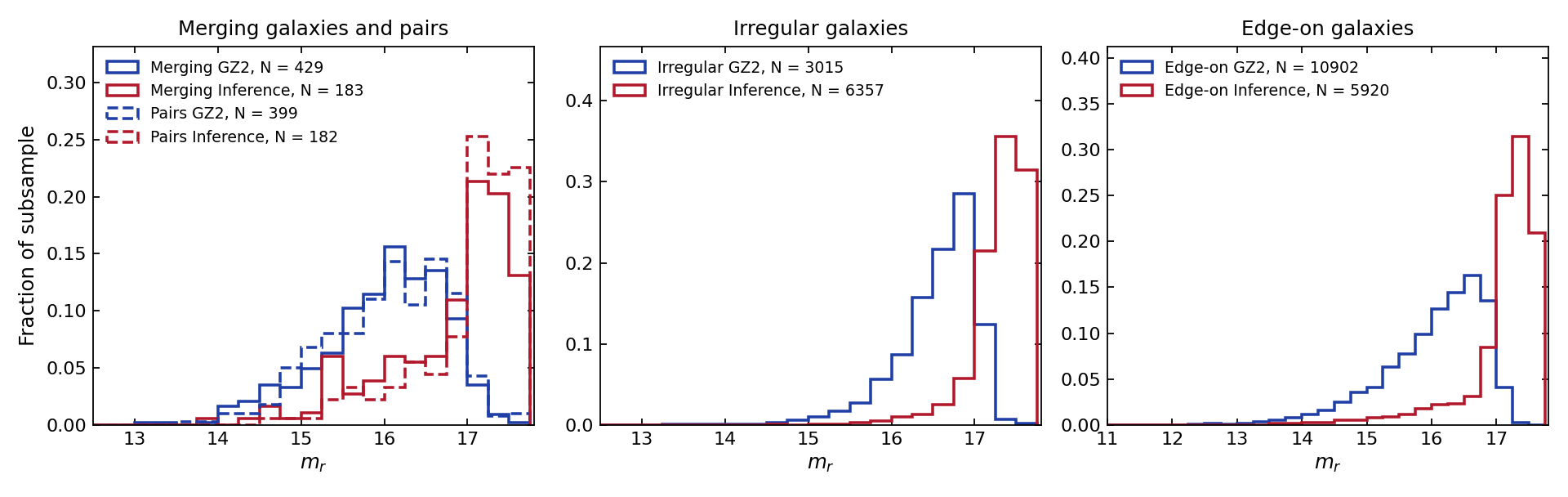}
    \caption{Distribution of apparent magnitude in the r-band of SDSS after visual inspection for: merging galaxies (left); irregular galaxies (centre); edge-on galaxies (right). Blue and red histograms correspond to the GZ2-matched and inference subsamples; in the left panel, solid and dashed lines distinguish merging galaxies from members of close pairs. Each histogram is normalised to the number of galaxies in the corresponding subsample, given in the legend; the bin width is 0.25~mag.}
    \label{fig:segm_showcase1}
\end{figure*}

The distributions of galaxies in these samples by visual magnitude are presented in \autoref{fig:segm_showcase1}, \autoref{fig:segm_showcase2}. Description of \autoref{tab:VI}: 
Column 1 gives the morphological type or feature assigned during visual inspection; Columns 2–4 list the total number of galaxies $N_{total}$, the GZ2-matched objects $N_{GZ2}$, and the galaxies originating from the inference sample $N_{Inf}$, respectively.

 \begin{table}[h!]
    \caption{Visual classification of SDSS galaxies with morphological features at $z$<0.1 initially assigned by the CNN model \citep{Vavilova2022Cat}.}
    \centering
    \begin{tabular}{|l|>{\centering\arraybackslash}p{1.4cm}|>{\centering\arraybackslash}p{1.4cm}|>{\centering\arraybackslash}p{1.4cm}|}
        \hline
        Type or       &  $N_{total}$ & $N_{GZ2}$  & $N_{Inf}$   \\
        feature       &              &           &     \\ \hline
\multicolumn{4}{|l|}{Panel A. Merging, $CNN_{total}$ score $5-95.49$ \%}    \\ \hline
merging       & 612          &   429     &  183    \\
pair          & 581          &   399     &  182    \\
no merging    & 1381         &   1008     &  373    \\ \hline
\multicolumn{4}{|l|}{Panel B. Irregular, $CNN_{total}$ score $5-70.09$ \%}    \\ \hline
        irregular    & 9372          &  3015     &  6357    \\
        spiral       & 58            &  47       &  11    \\
        elliptical   & 1             &  1        &  -    \\
        ring         & 1             &  -        &  1    \\ \hline
\multicolumn{4}{|l|}{Panel C. Edge-on, $CNN_{total}$ score $22.30-98.21$ \%}    \\ \hline
        edge-on      & 16822         & 10902     &  5920    \\
        irregular    & 68            &  31       &  37    \\
        elliptical   & 66            &  39       &  27    \\
        spiral       & 22            &  18       &  4    \\
        merging      & 10            &  6        &   4   \\
        no edge-on   & 8             &  4        &  4    \\
        artifacts    & 3             &  -        &  3    \\
        PRG          & 1             &  1        &    -  \\ \hline
\multicolumn{4}{|l|}{Panel D. Dust lane, $CNN_{total}$ score $5.03-85.59$ \%}    \\ \hline
        dust lane    & 575           &  512      &  63   \\
        no dust lane & 13            &  8        &  5    \\ \hline
\multicolumn{4}{|l|}{Panel E. Bar, $CNN_{total}$ score $20.85-99.23$ \%}    \\ \hline
        strong bar   & 332           & 300       & 32  \\
        bar          & 479           & 427       & 52   \\
        elliptical   & 1060          & 963       & 97   \\
        irregular    & 1837          & 1397      & 440  \\
        PRG          & 4             & 4         & -  \\
        no bar       & 611           & 513       & 98 \\ \hline
\multicolumn{4}{|l|}{Panel F. Ring, $CNN_{total}$ score $5-5.28$  \%}    \\ \hline
        strong       & 360           & 343       & 17  \\
        good         & 1274          & 1191      & 83   \\
        thin         & 511           & 443       & 68  \\
        PRG          & 4             & 4         &  -  \\
        no ring      & 11733         & 10555     & 1178  \\ \hline
    \end{tabular}
 \begin{minipage}{\linewidth}
\smallskip
{The catalogue purity, corresponding contamination fraction, and  uncertainty in the purity of the CNN-selected catalogues are presented in Section  \ref{subsec:purity}. }
\end{minipage}
    \label{tab:VI}
\end{table}

To support our primary goal, we verify these galaxy samples and create new catalogues of galaxies with assigned morphological features by discussing the results of visual inspection and the specific errors the CNN makes in recognising these features during image processing. Every galaxy in the six selected inspection samples was independently examined by at least two authors, irrespective of its feature score within the corresponding inspected score range; disagreements were resolved jointly to reach a consensus classification.

The visually verified samples obtained in this way constitute the basis for all analyses presented in the following sections.

\subsection{Optical emission-line diagnostics for CNN catalogues}
\label{subsec:spectral_diagnostics}

To characterise the ionised-gas and nuclear properties of the visually confirmed galaxies, we matched the morphological catalogues to the MPA--JHU emission-line measurements \citep{Brinchmann2004,  Tremonti2004}%
\footnote{\url{https://www.sdss4.org/dr17/spectro/galaxy_mpajhu/}} accessed through SDSS DR17 \citep{Abdurrouf2022}. The extended data set contains 49\,032 catalogue entries across the six visually inspected CNN candidate samples, the close-pair catalogue, and the companion table; these entries are not necessarily unique galaxies because objects may occur in more than one morphological sample. A spectrum was accepted when it lay within 2~arcsec of the catalogue position, was the primary science spectrum of an object classified as a galaxy with \texttt{zWarning}=0, and had a redshift consistent with the catalogue value. Throughout this paper, a spectrum is called \emph{usable} for a given diagnostic when every emission line that diagnostic requires has a finite positive flux measured with S/N$\geq3$; this is the definition behind each subsample quoted in Sect.~\ref{sec:nuclear_spectral}.

We applied two complementary diagnostics. The BPT diagram \citep{Baldwin1981} combines $\log([\mathrm{N\,II}]\lambda6584/\mathrm{H}\alpha)$ and $\log([\mathrm{O\,III}]\lambda5007/\mathrm{H}\beta)$ and therefore requires four emission lines. The fluxes were corrected for internal attenuation using the observed Balmer decrement, and star-forming, composite, Seyfert, and LINER-like classes were separated following \citet{Kauffmann2003}, \citet{Kewley2001}, and \citet{Schawinski2007}. The WHAN diagram \citep{Fernandes2011} uses the same abscissa, rest-frame H$\alpha$ equivalent width, and therefore requires only two lines, recovering galaxies with weak H$\beta$ or [O\,\textsc{iii}] emission that the BPT selection discards. Its retired-like class identifies objects whose low equivalent widths are consistent with ionisation by evolved stellar populations, which can reproduce LINER-like emission-line ratios without an active nucleus \citep{Belfiore2016}.

Classification uncertainties were assessed with 3000 Monte Carlo realisations of the measured emission-line fluxes, and the whole classification was repeated with a stricter S/N$\geq5$ threshold as a sensitivity test. The complete specification --- the spectroscopic matching procedure and its quality flags, the attenuation correction and its coefficients, the Monte Carlo and error-propagation scheme, the equivalent-width convention, and the dependence of the spectroscopic coverage on apparent magnitude and redshift --- is given in Appendix~\ref{app:spectra}.

\section{CNN-assisted expert validation for merging, irregular, and edge-on galaxies.}

\subsection{Merging galaxies}
The CNN classifier identified 2\,574 candidate merging galaxies with feature scores in the range $5\%<CNN< 95.49\%$. All candidate galaxies were subjected to complete visual inspection (\autoref{tab:VI}, Panel~A). Following visual examination, the sample was divided into three categories: genuine merging galaxies, galaxy pairs potentially undergoing interaction, and galaxies showing no evidence of merging. In total, 612 galaxies were confirmed as merging systems, 581 galaxies were classified as members of close pairs or small groups, and 1\,381 galaxies showed no signatures of merging.

\textbf{Genuine merging galaxies.} Visual inspection confirmed 612 galaxies as genuine merging systems. Their feature scores span the entire interval $ 5.02\%<CNN< 95.49\%$, whereas only nine galaxies have feature scores exceeding 60 \%. Among the confirmed mergers, 429 objects were already recognised by the GZ2 project, while 183 galaxies are absent from the GZ2. The distribution of apparent SDSS $r$-band magnitudes for the confirmed mergers is presented in the left-hand panel of \autoref{fig:segm_showcase1}.

\begin{figure}[h]
    \centering
    \subfloat[\centering $RA:162.7513$, $DEC:12.2873$, $z=0.036$, $CNN=26.52\%$, GZ;  bright foreground star is projected onto galaxy image]
    {\includegraphics[width=4cm]{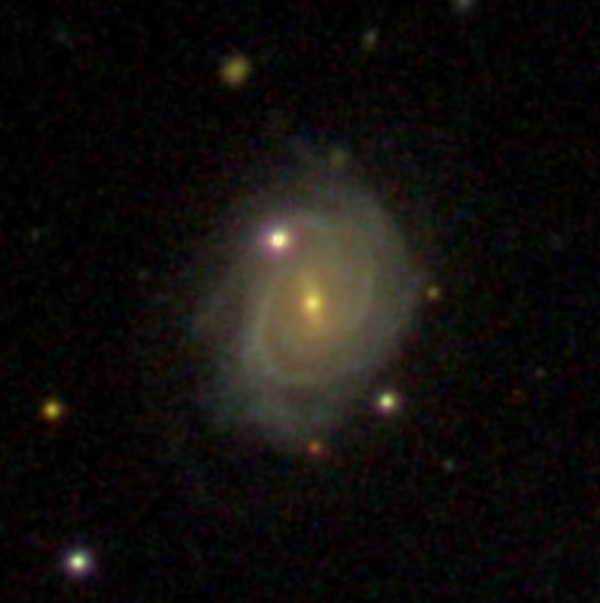}}
    \qquad
    \subfloat[\centering $RA:196.4065$, $DEC:60.1726$, $z=0.060$, $CNN=34.04\%$, GZ;  galaxy in projection $RA:196.4015$, $DEC:60.1751$, $z_{photo}=0.237$]
    {\includegraphics[width=4cm]{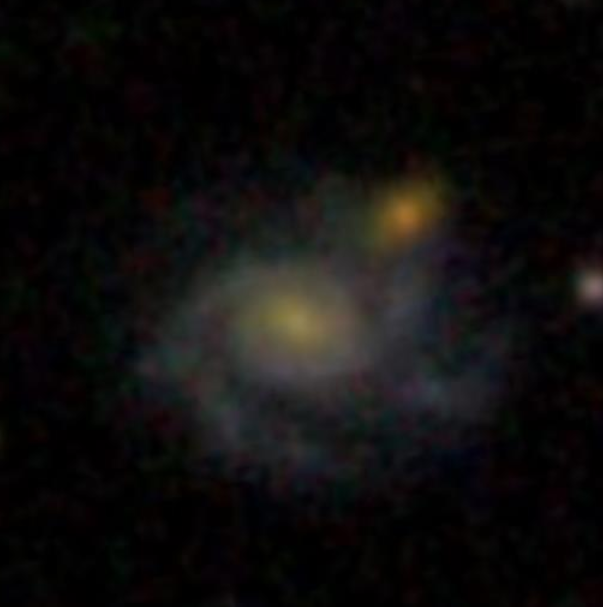}}
    \caption{Examples of galaxies classified by the CNN as mergers but rejected after visual inspection. The photometric redshift quoted in panel (b) is the SDSS KD-tree estimate \citep{Beck2016}.}
    \label{fig:no_merging}
\end{figure}

\textbf{Galaxy pairs.} Visual inspection additionally identified 581 galaxies appearing as members of close pairs or small groups. These objects comprise both spectroscopically confirmed systems and pair candidates for which spectroscopic confirmation of the companions is unavailable (Table~\ref{tab:VI}, Panel~A). A companion was accepted when its rest-frame velocity difference from the primary, $\Delta V = c\,|z_1 - z_2|/(1+\bar{z})$, satisfies $|\Delta V| \le 500$~km\,s$^{-1}$, the threshold standard in spectroscopic close-pair studies \citep{Patton2000, Lambas2003, Ellison2008}. All primary galaxies have SDSS spectroscopic redshifts by construction; the parent sample of \citet{Vavilova2021} is selected in $0.003 < z < 0.1$, and 54.7 per cent of the companions also have spectroscopic redshifts. For the other companions, the criterion cannot be applied, since $|\Delta V| \le 500$~km\,s$^{-1}$ corresponds to $\Delta z \simeq 0.0017$, an order of magnitude below the median photometric-redshift uncertainty of 0.013 \citep{Beck2016}; their photometric redshifts were used only to reject gross line-of-sight superpositions such as the galaxy in Fig.~\ref{fig:no_merging}b, and systems whose companions all lack spectroscopy are treated as pair candidates.
 
The 581 galaxies form 539 systems. Of these, 322 are spectroscopically confirmed: 281 pairs, 37 triplets, one quartet and three quintets. The remaining 217 are pair candidates: 196 pairs, 20 triplets and one quartet. A system with several companions is counted as confirmed when at least one of them satisfies the velocity criterion. The number of galaxies exceeds the number of systems because both members are listed when both satisfy the parent-sample selection. Companions lie at a median projected distance of 14.4~kpc from their primaries (interquartile range 10.8--18.2~kpc), and for the 334 companions with spectra the median velocity difference is 156~km\,s$^{-1}$. The catalogue lists $\theta$, $r_{\rm p}$ and $\Delta V$ for every companion, so that other pair definitions can also be applied.

Combining the galaxy-pair and non-merging categories, 1\,407 GZ2-matched galaxies selected as merger candidates by the CNN were not confirmed as genuine mergers during our visual inspection.

\textbf{Principal sources of CNN misclassification.} The remaining 1\,381 galaxies show no observable merger signatures despite being selected by the CNN. Most have merger-feature scores in the interval $5.01\%<CNN<48\%$. We note that 1\,008 of these non-merging galaxies belong to the GZ2-matched subset, illustrating that CNN-selected merger candidates may lack visually detectable merger signatures even when GZ2 classifications are available. Examples of typical false-positive classifications are presented in \autoref{fig:no_merging}. The most common failure modes are:

-- bright foreground stars projected onto galaxy images, producing structures that resemble interacting systems (\autoref{fig:no_merging}a);
-- projection of a distant background galaxy onto a spiral arm, creating an apparent merger morphology (\autoref{fig:no_merging}b);
-- spiral galaxies whose arm geometry mimics tidal structures, particularly when the detectability of low-surface-brightness tidal features in the SDSS images is limited.

Our results agree well with \citet{Bickley2024}, who showed that many CNN classification errors arise from sensitivity to faint tidal-like structures and projection effects. In relatively shallow surveys such as SDSS, post-merger signatures often become indistinguishable from ordinary galaxy morphology, leading to intrinsic ambiguity. These authors showed that increasing image depth beyond a limiting $5\sigma$ point-source magnitude of about $25^{m}$ yields only marginal improvement, suggesting that CNN merger identification is ultimately limited by morphological degeneracy rather than by the limited detectability of low-surface-brightness tidal features in SDSS images alone.

The statistics above show that expert visual inspection remains essential for building reliable catalogues of interacting galaxies and for understanding the physical origin of CNN classification errors (see also \autoref{fig:segm_showcase1}).

\subsection{Irregular galaxies} 
The CNN classifier identified 9\,432 candidate irregular galaxies with feature scores in the range $5\%<CNN<70.09\%$. All candidate galaxies were subjected to complete visual inspection (\autoref{tab:VI}, Panel~B). Following visual inspection, 9\,372 galaxies were confirmed as genuinely irregular, whereas only 60 objects were reclassified as non-irregular galaxies. This corresponds to a confirmation rate of 99.4 \%, making irregular galaxies the most reliably recognised morphological class among all six categories analysed in this work.

\textbf{Confirmed irregular galaxies.} Visual inspection confirmed 9\,372 galaxies as irregular systems, demonstrating the excellent performance of the CNN for this morphological class. Of the 9\,372 confirmed irregular galaxies, 3\,015 belong to the GZ2-matched subset, while 6\,357 are absent from the GZ2 catalogue. The distribution of apparent SDSS $r$-band magnitudes for the visually confirmed irregular galaxies is shown in the central panel of \autoref{fig:segm_showcase1}. The remarkably high confirmation rate indicates that the CNN successfully recognises the complex, diverse morphologies characteristic of irregular galaxies despite their wide structural diversity.

\textbf{Misclassified galaxies and CNN limitations.} Only 60 galaxies selected by the CNN were rejected after visual inspection. All rejected objects have irregular-feature scores in the interval $5.0\%<s<26.0\%$. Among the rejected galaxies, one is elliptical, one is a ringed spiral, and the rest are spirals. The most frequent source of misclassification is asymmetric spiral arms and bright star-forming regions that resemble the chaotic morphology typical of irregular galaxies. Such features occasionally confuse the CNN, particularly when they dominate the optical appearance of otherwise regular spiral galaxies.

\textbf{Comparison with previous studies.} \citet{Cavanagh2021} showed that irregular galaxies remain substantially more difficult to distinguish from late-type spirals than elliptical and spiral galaxies, while \citet{Aguilar2025} likewise reported increased classification uncertainty for transitional Sd–Irr systems. Our visual verification demonstrates that, despite these known challenges, the CNN identifies irregular galaxies with high reliability in the present sample.

\begin{figure}[h!]
    \centering
    \subfloat[\centering $RA:130.2441$, $DEC:34.9938$, $z=0.077$, $CNN=18.03\%$]
    {{\includegraphics[width=4cm]{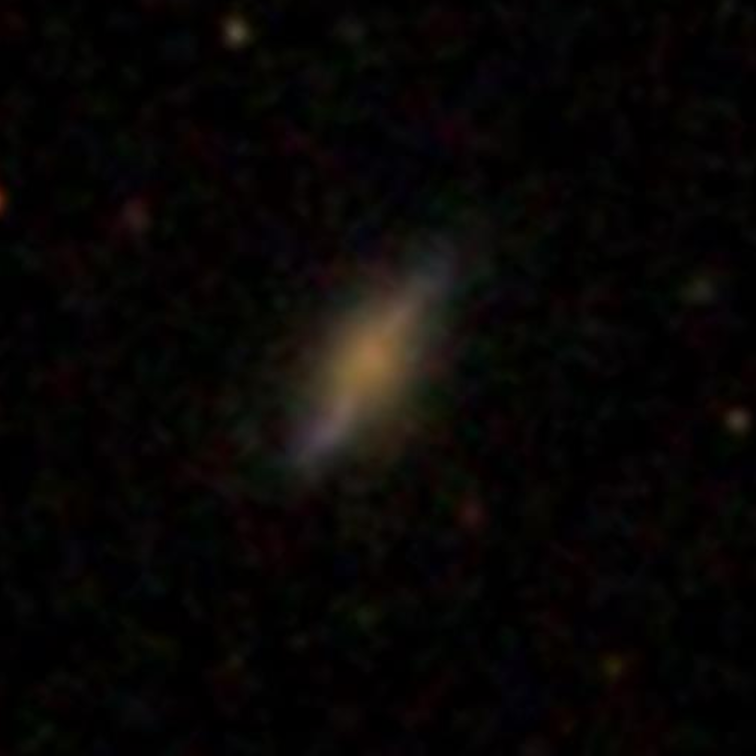}}}
    \caption{Polar-ring galaxy discovered during the visual inspection of galaxies classified by the CNN as irregular.}
    \label{fig:PRG_irr}
\end{figure}

Enlarging the training sample of irregular galaxies and incorporating additional photometric information, particularly colour indicators tracing recent star formation, may help reduce confusion between irregular and late-type spiral galaxies, although the effectiveness of these improvements remains to be tested. Among the visually inspected galaxies, we also identified a previously unrecognised polar-ring galaxy candidate, shown in \autoref{fig:PRG_irr}. This object illustrates that visual verification of CNN-selected samples may reveal morphologically rare systems beyond the primary target class. 

\subsection{Edge-on galaxies} 
The CNN classifier identified 34\,420 candidate edge-on galaxies with feature scores ranging from $5\%<CNN<98.21\%$. For the present study, we visually inspected the 17\,000 candidates with the highest edge-on feature scores ($22.30\%<CNN<98.21\%$), corresponding to the subset most likely to contain genuine edge-on systems (\autoref{tab:VI}, Panel~C).

\textbf{Confirmed edge-on galaxies.} Visual inspection confirmed 16\,822 galaxies as genuine edge-on systems, corresponding to a confirmation rate of almost 99\% for the inspected sample. Of these, 10\,902 galaxies belong to the GZ2-matched subset, while 5\,920 are absent from the GZ2 catalogue. The distribution of apparent SDSS $r$-band magnitudes for the confirmed edge-on galaxies is presented in the right-hand panel of \autoref{fig:segm_showcase1}. The high confirmation rate shows that the CNN reliably identifies edge-on morphology among the highest-scoring candidates, consistent with their distinctive projected geometry; however, this result should not be extrapolated to the full edge-on candidate catalogue without additional validation.

\textbf{Misclassified galaxies and CNN limitations.} Only 178 galaxies were rejected after visual inspection. These objects comprise several morphological categories that can mimic the elongated appearance of edge-on galaxies:

-- 68 irregular galaxies with elongated morphologies;
-- 66 flattened elliptical galaxies;
-- 22 spiral galaxies observed at intermediate inclinations between face-on and edge-on;
-- 10 merging galaxies forming elongated tidal structures;
-- 8 galaxies that are not edge-on;
-- 3 images affected by observational artefacts;
-- one previously known polar-ring galaxy included in the catalogue of \citet{Moiseev2011}.

These misclassifications show that the CNN primarily fails when different galaxy types have similar projected shapes or when observational effects alter the apparent morphology.

\textbf{Astrophysical importance of edge-on galaxies.} Edge-on galaxies provide a unique opportunity to investigate the vertical structure of galactic discs. Their orientation allows direct observations of the thin disc containing young stars, gas, and active star-forming regions, together with the thicker stellar disc composed predominantly of older, $\alpha$-enhanced stellar populations \citep{Yoachim2005, Somawanshi2024}. Combining edge-on galaxies with face-on galaxies of similar morphological type enables a more complete reconstruction of disc formation and evolutionary history.

Recent observations by \citet{Tsukui2025} identified both thick and thin stellar discs already at look-back times exceeding 10 Gyr. Their analysis shows that disc radial size and vertical scale height correlate with galaxy mass and suggests that thick discs formed first, followed by the growth of embedded thin discs. These results support a scenario of prolonged disc evolution and highlight the importance of homogeneous samples of edge-on galaxies for investigating the formation history of galactic discs.

Our visually verified catalogue therefore provides a homogeneous sample of edge-on galaxies suitable for future studies of disc structure, stellar populations, and the formation history of galactic discs.

\section{CNN-assisted expert validation for galaxies with dust lanes, bars, and rings. Results and Discussion.} 

\subsection{Galaxies with dust lanes}
\begin{figure*}[h]
\centering    
    \includegraphics[width=\textwidth]{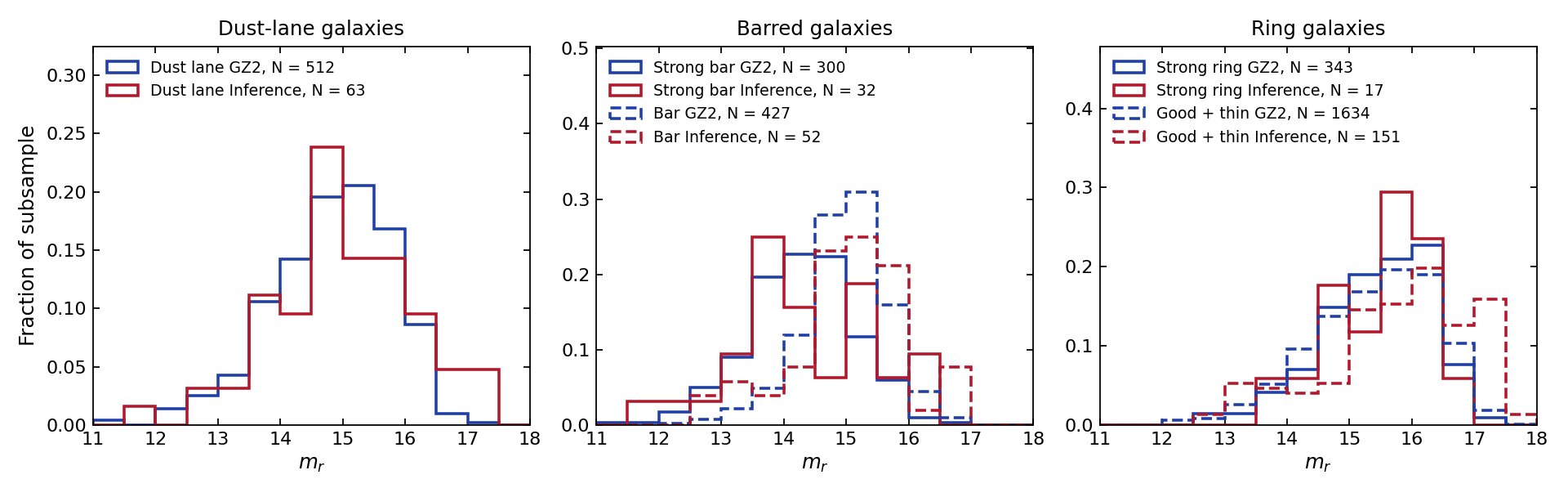}
\caption{Distribution of the apparent SDSS $r$-band magnitudes of visually verified galaxies with dust lanes (left), bars (centre), and rings (right). Blue and red histograms correspond to the GZ2-matched and inference subsamples; solid and dashed lines distinguish strong bars from bars and strong rings from good and thin rings. Each histogram is normalised to the number of galaxies in the corresponding subsample, given in the legend; the bin width is 0.5~mag.}
\label{fig:segm_showcase2}
\end{figure*}

The CNN classifier identified 588 candidate galaxies with dust lanes having feature scores in the range $5.03\%<CNN<85.59\%$. All candidates were subjected to complete visual inspection (\autoref{tab:VI}, Panel~D). Following visual inspection, 575 galaxies were confirmed as genuine dust lanes, whereas only 13 galaxies were reclassified as not containing dust lanes.

\textbf{Confirmed galaxies with dust lanes.} Visual inspection confirmed 575 galaxies with dust lanes, corresponding to a confirmation rate of approximately 98\%. Remarkably, 543 of the 575 confirmed galaxies have dust-lane feature scores in the interval $5.03\%<CNN<50\%$. Thus, a high score threshold would exclude many genuine dust-lane systems, demonstrating the importance of visually inspecting candidates across a broad range of CNN scores. This result indicates that the CNN confidence score is not necessarily a direct measure of classification reliability for dust lanes. Confirmation is therefore not a monotonic function of the score for this feature, and a score-based cut would have removed a large number of genuine dust-lane systems. This behaviour demonstrates that CNN confidence should not be interpreted as a direct measure of classification reliability for morphologically complex features.

\begin{figure}[h!]%
    \centering
    \subfloat[\centering $RA:233.6886$, $DEC:5.8294$, $z=0.053$, $CNN=85.59\%$, GZ]{{\includegraphics[width=4cm]{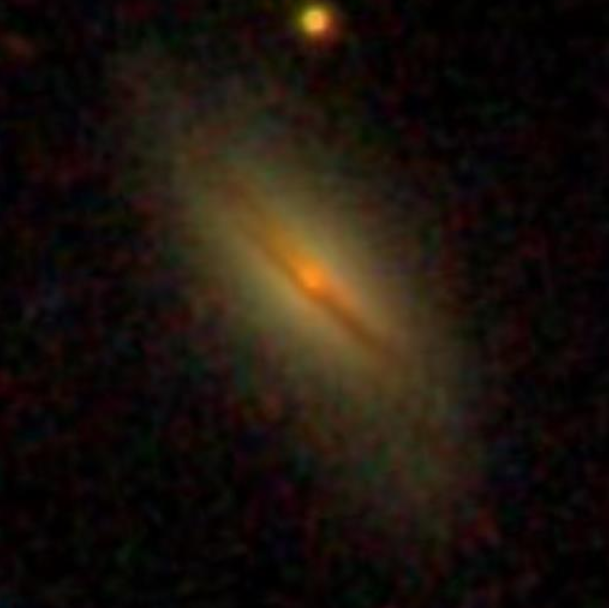} }}%
    \qquad
    \subfloat[\centering $RA:19.5182$, $DEC:14.5563$, $z=0.053$, $CNN=5.03\%$, Inf.]{{\includegraphics[width=4cm]{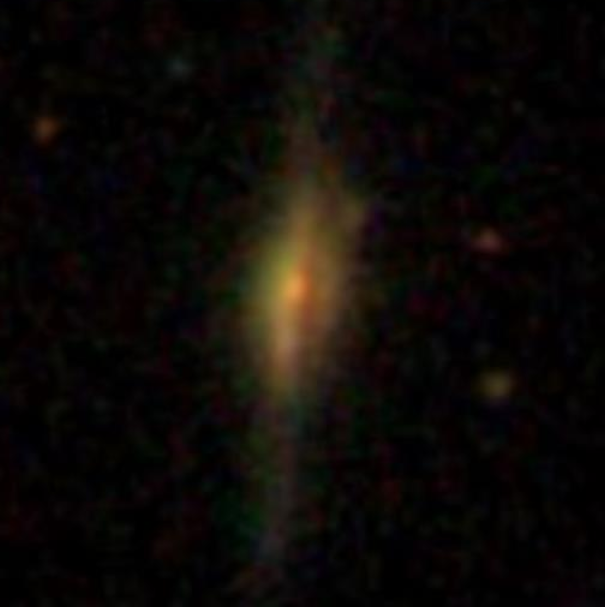} }}%
    \qquad
    \subfloat[\centering $RA:204.7691$, $DEC:2.1637$, $z=0.023$, $CNN=5.98\%$, GZ  \cite{Moiseev2011}]{{\includegraphics[width=4cm]{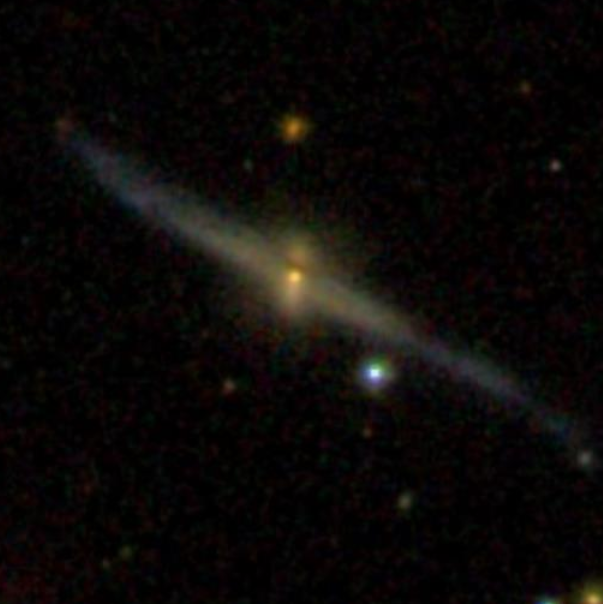} }}%
    \qquad
        \subfloat[\centering
    $RA:181.1798$, $DEC:60.6724$,
    $z=0.053$, $CNN=5.45\%$, GZ]
    {{\includegraphics[width=4cm]{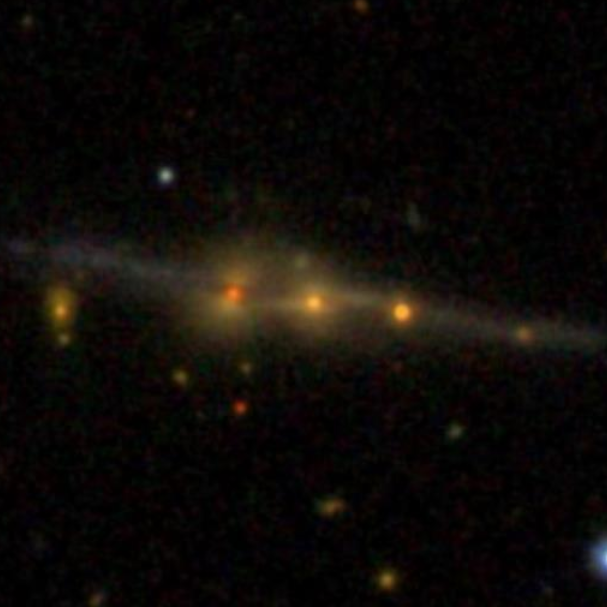}}}%
    \caption{SDSS \textit{gri} composite images of four
    galaxies confirmed as dust-lane systems, with their
    coordinates, redshifts, dust-lane feature scores and
    subsample flags. Panel (c) is a known polar-ring
    galaxy from \citet{Moiseev2011}; panel (d) is a
    new polar-ring candidate.}    
    \label{fig:dustlane}
\end{figure}

Of the 575 confirmed dust-lane galaxies, 512 are in the GZ2-matched subset, while 63 are absent from the GZ2 catalogue. The distribution of apparent SDSS $r$-band magnitudes for the confirmed sample is presented in the left-hand panel of \autoref{fig:segm_showcase2}.

Two particularly interesting objects deserve special attention. One galaxy (\autoref{fig:dustlane}c) is a previously known polar-ring galaxy included in the catalogue of \citet{Moiseev2011}, whereas another object (\autoref{fig:dustlane}d) appears to be a promising new candidate polar-ring galaxy proposed by \citet{Dobrycheva2025}. These examples illustrate that visual verification of CNN-selected samples can reveal morphologically rare systems extending beyond the primary target class.

\textbf{Misclassified galaxies and CNN limitations.} Only 13 galaxies selected by the CNN were rejected after visual inspection. All of them have very low feature score values ($5.05\%<CNN<12.25\%$), demonstrating that the network assigned low confidence to these ambiguous cases. In five of these, projected foreground red stars from the Milky Way overlap the galaxy image, producing structures that mimic dust lanes and likely causing the incorrect CNN classification. The low false-positive fraction indicates that the CNN-selected dust-lane sample has high confirmation purity within the adopted feature-score range, although this does not establish the completeness of dust-lane identification in the parent galaxy sample.

\textbf{Astrophysical importance of dust lanes.} Dust lanes are not merely morphological features but provide important diagnostics of disc stability, gas dynamics, and the secular evolution of galaxies. In edge-on spiral galaxies, dust lanes reflect the gravitational stability of galactic discs by tracing the balance between turbulent energy generated by star formation and the gravitational force acting toward the disc mid-plane \citep{Holwerda2019}. These authors demonstrated that prominent dust lanes become common above stellar masses of approximately $M_\ast\sim10^9\,M_\odot$. Their occurrence shows little dependence on disc thickness or Sérsic index but correlates with bulge morphology.

The origin of dust lanes is closely linked to the multiphase structure of the interstellar medium. \citet{Anderson2020} showed that the material responsible for the observed dust lanes is not associated exclusively with either the cold diffuse medium or the heated gas within H\,{\sc ii} regions, but instead results from a combination of both components along the line of sight.

Dust lanes also play an important role in nuclear activity. Although \citet{Butterfield2024} concluded that nuclear dust lanes do not significantly influence the growth rates of nearby active galactic nuclei ($z<0.01$), they nevertheless constitute efficient channels for transporting gas toward galactic centres. This mechanism is particularly well illustrated in the Milky Way. Numerical and observational studies by \citet{Sormani2019, Sormani2019b} demonstrated that the Galactic bar funnels gas through two prominent dust lanes from radii of approximately $R\simeq3$~kpc toward the Central Molecular Zone at $R\simeq200$~pc. Part of this gas accumulates in the central region, whereas another fraction overshoots the nucleus and collides with the dust lane on the opposite side. Such processes highlight the fundamental role of dust lanes in redistributing interstellar matter within barred galaxies and sustaining secular evolution \citep{Su2024}. Consequently, galaxies with dust lanes constitute valuable laboratories for studying gas transport, secular evolution, and the fuelling of galactic nuclei.

\subsection{Galaxies with bars}
The CNN classifier identified 29\,892 candidate barred galaxies with feature scores in the range $5\%<CNN<99.23\%$ \citep{Khramtsov2022}. Candidates were ranked by their independent bar-feature score and inspected in decreasing-score order. In the present study we visually inspected 4\,323 candidates within the bar-feature score interval $20.85\% < CNN < 99.23\%$ (\autoref{tab:VI}, Panel~E).

\textbf{Visual morphological classification.} During visual inspection, galaxies were assigned to five principal categories according to their observed morphology and the presence or absence of a stellar bar; four additional objects were identified as polar-ring galaxy candidates and recorded separately in Table~\ref{tab:VI} (Panel~E):

-- Strong bar --- spiral galaxies exhibiting a prominent, well-developed stellar bar;
-- Bar --- spiral galaxies with a clearly identifiable bar of moderate strength;
-- Elliptical --- elliptical galaxies in which the central light concentration was mistakenly recognised by the CNN as a bar;
-- Irregular --- irregular galaxies whose asymmetric central morphology resembles a bar-like structure;
-- No bar --- galaxies without convincing evidence for a stellar bar or systems in which the presence of a bar cannot be established reliably.

\textbf{Confirmed barred galaxies.} The primary objective of this analysis is to construct a reliable catalogue of barred galaxies, which is particularly important for the subsequent identification of Milky Way analogues, since the Galactic stellar bar represents one of the principal structural components that must be reproduced by close Milky Way counterparts such as Milky Way Near Twins (MWNeTs, \cite{Vavilova2026, Kompaniiets2026}). Of the 811 confirmed barred galaxies, 727 belong to the GZ2-matched subset, whereas 84 are absent from the GZ2 catalogue. The complete statistics are summarised in \autoref{tab:VI}. The distribution of apparent SDSS $r$-band mag. for the confirmed barred galaxies is shown in the panel of \autoref{fig:segm_showcase2}.

\textbf{CNN behaviour and classification limitations.} As the visual inspection progressed toward lower eature score values, approaching the limit of the inspected sample, we observed a gradual increase in the number of irregular galaxies mistakenly identified as barred systems. This trend suggests that weak or asymmetric central structures present in irregular galaxies are occasionally interpreted by the CNN as remnants of stellar bars. These results suggest that the CNN responds not only to genuine stellar bars but also to elongated or asymmetric central structures in morphologically disturbed galaxies. Such features may resemble weak bars in projection, although their physical nature cannot be established from the SDSS images alone.

\begin{figure}[h!]
    \centering
    \subfloat[\centering $RA:222.7625$, $DEC:-0.5271$, $z=0.043$, $CNN=33.66\%$]
    {{\includegraphics[width=4cm]{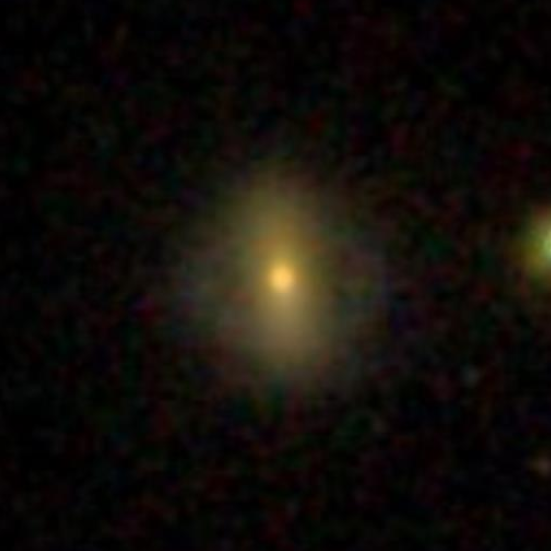}}}
    \qquad
    \subfloat[\centering $RA:203.443$, $DEC:29.9107$, $z=0.037$, $CNN=25.67\%$]
    {{\includegraphics[width=4cm]{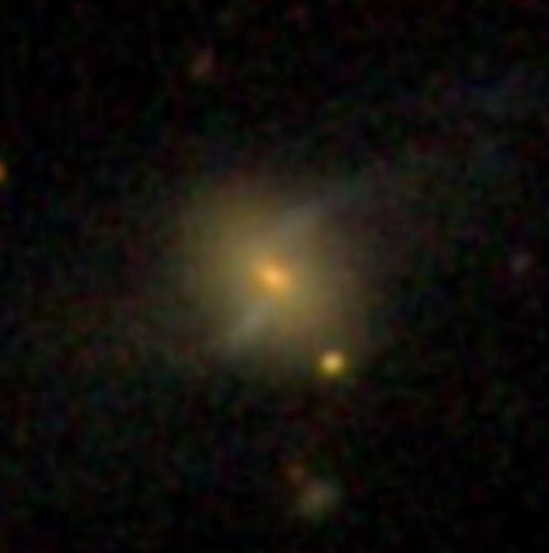}}}
    \qquad
    \subfloat[\centering $RA:15.3398$, $DEC:0.3842$, $z=0.093$, $CNN=21.99\%$]
    {{\includegraphics[width=4cm]{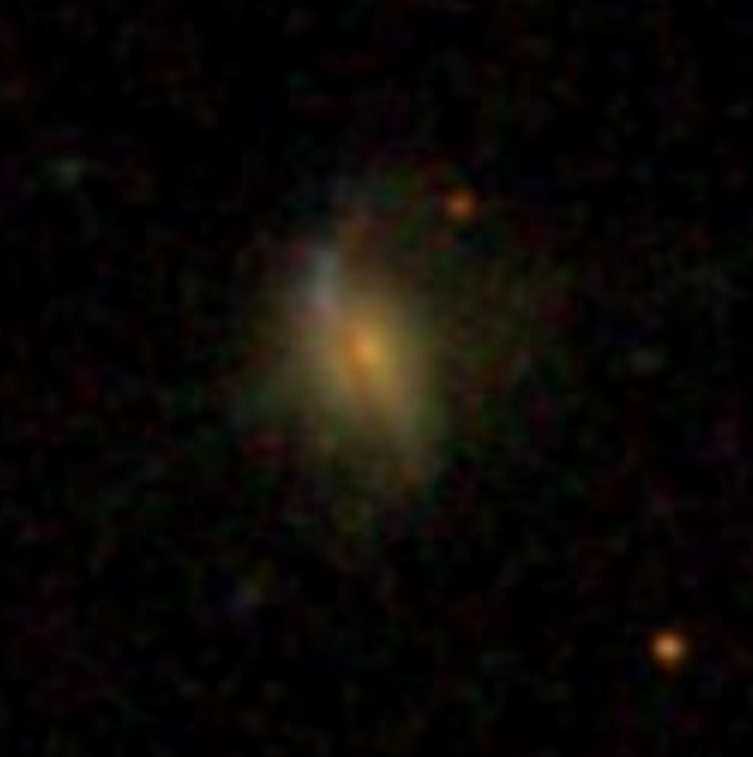}}}
    \qquad
    \subfloat[\centering $RA:208.1111$, $DEC:14.491$, $z=0.041$, $CNN=28.91\%$]
    {{\includegraphics[width=4cm]{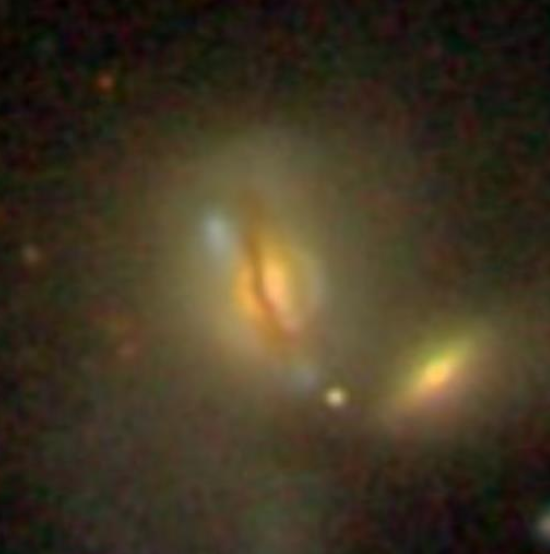}}}
    \qquad
    \subfloat[\centering $RA:238.2876$, $DEC:54.1472$, $z=0.047$, $CNN=29.22\%$]
    {{\includegraphics[width=4cm]{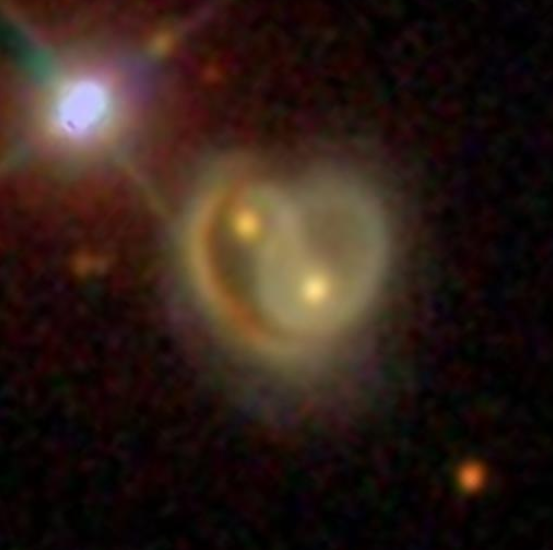}}}
    \qquad
    \subfloat[\centering $RA:227.7706$, $DEC:4.2939$, $z=0.042$, $CNN=22.17\%$]
    {{\includegraphics[width=4cm]{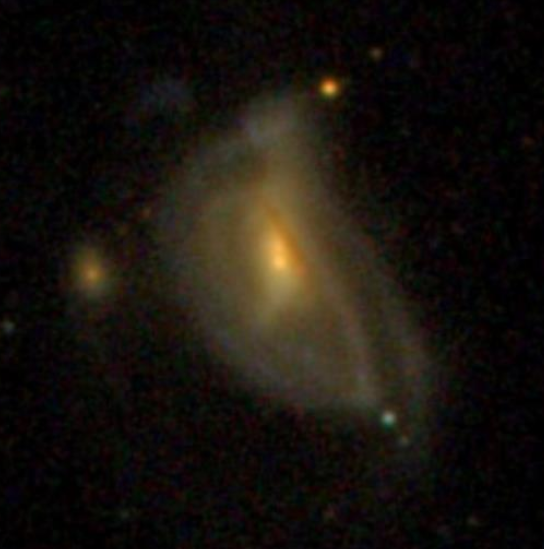}}}
    \caption{Representative examples of galaxies incorrectly classified by the CNN as barred systems.}
    \label{fig:bar}
\end{figure}

Examples illustrating the most common misclassifications are presented in Fig.~\ref{fig:bar}.

\subsection{Galaxies with rings}
Early attempts to identify ring galaxies in modern imaging surveys were largely based on automatic image analysis followed by human verification. In particular, \citet{Timmis2017} developed a computer-vision method for the first Pan-STARRS1 (PS1) data release by combining image smoothing, dynamic thresholding, and the search for \emph{closed regions} in binary masks. Applying this procedure to approximately $3\times10^{6}$ PS1 galaxy images produced a catalogue of 185 ring-galaxy candidates. The authors noted that the final sample strongly depends on image artefacts and on the strictness of the closed-region criterion, illustrating the need for more flexible machine-learning approaches to morphological classification. In our study \citep{Vavilova2022, Vavilova2022Cat, Khramtsov2022}, the CNN classifier identified 13\,882 candidate ring galaxies with feature scores in the range $5\%<CNN<95.28\%$. All candidates were subjected to complete visual inspection (\autoref{tab:VI}, Panel~F).

\begin{figure}[h]
    \centering
    \subfloat[\centering $RA:215.0675$, $DEC:30.2846$, $z=0.067$, $CNN=95.28\%$]
    {{\includegraphics[width=3cm]{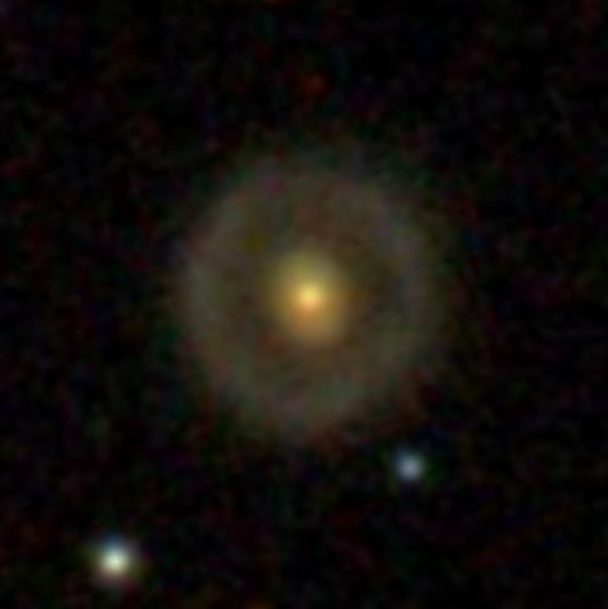}}}
    \subfloat[\centering $RA:327.7122$, $DEC:-0.8479$, $z=0.027$, $CNN=90.02\%$]
    {{\includegraphics[width=3cm]{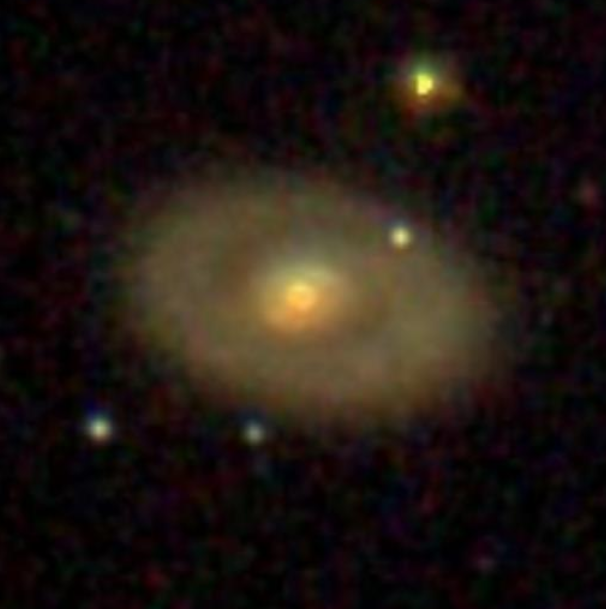}}}
    \subfloat[\centering $RA:206.6538$, $DEC:27.5982$, $z=0.078$, $CNN=88\%$]
    {{\includegraphics[width=3cm]{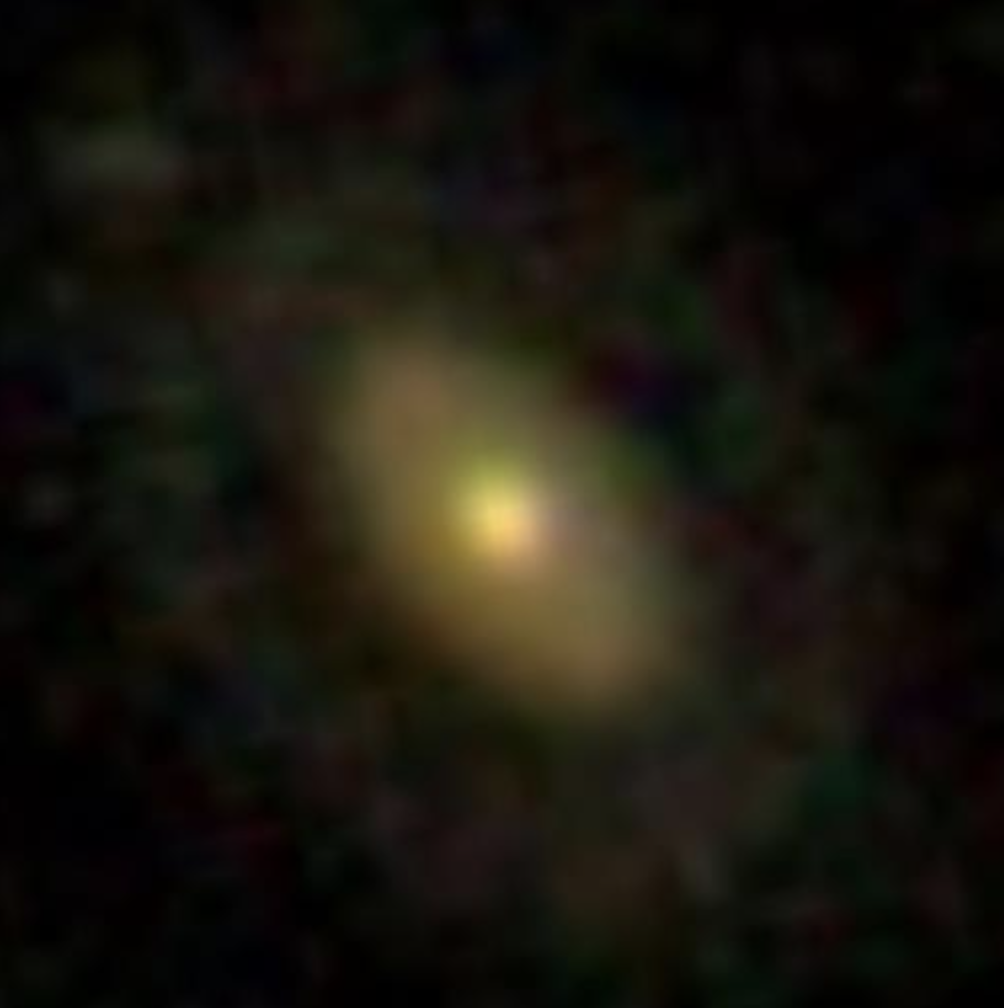}}}
    \caption{Examples of visually confirmed ring galaxies.}
    \label{fig:strong}
\end{figure}

\textbf{Confirmed ring galaxies.} Visual inspection confirmed 2\,145 galaxies as genuine ring systems, while 11\,733 were rejected as non-ring objects and four additional galaxies were classified separately as polar-ring galaxy candidates (Table~\ref{tab:VI}, Panel~F). Of the confirmed ring galaxies, 1\,977 belong to the GZ2-matched subset, whereas 168 are absent from the GZ2 catalogue. The distribution of apparent SDSS $r$-band magnitudes for the confirmed ring galaxies is presented in the right-hand panel of \autoref{fig:segm_showcase2}. Representative examples of visually confirmed ring galaxies are shown in Fig.~\ref{fig:strong}.

CNN behaviour and classification reliability. Among the 11\,733 galaxies rejected after visual inspection, only 235 objects have feature score exceeding $50\%$. This result shows that high ring-feature scores reliably indicate genuine ring morphology, whereas most false-positive identifications occur below $50\%$.

\textbf{Ring galaxies hosting active galactic nuclei.} 
As an illustrative application of the
visually verified ring-galaxy catalogue, we
examined positional associations with published
optical AGN and X-ray source catalogues. This
exploratory investigation is limited to the
ring-galaxy sample and is not intended to
establish differences in AGN incidence among
morphological classes. To investigate the occurrence of nuclear activity, we cross-matched the catalogue of visually confirmed ring galaxies with the Uniformly Selected All-sky Optical AGN Catalogue at $z<0.09$ \citep{Zaw2019}, which contains 1\,929 broad-line AGNs and 6\,562 narrow-line AGNs identified from optical spectroscopy. A total of 203 ring galaxies were found to host optical AGNs. These 203 positional matches provide an independent AGN-catalogue comparison rather than a measure of the AGN fraction derived from our BPT and WHAN diagnostics in Sect.~\ref{sec:nuclear_spectral}. The apparent decrease in the number of matched AGNs beyond $z=0.03$ may reflect selection effects and spectroscopic incompleteness, and should not be interpreted as evidence for an intrinsic redshift dependence without further analysis.

\textbf{X-ray counterparts.} To examine whether the optically identified AGNs exhibit high-energy emission, we cross-matched the sample with the 105-month {\textit Swift}/BAT hard X-ray survey \citep{Oh2018} and with the first SRG/eROSITA all-sky X-ray catalogue \citep{Merloni2024}. No X-ray counterparts were found within a search radius of 5 arcsec. A further cross-match with the Chandra Source Catalogue Release~2.0 \citep{Evans2019}, containing 317\,167 X-ray sources, yielded five positional matches: J173809.3+584253, J135317.7+332927, J161725.6+350809, J082705.9+213842, and J114612.1+20232. These five positional matches are promising targets for further investigation, although their association with nuclear activity requires verification of the X-ray source positions, host-galaxy centres, and available spectral information. The absence of counterparts in the other X-ray catalogues should be interpreted in the context of their different sensitivities, sky coverage, and selection functions \citep{Kompan2026}.

\subsection{Polar-ring galaxies}
One valuable outcome of the visual inspection is identifying several previously recognised and new candidate polar-ring galaxies (PRGs) across different morphological classes. These discoveries demonstrate that visual verification of CNN-selected samples is capable of revealing rare and astrophysically important galaxy types beyond the primary classification target. Among the galaxies classified by the CNN as irregulars, we identified one confirmed PRG (SDSS~J084058.57+345937.5; $RA=130.2441^\circ$, $DEC=34.9938^\circ$, $z=0.077$, $CNN=18.03\%$; \autoref{fig:PRG_irr}).

Among the galaxies with dust lanes, one previously known PRG (SDSS~J133904.58+020949.5; $RA=204.7691^\circ$, $DEC=2.1637^\circ$, $z=0.023$, $CNN=5.98\%$; \autoref{fig:dustlane}c) was recovered. This object is included in the PRG catalogue of \citet{Moiseev2011}. In addition, another galaxy (SDSS~J120443.17+604020.9; $RA=181.1798^\circ$, $DEC=60.6724^\circ$, $z=0.053$, $CNN=5.45\%$; \autoref{fig:dustlane}d) exhibits morphological characteristics that make it a promising new PRG candidate. 

Additional polar-ring galaxy candidates were identified among objects initially selected by the CNN as barred galaxies but subsequently assigned to the separate PRG category during expert visual inspection (\autoref{fig:bar}a--c; Table~\ref{tab:VI}, Panel~E). These objects will be incorporated into our catalogue of visually verified polar-ring galaxies compiled using a combination of visual inspection and machine-learning techniques \citep{Dobrycheva2025, Dobrycheva2025cat}. An especially interesting object was found among galaxies initially classified by the CNN as ring galaxies. Visual inspection revealed that this system is, in fact, a merging galaxy (\autoref{fig:bar}d--f). Its morphology appears consistent with a possible intermediate stage of polar-ring formation, making it a particularly attractive target for future observational studies of PRG formation and evolution.

The discovery of confirmed and candidate polar-ring galaxies across several independently selected morphological classes highlights a major advantage of combining CNN-based classification with expert visual inspection. Beyond improving catalogue reliability, this strategy naturally reveals rare, astrophysically important objects that would otherwise be overlooked.

\subsection{Overall purity and contamination of the CNN-selected catalogues.}
\label{subsec:purity}
Based on expert visual inspection of the CNN-selected morphological samples, including score-selected subsets for edge-on and barred galaxies, we quantified the confirmation purity and contamination of the inspected candidates in each morphological class. Namely, we calculated the catalogue purity as $P=TP/(TP+FP)=N_{\rm confirmed}/N_{\rm CNN}$, where $TP$ is the number of CNN-selected galaxies confirmed by visual inspection, $FP$ is the number of candidates for which the assigned morphological feature was not visually confirmed, and $N_{\rm CNN}=TP+FP$ is the total number of visually inspected CNN-selected candidates. The corresponding contamination fraction was calculated as $f_{\rm cont}=FP/(TP+FP)=1-P$. Assuming binomial statistics, the uncertainty in the purity was estimated as $\sigma_P=[P(1-P)/N_{\rm CNN}]^{1/2}$. Thus, $P$ measures the fraction of CNN-selected galaxies whose predicted morphological class was confirmed by expert visual inspection, whereas $f_{\rm cont}$ represents the fraction of CNN-selected candidates whose assigned morphology was not confirmed by expert inspection. Since only CNN-selected candidates were visually inspected, these quantities characterise catalogue purity and contamination but do not, by themselves, provide an estimate of absolute completeness. The resulting class-specific values are summarised in \autoref{tab:purity}.

Although the CNN was applied uniformly to the parent SDSS sample, the reported purity estimates refer to the visually inspected candidates and should not be assumed to be identical for the GZ2-matched and inference subsets without a separate comparison.

\begin{table*}[t]
\caption{Purity and contamination of the CNN-selected catalogues based on
expert visual inspection.}
\label{tab:purity}
\centering
\begin{tabular}{lrrrrrr}
\hline
CNN class &
$N_{\rm CNN}$ &
$TP$ &
$FP$ &
$P$ (\%) &
$f_{\rm cont}$ (\%) &
$\sigma_P$ (\%) \\
\hline
Merging (strict)  & 2\,574  & 612    & 1\,962  & 23.78 & 76.22 & 0.84 \\
Merging + pairs & 2\,574 & 1\,193 & 1\,381 & 46.35 & 53.65 & 0.98 \\
Irregular         & 9\,432  & 9\,372 & 60      & 99.36 & 0.64  & 0.08 \\
Edge-on           & 17\,000 & 16\,822& 178     & 98.95 & 1.05  & 0.08 \\
Dust lane         & 588     & 575    & 13      & 97.79 & 2.21  & 0.61 \\
Bar               & 4\,323  & 811    & 3\,512 & 18.76 & 81.24 & 0.59 \\
Ring              & 13\,882 & 2\,145 & 11\,737 & 15.45 & 84.55 & 0.31 \\
\hline
\end{tabular}

\begin{minipage}{0.96\linewidth}
\smallskip
\footnotesize
For the strict merger estimate, only visually confirmed genuine mergers were counted as true positives. The \textit{Merging + pairs} row gives an alternative broader estimate in which both genuine mergers and close galaxy pairs were treated as potentially interacting systems. These two rows are alternative interpretations of the same CNN-selected sample and should not be added together. For the bar class, the \textit{strong bar} and \textit{bar} categories were combined. For the ring class, the \textit{strong}, \textit{good} and \textit{thin} categories were combined; the four polar-ring candidates are catalogued separately and are not counted as confirmed rings here. The quoted uncertainties assume binomial statistics.
\end{minipage}
\end{table*}

The confirmation purity varies substantially among morphological classes and depends on the adopted candidate-selection strategy. The irregular and dust-lane samples, together with the high-score edge-on subset, show confirmation purities exceeding $97\%$. In contrast, the strict merger, bar, and ring samples have substantially lower confirmation purities, reflecting both morphological ambiguity and differences in the CNN-score distributions and inspected samples. Even for mergers, including close galaxy pairs as potentially interacting systems increases purity from $(23.78 \pm 0.84)\%$ to $(46.35 \pm 0.98)\%$.

\section{Types of Activity in Visually Confirmed Galaxies from the CNN Morphological Catalogues}
\label{sec:nuclear_spectral} 

We analysed the optical emission-line properties of all seven visually confirmed morphological samples: ringed, barred, dust-lane, edge-on, merging, irregular, and close-pair galaxies. The spectral classifications were obtained using the BPT and WHAN diagnostics described in Sect.~\ref{subsec:spectral_diagnostics}. Table~\ref{tab:bpt_whan_coverage} summarises the number of visually confirmed galaxies and the corresponding spectroscopic subsamples satisfying the adopted S/N$\geq3$ criteria.

\begin{table}[t]
\centering
\caption{Spectroscopic coverage of the seven visually confirmed morphological samples.}
\label{tab:bpt_whan_coverage}

\begin{tabular}{lrrr}
\hline\hline
Sample &
$N_{\rm visual}$ &
$N_{\rm BPT}$ (\%) &
$N_{\rm WHAN}$ (\%) \\
\hline
Ring       & 2\,145  & 1\,620 (75.5)  & 2\,062 (96.1) \\
Bar        & 811     & 702 (86.6)    & 775 (95.6)   \\
Dust lane  & 575     & 436 (75.8)    & 566 (98.4)   \\
Edge-on    & 16\,822 & 12\,276 (73.0) & 15\,613 (92.8) \\
Merging    & 612     & 458 (74.8)    & 520 (85.0)   \\
Irregular  & 9\,372  & 8\,409 (89.7) & 8\,728 (93.1) \\
Pairs      & 581     & 328 (56.5)    & 423 (72.8)   \\
\hline
Total      & 30\,918 & 24\,229        & 28\,687       \\
\hline
\end{tabular}

\vspace{1mm}
\parbox{\columnwidth}{%
\footnotesize\raggedright
\textit{Note.} $N_{\rm visual}$ is the number of visually confirmed galaxies; $N_{\rm BPT}$ and $N_{\rm WHAN}$ give the numbers satisfying the corresponding emission-line quality criteria at S/N$\geq3$.
Percentages are calculated relative to $N_{\rm visual}$ for each morphological sample. The morphological samples may overlap.
}
\end{table}

The totals in Table~\ref{tab:bpt_whan_coverage} represent morphological sample entries, with galaxies belonging to multiple morphological samples counted separately. Of the 24\,229 BPT-classified entries, 42.6\% belong to the CNN inference subsample and therefore carry no GZ2 morphological label; the inference contribution is particularly substantial for irregular (66.2\%) and edge-on galaxies (35.0\%). The uniform expert verification of both subsets enables a consistent analysis of excitation properties, including galaxies without GZ2 morphological labels that would be omitted from an analysis restricted to GZ2-matched objects.

The BPT distributions are shown in Fig.~\ref{fig:BPT_all}. LINER-like excitation is the most frequent class among ringed (52.5\%) and barred galaxies (43.6\%), whereas composite spectra dominate the dust-lane sample (47.2\%). Star-forming excitation prevails among edge-on galaxies (56.3\%), merging systems (42.8\%), and close pairs (45.7\%), and irregular galaxies show the strongest concentration in the star-forming region (99.6\%).

The complementary WHAN diagnostic (Fig.~\ref{fig:WHAN_all}) reveals substantial retired-like populations among ringed (64.0\%), barred (45.5\%), and dust-lane galaxies (42.9\%), while strong-AGN-like is the largest WHAN category among merging galaxies (42.5\%) and irregular galaxies remain overwhelmingly star-forming-like (99.2\%). The comparison between the two diagnostics is most informative for the galaxies with LINER-like BPT spectra: 87.8\% of the ringed and 83.0\% of the barred LINER-like galaxies are classified as retired-like by WHAN. These results indicate that a substantial fraction of LINER-like BPT classifications may arise from ionisation by evolved stellar populations rather than accretion-powered nuclear activity, emphasising that LINER-like excitation should not automatically be interpreted as evidence of an AGN.

Because both diagnostics require emission lines to be detected, the reported fractions characterise the spectroscopically selected subsamples and cannot be interpreted as intrinsic AGN incidence rates in the underlying galaxy populations. Their sensitivity to this selection is quantified in Appendix~\ref{app:completeness}: raising the threshold to S/N$\geq5$, for example, lowers the LINER-like fraction of ringed galaxies from 52.5\% to 40.2\% and that of barred galaxies from 43.6\% to 33.3\%.

\section{Milky Way galaxy analogues and Milky Way Near Twins.}

The Milky Way is a barred spiral galaxy with a thin and thick stellar disc, a boxy/peanut bulge, prominent dust lanes associated with the Galactic bar, and at least one inner ring or pseudo-ring structure \citep{Vavilova2026}. Placing the Milky Way into the broader population of disc galaxies requires identifying external systems that reproduce not only its global properties but also its characteristic structural components and evolutionary history. Such Milky Way analogues (MWAs) provide fundamental benchmarks for understanding how typical the Milky Way is among present-day spiral galaxies and for constraining its formation through mergers and secular evolution.

This section does not define a final sample of MWAs, but demonstrates how the visually verified catalogues presented here provide the morphological framework for future multi-parameter searches based on structural, photometric, spectroscopic, kinematic, and evolutionary diagnostics. 

\textbf{Merging galaxies.} Although the Milky Way is currently a relatively isolated galaxy, its structure reflects the cumulative effects of past merger events. Therefore, understanding the merger history of Milky Way-mass galaxies is essential for identifying systems that have followed evolutionary pathways similar to that of our Galaxy. \citet{Li2024} investigated how galaxy interactions affect star formation in Milky Way-mass galaxies using FIRE-2 simulations. They found significant torque--SFR correlations only for near-equal-mass interactions at $z=1.2$--3.6, whereas most interactions neither trigger starbursts nor substantially affect the long-term star-formation history.

Similarly, \citet{Ciucua2024} suggested that the Milky Way likely underwent a gas-rich merger that triggered a starburst episode, enriched its chemistry, and formed the Galactic thick disc. Their interpretation of the age--metallicity relation agrees with cosmological Auriga simulations. The possible role of the Gaia--Enceladus/Sausage merger in the formation of the Galactic bar has also been discussed by \citet{Merrow2024}. Consequently, selecting galaxies with comparable merger histories constitutes an important evolutionary constraint in future searches for physically meaningful MWAs.

\textbf{Edge-on galaxies} provide the closest observational proxy for how the Milky Way would appear to an external observer viewed edge-on. They allow direct comparison of the thin and thick discs, dust distribution, bulge prominence, and vertical galaxy structure. Our catalogue of 16\,822 visually inspected edge-on galaxies, including 10\,902 objects matched with GZ2 and 5\,920 galaxies absent from it, provides a statistically robust basis for selecting galaxies with Milky Way-like disc thicknesses, bulge properties, and dust extinction patterns \citep{Vavilova2024b}. These galaxies provide an important reference sample for comparing the projected vertical structure of disc galaxies with that expected for the Milky Way, although their edge-on orientation alone doesn't establish their MWA status.

\textbf{Galaxies with bars and dust lanes.} The Milky Way hosts a strong stellar bar accompanied by prominent dust lanes that efficiently channel gas toward the Galactic Centre \citep{Sormani2019, Sormani2019b, Su2024}. The simultaneous presence of these two structural components is one of our Galaxy's most distinctive morphological characteristics. Our visually verified catalogue contains 811 barred spiral galaxies (332 strongly barred and 479 barred systems) together with 575 galaxies exhibiting well-defined dust lanes. Cross-matching these visually verified catalogues may identify galaxies exhibiting a Milky Way-like bar--dust-lane configuration, providing morphological candidates for subsequent evaluation within the multi-stage MWNeTs framework.

\textbf{Galaxies with rings.} Dynamical models and observations indicate that the Milky Way hosts at least one inner ring or pseudo-ring associated with the Galactic bar and the molecular-gas distribution. Our catalogue of 2\,145 visually confirmed ring galaxies, including 850 LINER-like systems identified through BPT diagnostics and 218 weak-AGN-like systems classified through WHAN, enables us to investigate how common such structures are among nearby disc galaxies. Joint analysis of galaxies exhibiting bars, rings, and dust lanes provides additional constraints on the evolutionary stage occupied by the Milky Way within the general population of barred spirals. Unlike studies of our own Galaxy, external galaxies allow all these structural components to be observed simultaneously in projection, facilitating direct comparisons of their mutual geometry and evolutionary relationships.

Taken together, the visually verified catalogues presented in this work provide a morphological basis for future searches for Milky Way analogues and Milky Way Near Twins (MWNeTs) \citep{Vavilova2026, Vav2026, Izvuekova2026}. Within the seven-stage MWNeTs framework, morphological and structural compatibility is a mandatory selection stage and must be considered alongside environmental, nuclear, and global physical constraints. Additional multiwavelength, kinematic, and evolutionary diagnostics are required to assess evolutionary similarity to the Milky Way.

\section{Conclusions.}
This paper presents the final stage of our CNN-based morphological classification of SDSS galaxies: comprehensive visual verification of galaxies in six astrophysically important morphological classes: merging, irregular, edge-on, barred, dust-lane, and ring. The resulting catalogues substantially improve the reliability of automated classifications and provide homogeneous datasets for future studies of galaxy morphology, evolution, and nuclear activity. The principal results of this work can be summarised as follows.

-- We performed a complete visual inspection of all CNN-selected candidates and constructed homogeneous catalogues of visually verified galaxies. This two-stage procedure combines the scalability of automated selection with the reliability of expert validation, allowing false-positive classifications to be identified and removed from the final catalogues.

--  The quantitative assessment reveals that catalogue reliability is strongly class-dependent, with purity spanning from $(15.45 \pm 0.31)\%$ for ring galaxies to $(99.36 \pm 0.08)\%$ for irregular galaxies. Edge-on and dust-lane galaxies are also recovered with high purities of $(98.95 \pm 0.08)\%$ and $(97.79 \pm 0.61)\%$, respectively, whereas barred galaxies are the second least reliable class, $(18.76 \pm 0.59)\%$. For mergers, the result depends substantially on the adopted physical definition: the strict purity based only on visually confirmed mergers is $(23.78 \pm 0.84)\%$, but increases to $(46.35 \pm 0.98)\%$ when close galaxy pairs are treated as potentially interacting systems. This difference shows that automated morphological classification depends not only on classifier performance but also on the operational definition of the target class. The main sources of contamination are projection effects, foreground stars and saturated sources, low-surface-brightness structures, close companions, and intrinsically ambiguous morphologies. The complete class-specific statistics are presented in \autoref{tab:purity}.

-- Visual inspection revealed that the main CNN failure modes depend on the morphological class. The inspected high-score subset reliably recognises distinctive projected structures, particularly edge-on discs, whereas mergers, bars, and rings are harder to distinguish from morphologically similar systems. These results highlight the importance of expert verification when constructing catalogues of complex or dynamically evolving galaxy structures. 

-- Comparison with GZ2 demonstrates that CNN-assisted expert validation provides morphological classifications both for GZ2-matched galaxies and for additional SDSS galaxies without GZ2 morphological labels. The resulting catalogues therefore extend the available samples of visually verified galaxies at $z<0.1$.

-- Both diagnostics were applied homogeneously to all seven visually confirmed morphological samples --- 24\,229 BPT and 28\,687 WHAN classifications at S/N$\geq3$, with overlapping morphological memberships counted separately --- so that the excitation properties of the different classes are measured on a common basis. The BPT analysis reveals a substantial LINER-like contribution among ringed (52.5\%) and barred (43.6\%) galaxies, while irregular galaxies are predominantly star-forming (99.6\%). Among ringed galaxies, WHAN classifies 746 of 850 BPT-LINER-like systems (87.8\%) as retired-like, indicating a substantial contribution from evolved stellar populations to the observed low-ionisation emission. The reported excitation-class fractions refer to spectroscopically selected subsamples and do not directly measure the incidence of accretion-powered nuclear activity. The spectral classifications and their associated uncertainties are included in the extended catalogues.

-- During the visual inspection we identified five strong candidates for polar-ring galaxies: SDSS J084058.57+345937.5, SDSS J120443.17+604020.9, SDSS J145102.98$-$003137.4, SDSS J133346.32+295438.6, and SDSS J010121.58+002305.3. These objects will be included as morphological candidates in the updated polar-ring galaxy catalogue \citep{Dobrycheva2025cat}, pending further observational confirmation.

-- The visually verified catalogues presented here provide statistically significant samples for future investigations of galaxy morphology, secular evolution, nuclear activity, and environmental effects. In particular, the catalogues of edge-on galaxies, barred galaxies with dust lanes, and ring galaxies establish an important morphological basis for future searches for Milky Way analogues using combined structural, photometric, spectroscopic, kinematic, and multiwavelength diagnostics. 

-- The complete catalogues of visually verified SDSS galaxies presented in this work will be made publicly available through the VizieR database and are currently accessible through the Ukrainian Virtual Observatory (UkrVO)\footnote{\href{https://mao.kiev.ua/index.php/ua/vpaai-labvmsv-catalogues}{UkrVO}.}. Beyond their immediate astrophysical applications, these catalogues constitute valuable benchmark datasets for training and validating future machine-learning algorithms and for improving automated galaxy classification methods in the era of Rubin LSST, \textit{Euclid}, the \textit{Nancy Grace Roman Space Telescope}, and SKA.

The present work demonstrates that expert visual validation remains an indispensable component of modern machine-learning approaches to galaxy morphology. Rather than replacing human expertise, convolutional neural networks and visual inspection complement one another, providing an efficient and reliable strategy for constructing large, homogeneous, and astrophysically meaningful morphological catalogues for contemporary and future extragalactic surveys.

Together with the previous papers of this series, the present work completes a homogeneous framework for the photometry- and image-based morphological classification of nearby SDSS galaxies and provides a reliable observational foundation for future astrophysical studies based on these catalogues.

\begin{acknowledgements}
The research was supported by the National Research Foundation of Ukraine (project 2023.03/0188). I.B. Vavilova, D.V. Dobrycheva, and O.V. Kompaniiets are members of the LSST Galaxies Science Collaboration. Funding for the Sloan Digital Sky Survey IV (SDSS-IV) has been provided by the Alfred P. Sloan Foundation, the U.S. Department of Energy Office of Science, and the Participating Institutions. SDSS acknowledges support and resources from the Center for High-Performance Computing at the University of Utah. The SDSS website is \url{https://www.sdss.org}. This research has made use of the resources of the Ukrainian Virtual Observatory (UkrVO), developed within the framework of the International Virtual Observatory Alliance (IVOA). This research has made use of the VizieR catalogue access tool and the SIMBAD database, operated at CDS, Strasbourg, France. The authors acknowledge the Centre de Données astronomiques de Strasbourg (CDS) for providing these astronomical data services.

The authors thank the reviewer for carefully reading the manuscript and for constructive comments and suggestions that improved its clarity and scientific presentation.
\end{acknowledgements}

\bibliographystyle{aa} 
\bibliography{library} 

@ARTICLE{Abraham2025,
       author = {{Abraham}, Linn and {Abraham}, Sheelu and {Kembhavi}, Ajit K. and {Philip}, N.~S. and {Aniyan}, A.~K. and {Barway}, Sudhanshu and {Kumar}, Harish},
        title = "{Automated Detection of Galactic Rings from Sloan Digital Sky Survey Images}",
      journal = {\apj},
         year = 2025,
        month = jan,
       volume = {978},
       number = {2},
          eid = {137},
        pages = {137},
          doi = {10.3847/1538-4357/ad856d},
archivePrefix = {arXiv},
       eprint = {2404.04484},
 primaryClass = {astro-ph.GA},
       adsurl = {https://ui.adsabs.harvard.edu/abs/2025ApJ...978..137A}
}

@ARTICLE{Abdurrouf2022,
       author = {{Abdurro'uf} and {Accetta}, Katherine and {Aerts}, Conny and {Silva Aguirre}, V{\'\i}ctor and {Ahumada}, Romina and {Ajgaonkar}, Nikhil and {Filiz Ak}, N. and {Alam}, Shadab and {Allende Prieto}, Carlos and {Almeida}, Andr{\'e}s and {Anders}, Friedrich and {Anderson}, Scott F. and {Andrews}, Brett H. and {Anguiano}, Borja and {Aquino-Ort{\'\i}z}, Erik and {Arag{\'o}n-Salamanca}, Alfonso and {Argudo-Fern{\'a}ndez}, Maria and {Ata}, Metin and {Aubert}, Marie and {Avila-Reese}, Vladimir and {Badenes}, Carles and {Barb{\'a}}, Rodolfo H. and {Barger}, Kat and {Barrera-Ballesteros}, Jorge K. and {Beaton}, Rachael L. and {Beers}, Timothy C. and {Belfiore}, Francesco and {Bender}, Chad F. and {Bernardi}, Mariangela and {Bershady}, Matthew A. and {Beutler}, Florian and {Bidin}, Christian Moni and {Bird}, Jonathan C. and {Bizyaev}, Dmitry and {Blanc}, Guillermo A. and {Blanton}, Michael R. and {Boardman}, Nicholas Fraser and {Bolton}, Adam S. and {Boquien}, M{\'e}d{\'e}ric and {Borissova}, Jura and {Bovy}, Jo and {Brandt}, W.~N. and {Brown}, Jordan and {Brownstein}, Joel R. and {Brusa}, Marcella and {Buchner}, Johannes and {Bundy}, Kevin and {Burchett}, Joseph N. and {Bureau}, Martin and {Burgasser}, Adam and {Cabang}, Tuesday K. and {Campbell}, Stephanie and {Cappellari}, Michele and {Carlberg}, Joleen K. and {Wanderley}, F{\'a}bio Carneiro and {Carrera}, Ricardo and {Cash}, Jennifer and {Chen}, Yan-Ping and {Chen}, Wei-Huai and {Cherinka}, Brian and {Chiappini}, Cristina and {Choi}, Peter Doohyun and {Chojnowski}, S. Drew and {Chung}, Haeun and {Clerc}, Nicolas and {Cohen}, Roger E. and {Comerford}, Julia M. and {Comparat}, Johan and {da Costa}, Luiz and {Covey}, Kevin and {Crane}, Jeffrey D. and {Cruz-Gonzalez}, Irene and {Culhane}, Connor and {Cunha}, Katia and {Dai}, Y. Sophia and {Damke}, Guillermo and {Darling}, Jeremy and {Davidson}, Jr., James W. and {Davies}, Roger and {Dawson}, Kyle and {De Lee}, Nathan and {Diamond-Stanic}, Aleksandar M. and {Cano-D{\'\i}az}, Mariana and {S{\'a}nchez}, Helena Dom{\'\i}nguez and {Donor}, John and {Duckworth}, Chris and {Dwelly}, Tom and {Eisenstein}, Daniel J. and {Elsworth}, Yvonne P. and {Emsellem}, Eric and {Eracleous}, Mike and {Escoffier}, Stephanie and {Fan}, Xiaohui and {Farr}, Emily and {Feng}, Shuai and {Fern{\'a}ndez-Trincado}, Jos{\'e} G. and {Feuillet}, Diane and {Filipp}, Andreas and {Fillingham}, Sean P. and {Frinchaboy}, Peter M. and {Fromenteau}, Sebastien and {Galbany}, Llu{\'\i}s and {Garc{\'\i}a}, Rafael A. and {Garc{\'\i}a-Hern{\'a}ndez}, D.~A. and {Ge}, Junqiang and {Geisler}, Doug and {Gelfand}, Joseph and {G{\'e}ron}, Tobias and {Gibson}, Benjamin J. and {Goddy}, Julian and {Godoy-Rivera}, Diego and {Grabowski}, Kathleen and {Green}, Paul J. and {Greener}, Michael and {Grier}, Catherine J. and {Griffith}, Emily and {Guo}, Hong and {Guy}, Julien and {Hadjara}, Massinissa and {Harding}, Paul and {Hasselquist}, Sten and {Hayes}, Christian R. and {Hearty}, Fred and {Hern{\'a}ndez}, Jes{\'u}s and {Hill}, Lewis and {Hogg}, David W. and {Holtzman}, Jon A. and {Horta}, Danny and {Hsieh}, Bau-Ching and {Hsu}, Chin-Hao and {Hsu}, Yun-Hsin and {Huber}, Daniel and {Huertas-Company}, Marc and {Hutchinson}, Brian and {Hwang}, Ho Seong and {Ibarra-Medel}, H{\'e}ctor J. and {Chitham}, Jacob Ider and {Ilha}, Gabriele S. and {Imig}, Julie and {Jaekle}, Will and {Jayasinghe}, Tharindu and {Ji}, Xihan and {Johnson}, Jennifer A. and {Jones}, Amy and {J{\"o}nsson}, Henrik and {Katkov}, Ivan and {Khalatyan}, Dr., Arman and {Kinemuchi}, Karen and {Kisku}, Shobhit and {Knapen}, Johan H. and {Kneib}, Jean-Paul and {Kollmeier}, Juna A. and {Kong}, Miranda and {Kounkel}, Marina and {Kreckel}, Kathryn and {Krishnarao}, Dhanesh and {Lacerna}, Ivan and {Lane}, Richard R. and {Langgin}, Rachel and {Lavender}, Ramon and {Law}, David R. and {Lazarz}, Daniel and {Leung}, Henry W. and {Leung}, Ho-Hin and {Lewis}, Hannah M. and {Li}, Cheng and {Li}, Ran and {Lian}, Jianhui and {Liang}, Fu-Heng and {Lin}, Lihwai and {Lin}, Yen-Ting and {Lin}, Sicheng and {Lintott}, Chris and {Long}, Dan and {Longa-Pe{\~n}a}, Pen{\'e}lope and {L{\'o}pez-Cob{\'a}}, Carlos and {Lu}, Shengdong and {Lundgren}, Britt F. and {Luo}, Yuanze and {Mackereth}, J. Ted and {de la Macorra}, Axel and {Mahadevan}, Suvrath and {Majewski}, Steven R. and {Manchado}, Arturo and {Mandeville}, Travis and {Maraston}, Claudia and {Margalef-Bentabol}, Berta and {Masseron}, Thomas and {Masters}, Karen L. and {Mathur}, Savita and {McDermid}, Richard M. and {Mckay}, Myles and {Merloni}, Andrea and {Merrifield}, Michael and {Meszaros}, Szabolcs and {Miglio}, Andrea and {Di Mille}, Francesco and {Minniti}, Dante and {Minsley}, Rebecca and {Monachesi}, Antonela},
        title = "{The Seventeenth Data Release of the Sloan Digital Sky Surveys: Complete Release of MaNGA, MaStar, and APOGEE-2 Data}",
      journal = {\apjs},
         year = 2022,
        month = apr,
       volume = {259},
       number = {2},
          eid = {35},
        pages = {35},
          doi = {10.3847/1538-4365/ac4414},
archivePrefix = {arXiv},
       eprint = {2112.02026},
 primaryClass = {astro-ph.GA},
       adsurl = {https://ui.adsabs.harvard.edu/abs/2022ApJS..259...35A}
}

@ARTICLE{Aihara2011,
       author = {{Aihara}, Hiroaki and {Allende Prieto}, Carlos and {An}, Deokkeun and {Anderson}, Scott F. and {Aubourg}, {\'E}ric and {Balbinot}, Eduardo and {Beers}, Timothy C. and {Berlind}, Andreas A. and {Bickerton}, Steven J. and {Bizyaev}, Dmitry and {Blanton}, Michael R. and {Bochanski}, John J. and {Bolton}, Adam S. and {Bovy}, Jo and {Brandt}, W.~N. and {Brinkmann}, J. and {Brown}, Peter J. and {Brownstein}, Joel R. and {Busca}, Nicolas G. and {Campbell}, Heather and {Carr}, Michael A. and {Chen}, Yanmei and {Chiappini}, Cristina and {Comparat}, Johan and {Connolly}, Natalia and {Cortes}, Marina and {Croft}, Rupert A.~C. and {Cuesta}, Antonio J. and {da Costa}, Luiz N. and {Davenport}, James R.~A. and {Dawson}, Kyle and {Dhital}, Saurav and {Ealet}, Anne and {Ebelke}, Garrett L. and {Edmondson}, Edward M. and {Eisenstein}, Daniel J. and {Escoffier}, Stephanie and {Esposito}, Massimiliano and {Evans}, Michael L. and {Fan}, Xiaohui and {Femen{\'\i}a Castell{\'a}}, Bruno and {Font-Ribera}, Andreu and {Frinchaboy}, Peter M. and {Ge}, Jian and {Gillespie}, Bruce A. and {Gilmore}, G. and {Gonz{\'a}lez Hern{\'a}ndez}, Jonay I. and {Gott}, J. Richard and {Gould}, Andrew and {Grebel}, Eva K. and {Gunn}, James E. and {Hamilton}, Jean-Christophe and {Harding}, Paul and {Harris}, David W. and {Hawley}, Suzanne L. and {Hearty}, Frederick R. and {Ho}, Shirley and {Hogg}, David W. and {Holtzman}, Jon A. and {Honscheid}, Klaus and {Inada}, Naohisa and {Ivans}, Inese I. and {Jiang}, Linhua and {Johnson}, Jennifer A. and {Jordan}, Cathy and {Jordan}, Wendell P. and {Kazin}, Eyal A. and {Kirkby}, David and {Klaene}, Mark A. and {Knapp}, G.~R. and {Kneib}, Jean-Paul and {Kochanek}, C.~S. and {Koesterke}, Lars and {Kollmeier}, Juna A. and {Kron}, Richard G. and {Lampeitl}, Hubert and {Lang}, Dustin and {Le Goff}, Jean-Marc and {Lee}, Young Sun and {Lin}, Yen-Ting and {Long}, Daniel C. and {Loomis}, Craig P. and {Lucatello}, Sara and {Lundgren}, Britt and {Lupton}, Robert H. and {Ma}, Zhibo and {MacDonald}, Nicholas and {Mahadevan}, Suvrath and {Maia}, Marcio A.~G. and {Makler}, Martin and {Malanushenko}, Elena and {Malanushenko}, Viktor and {Mandelbaum}, Rachel and {Maraston}, Claudia and {Margala}, Daniel and {Masters}, Karen L. and {McBride}, Cameron K. and {McGehee}, Peregrine M. and {McGreer}, Ian D. and {M{\'e}nard}, Brice and {Miralda-Escud{\'e}}, Jordi and {Morrison}, Heather L. and {Mullally}, F. and {Muna}, Demitri and {Munn}, Jeffrey A. and {Murayama}, Hitoshi and {Myers}, Adam D. and {Naugle}, Tracy and {Neto}, Angelo Fausti and {Nguyen}, Duy Cuong and {Nichol}, Robert C. and {O'Connell}, Robert W. and {Ogando}, Ricardo L.~C. and {Olmstead}, Matthew D. and {Oravetz}, Daniel J. and {Padmanabhan}, Nikhil and {Palanque-Delabrouille}, Nathalie and {Pan}, Kaike and {Pandey}, Parul and {P{\^a}ris}, Isabelle and {Percival}, Will J. and {Petitjean}, Patrick and {Pfaffenberger}, Robert and {Pforr}, Janine and {Phleps}, Stefanie and {Pichon}, Christophe and {Pieri}, Matthew M. and {Prada}, Francisco and {Price-Whelan}, Adrian M. and {Raddick}, M. Jordan and {Ramos}, Beatriz H.~F. and {Reyl{\'e}}, C{\'e}line and {Rich}, James and {Richards}, Gordon T. and {Rix}, Hans-Walter and {Robin}, Annie C. and {Rocha-Pinto}, Helio J. and {Rockosi}, Constance M. and {Roe}, Natalie A. and {Rollinde}, Emmanuel and {Ross}, Ashley J. and {Ross}, Nicholas P. and {Rossetto}, Bruno M. and {S{\'a}nchez}, Ariel G. and {Sayres}, Conor and {Schlegel}, David J. and {Schlesinger}, Katharine J. and {Schmidt}, Sarah J. and {Schneider}, Donald P. and {Sheldon}, Erin and {Shu}, Yiping and {Simmerer}, Jennifer and {Simmons}, Audrey E. and {Sivarani}, Thirupathi and {Snedden}, Stephanie A. and {Sobeck}, Jennifer S. and {Steinmetz}, Matthias and {Strauss}, Michael A. and {Szalay}, Alexander S. and {Tanaka}, Masayuki and {Thakar}, Aniruddha R. and {Thomas}, Daniel and {Tinker}, Jeremy L. and {Tofflemire}, Benjamin M. and {Tojeiro}, Rita and {Tremonti}, Christy A. and {Vandenberg}, Jan and {Vargas Maga{\~n}a}, M. and {Verde}, Licia and {Vogt}, Nicole P. and {Wake}, David A. and {Wang}, Ji and {Weaver}, Benjamin A. and {Weinberg}, David H. and {White}, Martin and {White}, Simon D.~M. and {Yanny}, Brian and {Yasuda}, Naoki and {Yeche}, Christophe and {Zehavi}, Idit},
        title = "{The Eighth Data Release of the Sloan Digital Sky Survey: First Data from SDSS-III}",
      journal = {\apjs},
         year = 2011,
        month = apr,
       volume = {193},
       number = {2},
          eid = {29},
        pages = {29},
          doi = {10.1088/0067-0049/193/2/29},
archivePrefix = {arXiv},
       eprint = {1101.1559},
 primaryClass = {astro-ph.IM},
       adsurl = {https://ui.adsabs.harvard.edu/abs/2011ApJS..193...29A}
}

@ARTICLE{Aguilar2025,
       author = {{Aguilar-Arg{\"u}ello}, G. and {Fuentes-Pineda}, G. and {Hern{\'a}ndez-Toledo}, H.~M. and {Mart{\'\i}nez-V{\'a}zquez}, L.~A. and {V{\'a}zquez-Mata}, J.~A. and {Brough}, S. and {Demarco}, R. and {Ghosh}, A. and {Jim{\'e}nez-Teja}, Y. and {Martin}, G. and {Pearson}, W.~J. and {Sif{\'o}n}, C.},
        title = "{Morphological classification of galaxies through structural and star formation parameters using machine learning}",
      journal = {\mnras},
         year = 2025,
        month = feb,
       volume = {537},
       number = {2},
        pages = {876-896},
          doi = {10.1093/mnras/staf085},
archivePrefix = {arXiv},
       eprint = {2501.06340},
 primaryClass = {astro-ph.GA},
       adsurl = {https://ui.adsabs.harvard.edu/abs/2025MNRAS.537..876A}
}

@ARTICLE{Anderson2020,
       author = {{Anderson}, L.~D. and {Sormani}, M.~C. and {Ginsburg}, Adam and {Glover}, Simon C.~O. and {Heywood}, I. and {Rammala}, I. and {Schuller}, F. and {Csengeri}, T. and {Urquhart}, J.~S. and {Bronfman}, Leonardo},
        title = "{Unusual Galactic H II Regions at the Intersection of the Central Molecular Zone and the Far Dust Lane}",
      journal = {\apj},
         year = 2020,
        month = sep,
       volume = {901},
       number = {1},
          eid = {51},
        pages = {51},
          doi = {10.3847/1538-4357/abadf6},
archivePrefix = {arXiv},
       eprint = {2008.04258},
 primaryClass = {astro-ph.GA},
       adsurl = {https://ui.adsabs.harvard.edu/abs/2020ApJ...901...51A}
}

@ARTICLE{Argudo2024,
       author = {{Argudo-Fern{\'a}ndez}, M. and {G{\'o}mez Hern{\'a}ndez}, C. and {Verley}, S. and {Zurita}, A. and {Duarte Puertas}, S. and {Bl{\'a}zquez Calero}, G. and {Dom{\'\i}nguez-G{\'o}mez}, J. and {Espada}, D. and {Florido}, E. and {P{\'e}rez}, I. and {S{\'a}nchez-Menguiano}, L.},
        title = "{Morphologies of galaxies within voids}",
      journal = {\aap},
         year = 2024,
        month = dec,
       volume = {692},
          eid = {A258},
        pages = {A258},
          doi = {10.1051/0004-6361/202450809},
archivePrefix = {arXiv},
       eprint = {2411.02129},
 primaryClass = {astro-ph.GA},
       adsurl = {https://ui.adsabs.harvard.edu/abs/2024A&A...692A.258A}
}

@ARTICLE{Aussel2024,
       author = {{Euclid Collaboration} and {Aussel}, B. and {Kruk}, S. and {Walmsley}, M. and {Huertas-Company}, M. and {Castellano}, M. and {Conselice}, C.~J. and {Delli Veneri}, M. and {Dom{\'\i}nguez S{\'a}nchez}, H. and {Duc}, P. -A. and {Knapen}, J.~H. and {Kuchner}, U. and {La Marca}, A. and {Margalef-Bentabol}, B. and {Marleau}, F.~R. and {Stevens}, G. and {Toba}, Y. and {Tortora}, C. and {Wang}, L. and {Aghanim}, N. and {Altieri}, B. and {Amara}, A. and {Andreon}, S. and {Auricchio}, N. and {Baldi}, M. and {Bardelli}, S. and {Bender}, R. and {Bodendorf}, C. and {Bonino}, D. and {Branchini}, E. and {Brescia}, M. and {Brinchmann}, J. and {Camera}, S. and {Capobianco}, V. and {Carbone}, C. and {Carretero}, J. and {Casas}, S. and {Cavuoti}, S. and {Cimatti}, A. and {Congedo}, G. and {Conversi}, L. and {Copin}, Y. and {Courbin}, F. and {Courtois}, H.~M. and {Cropper}, M. and {Da Silva}, A. and {Degaudenzi}, H. and {Di Giorgio}, A.~M. and {Dinis}, J. and {Dubath}, F. and {Dupac}, X. and {Dusini}, S. and {Farina}, M. and {Farrens}, S. and {Ferriol}, S. and {Fotopoulou}, S. and {Frailis}, M. and {Franceschi}, E. and {Franzetti}, P. and {Fumana}, M. and {Galeotta}, S. and {Garilli}, B. and {Gillis}, B. and {Giocoli}, C. and {Grazian}, A. and {Grupp}, F. and {Haugan}, S.~V.~H. and {Holmes}, W. and {Hook}, I. and {Hormuth}, F. and {Hornstrup}, A. and {Hudelot}, P. and {Jahnke}, K. and {Keih{\"a}nen}, E. and {Kermiche}, S. and {Kiessling}, A. and {Kilbinger}, M. and {Kubik}, B. and {K{\"u}mmel}, M. and {Kunz}, M. and {Kurki-Suonio}, H. and {Laureijs}, R. and {Ligori}, S. and {Lilje}, P.~B. and {Lindholm}, V. and {Lloro}, I. and {Maiorano}, E. and {Mansutti}, O. and {Marggraf}, O. and {Markovic}, K. and {Martinet}, N. and {Marulli}, F. and {Massey}, R. and {Maurogordato}, S. and {Medinaceli}, E. and {Mei}, S. and {Mellier}, Y. and {Meneghetti}, M. and {Merlin}, E. and {Meylan}, G. and {Moresco}, M. and {Moscardini}, L. and {Munari}, E. and {Niemi}, S. -M. and {Padilla}, C. and {Paltani}, S. and {Pasian}, F. and {Pedersen}, K. and {Percival}, W.~J. and {Pettorino}, V. and {Pires}, S. and {Polenta}, G. and {Poncet}, M. and {Popa}, L.~A. and {Pozzetti}, L. and {Raison}, F. and {Rebolo}, R. and {Renzi}, A. and {Rhodes}, J. and {Riccio}, G. and {Romelli}, E. and {Roncarelli}, M. and {Rossetti}, E. and {Saglia}, R. and {Sapone}, D. and {Sartoris}, B. and {Schirmer}, M. and {Schneider}, P. and {Secroun}, A. and {Seidel}, G. and {Serrano}, S. and {Sirignano}, C. and {Sirri}, G. and {Stanco}, L. and {Starck}, J. -L. and {Tallada-Cresp{\'\i}}, P. and {Taylor}, A.~N. and {Teplitz}, H.~I. and {Tereno}, I. and {Toledo-Moreo}, R. and {Torradeflot}, F. and {Tutusaus}, I. and {Valentijn}, E.~A. and {Valenziano}, L. and {Vassallo}, T. and {Veropalumbo}, A. and {Wang}, Y. and {Weller}, J. and {Zacchei}, A. and {Zamorani}, G. and {Zoubian}, J. and {Zucca}, E. and {Biviano}, A. and {Bolzonella}, M. and {Boucaud}, A. and {Bozzo}, E. and {Burigana}, C. and {Colodro-Conde}, C. and {Di Ferdinando}, D. and {Farinelli}, R. and {Graci{\'a}-Carpio}, J. and {Mainetti}, G. and {Marcin}, S. and {Mauri}, N. and {Neissner}, C. and {Nucita}, A.~A. and {Sakr}, Z. and {Scottez}, V. and {Tenti}, M. and {Viel}, M. and {Wiesmann}, M. and {Akrami}, Y. and {Allevato}, V. and {Anselmi}, S. and {Baccigalupi}, C. and {Ballardini}, M. and {Borgani}, S. and {Borlaff}, A.~S. and {Bretonni{\`e}re}, H. and {Bruton}, S. and {Cabanac}, R. and {Calabro}, A. and {Cappi}, A. and {Carvalho}, C.~S. and {Castignani}, G. and {Castro}, T. and {Ca{\~n}as-Herrera}, G. and {Chambers}, K.~C. and {Coupon}, J. and {Cucciati}, O. and {Davini}, S. and {De Lucia}, G. and {Desprez}, G. and {Di Domizio}, S. and {Dole}, H. and {D{\'\i}az-S{\'a}nchez}, A. and {Escartin Vigo}, J.~A. and {Escoffier}, S. and {Ferrero}, I. and {Finelli}, F.},
        title = "{Euclid preparation: XLIII. Measuring detailed galaxy morphologies for Euclid with machine learning}",
      journal = {\aap},
         year = 2024,
        month = sep,
       volume = {689},
          eid = {A274},
        pages = {A274},
          doi = {10.1051/0004-6361/202449609},
archivePrefix = {arXiv},
       eprint = {2402.10187},
 primaryClass = {astro-ph.GA},
       adsurl = {https://ui.adsabs.harvard.edu/abs/2024A&A...689A.274E}
}

@dataset{Baillard2011,
       author = {{Baillard}, A. and {Bertin}, W. and {de Lapparent}, V. and {Fouque}, P. and {Arnouts}, S. and {Mellier}, Y. and {Pello}, R. and {Leborgne}, J. -F. and {Prugniel}, P. and {Makarov}, D. and {Marakova}, L. and {McCraken}, H.~J. and {Bijaoui}, A. and {Tasca}, L.},
        title = "{VizieR Online Data Catalog: EFIGI catalogue of 4458 nearby galaxies (de Lapparent+, 2011)}",
 howpublished = {VizieR On-line Data Catalog: J/A+A/532/A74. Originally published in: 2011A\&A...532A..74B},
         year = 2011,
        month = jun,
          eid = {J/A+A/532/A74},
          doi = {10.26093/cds/vizier.35320074},
       adsurl = {https://ui.adsabs.harvard.edu/abs/2011yCat..35320074B}
}

@ARTICLE{Banerji2010,
       author = {{Banerji}, Manda and {Lahav}, Ofer and {Lintott}, Chris J. and {Abdalla}, Filipe B. and {Schawinski}, Kevin and {Bamford}, Steven P. and {Andreescu}, Dan and {Murray}, Phil and {Raddick}, M. Jordan and {Slosar}, Anze and {Szalay}, Alex and {Thomas}, Daniel and {Vandenberg}, Jan},
        title = "{Galaxy Zoo: reproducing galaxy morphologies via machine learning}",
      journal = {\mnras},
         year = 2010,
        month = jul,
       volume = {406},
       number = {1},
        pages = {342-353},
          doi = {10.1111/j.1365-2966.2010.16713.x},
archivePrefix = {arXiv},
       eprint = {0908.2033},
 primaryClass = {astro-ph.CO},
       adsurl = {https://ui.adsabs.harvard.edu/abs/2010MNRAS.406..342B}
}

@ARTICLE{Beck2016,
    author  = {{Beck}, R. and {Dobos}, L. and {Budav{\'a}ri}, T. and
               {Szalay}, A.~S. and {Csabai}, I.},
    title   = "{Photometric redshifts for the SDSS Data Release 12}",
    journal = {MNRAS}, year = 2016, volume = {460}, pages = {1371}
 }

@ARTICLE{Bickley2024,
       author = {{Bickley}, Robert W. and {Wilkinson}, Scott and {Ferreira}, Leonardo and {Ellison}, Sara L. and {Bottrell}, Connor and {Jyoti}, Debarpita},
        title = "{The effect of image quality on galaxy merger identification with deep learning}",
      journal = {\mnras},
         year = 2024,
        month = nov,
       volume = {534},
       number = {3},
        pages = {2533-2550},
          doi = {10.1093/mnras/stae2246},
archivePrefix = {arXiv},
       eprint = {2409.17081},
 primaryClass = {astro-ph.GA},
       adsurl = {https://ui.adsabs.harvard.edu/abs/2024MNRAS.534.2533B}
}

@article{Baldwin1981,
  author = {Baldwin, J. A. and Phillips, M. M. and Terlevich, R.},
  title = {Classification parameters for the emission-line spectra of extragalactic objects},
  journal = {\pasp},
  volume = {93},
  pages = {5--19},
  year = {1981},
  doi = {10.1086/130766}
}

@article{Belfiore2016,
  author  = {Belfiore, F. and Maiolino, R. and Maraston, C. and et al.},
  title   = {SDSS IV MaNGA -- spatially resolved diagnostic diagrams: a proof that many galaxies are LIERs},
  journal = {\mnras},
  volume  = {461},
  number  = {3},
  pages   = {3111--3134},
  year    = {2016},
  doi     = {10.1093/mnras/stw1234}
}

@ARTICLE{Buta2015,
       author = {{Buta}, Ronald J. and {Sheth}, Kartik and {Athanassoula}, E. and {Bosma}, A. and {Knapen}, Johan H. and {Laurikainen}, Eija and {Salo}, Heikki and {Elmegreen}, Debra and {Ho}, Luis C. and {Zaritsky}, Dennis and {Courtois}, Helene and {Hinz}, Joannah L. and {Mu{\~n}oz-Mateos}, Juan-Carlos and {Kim}, Taehyun and {Regan}, Michael W. and {Gadotti}, Dimitri A. and {Gil de Paz}, Armando and {Laine}, Jarkko and {Men{\'e}ndez-Delmestre}, Kar{\'\i}n and {Comer{\'o}n}, S{\'e}bastien and {Erroz Ferrer}, Santiago and {Seibert}, Mark and {Mizusawa}, Trisha and {Holwerda}, Benne and {Madore}, Barry F.},
        title = "{A Classical Morphological Analysis of Galaxies in the Spitzer Survey of Stellar Structure in Galaxies (S4G)}",
      journal = {\apjs},
         year = 2015,
        month = apr,
       volume = {217},
       number = {2},
          eid = {32},
        pages = {32},
          doi = {10.1088/0067-0049/217/2/32},
archivePrefix = {arXiv},
       eprint = {1501.00454},
 primaryClass = {astro-ph.GA},
       adsurl = {https://ui.adsabs.harvard.edu/abs/2015ApJS..217...32B}
}

@INPROCEEDINGS{Butterfield2024,
       author = {{Butterfield}, Natalie and {Morgan}, Larry},
        title = "{BARFLYS: Investigating the Star Forming Potential of the Galactic Bar Dust Lanes}",
    booktitle = {AAS Meeting Abstracts},
         year = 2024,
       series = {American Astronomical Society Meeting Abstracts},
       volume = {243},
        month = feb,
          eid = {262.19},
        pages = {262.19},
       adsurl = {https://ui.adsabs.harvard.edu/abs/2024AAS...24326219B}
}

@ARTICLE{Brinchmann2004,
       author = {{Brinchmann}, J. and {Charlot}, S. and {White}, S.~D.~M. and {Tremonti}, C. and {Kauffmann}, G. and {Heckman}, T. and {Brinkmann}, J.},
        title = "{The physical properties of star-forming galaxies in the low-redshift Universe}",
      journal = {\mnras},
         year = 2004,
        month = jul,
       volume = {351},
       number = {4},
        pages = {1151-1179},
          doi = {10.1111/j.1365-2966.2004.07881.x},
archivePrefix = {arXiv},
       eprint = {astro-ph/0311060},
 primaryClass = {astro-ph},
       adsurl = {https://ui.adsabs.harvard.edu/abs/2004MNRAS.351.1151B}
}

@ARTICLE{Cardelli1989,
  author = {{Cardelli}, J.~A. and {Clayton}, G.~C. and {Mathis}, J.~S.},
  title = "{The Relationship between Infrared, Optical, and
            Ultraviolet Extinction}",
  journal = {ApJ}, year = 1989, volume = {345}, pages = {245}}

@ARTICLE{Cavanagh2021,
       author = {{Cavanagh}, Mitchell K. and {Bekki}, Kenji and {Groves}, Brent A.},
        title = "{Morphological classification of galaxies with deep learning: comparing 3-way and 4-way CNNs}",
      journal = {\mnras},
         year = 2021,
        month = sep,
       volume = {506},
       number = {1},
        pages = {659-676},
          doi = {10.1093/mnras/stab1552},
archivePrefix = {arXiv},
       eprint = {2106.01571},
 primaryClass = {astro-ph.GA},
       adsurl = {https://ui.adsabs.harvard.edu/abs/2021MNRAS.506..659C}
}

@ARTICLE{Cybulski2014,
  author  = {Cybulski, R. and Yun, M. S. and
             Fazio, G. G. and Gutermuth, R. A.},
  title   = {From voids to Coma: the prevalence
             of pre-processing in the local Universe},
  journal = {MNRAS},
  year    = {2014},
  volume  = {439},
  number  = {4},
  pages   = {3564--3586},
  doi     = {10.1093/mnras/stu200}
}

@article{Fernandes2011,
  author  = {Cid Fernandes, R. and Stasi{\'n}ska, G. and Mateus, A. and Vale Asari, N.},
  title   = {A comprehensive classification of galaxies in the Sloan Digital Sky Survey: how to tell true from fake AGN?},
  journal = {\mnras},
  volume  = {413},
  number  = {3},
  pages   = {1687--1699},
  year    = {2011},
  doi     = {10.1111/j.1365-2966.2011.18244.x}
}

@ARTICLE{Cheng2020,
       author = {{Cheng}, Ting-Yun and {Conselice}, Christopher J. and {Arag{\'o}n-Salamanca}, Alfonso and {Li}, Nan and {Bluck}, Asa F.~L. and {Hartley}, Will G. and {Annis}, James and {Brooks}, David and {Doel}, Peter and {Garc{\'\i}a-Bellido}, Juan and {James}, David J. and {Kuehn}, Kyler and {Kuropatkin}, Nikolay and {Smith}, Mathew and {Sobreira}, Flavia and {Tarle}, Gregory},
        title = "{Optimizing automatic morphological classification of galaxies with machine learning and deep learning using Dark Energy Survey imaging}",
      journal = {\mnras},
         year = 2020,
        month = apr,
       volume = {493},
       number = {3},
        pages = {4209-4228},
          doi = {10.1093/mnras/staa501},
archivePrefix = {arXiv},
       eprint = {1908.03610},
 primaryClass = {astro-ph.GA},
       adsurl = {https://ui.adsabs.harvard.edu/abs/2020MNRAS.493.4209C}
}

@ARTICLE{Chrobakova2025,
       author = {{Chrob{\'a}kov{\'a}}, {\v{Z}}. and {Kre{\v{s}}{\v{n}}{\'a}kov{\'a}}, V. and {Nagy}, R. and {Gazdov{\'a}}, J. and {Butka}, P.},
        title = "{Deep Learning-based Detection and Segmentation of Edge-on and Highly Inclined Galaxies}",
      journal = {\pasp},
         year = 2025,
        month = mar,
       volume = {137},
       number = {3},
          eid = {034101},
        pages = {034101},
          doi = {10.1088/1538-3873/adbcd6},
archivePrefix = {arXiv},
       eprint = {2406.15064},
 primaryClass = {astro-ph.GA},
       adsurl = {https://ui.adsabs.harvard.edu/abs/2025PASP..137c4101C}
}

@ARTICLE{Ciucua2024,
       author = {{Ciuc{\u{a}}}, Ioana and {Kawata}, Daisuke and {Ting}, Yuan-Sen and {Grand}, Robert J.~J. and {Miglio}, Andrea and {Hayden}, Michael and {Baba}, Junichi and {Fragkoudi}, Francesca and {Monty}, Stephanie and {Buder}, Sven and {Freeman}, Ken},
        title = "{Chasing the impact of the Gaia-Sausage-Enceladus merger on the formation of the Milky Way thick disc}",
      journal = {\mnras},
         year = 2024,
        month = feb,
       volume = {528},
       number = {1},
        pages = {L122-L126},
          doi = {10.1093/mnrasl/slad033},
archivePrefix = {arXiv},
       eprint = {2211.01006},
 primaryClass = {astro-ph.GA},
       adsurl = {https://ui.adsabs.harvard.edu/abs/2024MNRAS.528L.122C}
}

@ARTICLE{Clarke2020,
       author = {{Clarke}, A.~O. and {Scaife}, A.~M.~M. and {Greenhalgh}, R. and {Griguta}, V.},
        title = "{Identifying galaxies, quasars, and stars with machine learning: A new catalogue of classifications for 111 million SDSS sources without spectra}",
      journal = {\aap},
         year = 2020,
        month = jul,
       volume = {639},
          eid = {A84},
        pages = {A84},
          doi = {10.1051/0004-6361/201936770},
archivePrefix = {arXiv},
       eprint = {1909.10963},
 primaryClass = {astro-ph.GA},
       adsurl = {https://ui.adsabs.harvard.edu/abs/2020A&A...639A..84C}
}

@ARTICLE{Dieleman2015,
       author = {{Dieleman}, Sander and {Willett}, Kyle W. and {Dambre}, Joni},
        title = "{Rotation-invariant convolutional neural networks for galaxy morphology prediction}",
      journal = {\mnras},
         year = 2015,
        month = jun,
       volume = {450},
       number = {2},
        pages = {1441-1459},
          doi = {10.1093/mnras/stv632},
archivePrefix = {arXiv},
       eprint = {1503.07077},
 primaryClass = {astro-ph.IM},
       adsurl = {https://ui.adsabs.harvard.edu/abs/2015MNRAS.450.1441D}
}

@ARTICLE{Dobrycheva2025,
       author = {{Dobrycheva}, D.~V. and {Hetmantsev}, O.~O. and {Vavilova}, I.~B. and {Shportko}, A. and {Gugnin}, O. and {Kompaniiets}, O.~V.},
        title = "{Discovery of the polar ring galaxies with deep learning}",
      journal = {\aap},
         year = 2025,
        month = oct,
       volume = {702},
          eid = {A258},
        pages = {A258},
          doi = {10.1051/0004-6361/202555052},
archivePrefix = {arXiv},
       eprint = {2505.05890},
 primaryClass = {astro-ph.GA},
       adsurl = {https://ui.adsabs.harvard.edu/abs/2025A&A...702A.258D}
}

@dataset{Dobrycheva2025cat,
       author = {{Dobrycheva}, D.~V. and {Hetmantsev}, O.~O. and {Vavilova}, I.~B. and {Shportko}, A. and {Gugnin}, O. and {Kompaniiets}, O.~V.},
        title = "{VizieR Online Data Catalog: Catalogue of inspected PRGs - CIPRG (Dobrycheva+, 2025)}",
 howpublished = {VizieR On-line Data Catalog: J/A+A/702/A258. Originally published in: 2025A\&A...702A.258D},
         year = 2025,
        month = sep,
          eid = {J/A+A/702/A258},
       adsurl = {https://ui.adsabs.harvard.edu/abs/2025yCat..37020258D}
}

@ARTICLE{Dominguez2022,
       author = {{Dom{\'\i}nguez S{\'a}nchez}, H. and {Margalef}, B. and {Bernardi}, M. and {Huertas-Company}, M.},
        title = "{SDSS-IV DR17: final release of MaNGA PyMorph photometric and deep-learning morphological catalogues}",
      journal = {\mnras},
         year = 2022,
        month = jan,
       volume = {509},
       number = {3},
        pages = {4024-4036},
          doi = {10.1093/mnras/stab3089},
archivePrefix = {arXiv},
       eprint = {2110.10694},
 primaryClass = {astro-ph.GA},
       adsurl = {https://ui.adsabs.harvard.edu/abs/2022MNRAS.509.4024D}
}

@dataset{Evans2019,
       author = {{Evans}, I.~N. and {Primini}, F.~A. and {Glotfelty}, C.~S. and {Anderson}, C.~S. and {Bonaventura}, N.~R. and {Chen}, J.~C. and {Davis}, J.~E. and {Doe}, S.~M. and {Evans}, J.~D. and {Fabbiano}, G. and {Galle}, E.~C. and {Gibbs}, D.~G. and {Grier}, J.~D. and {Hain}, R.~M. and {Hall}, D.~M. and {Harbo}, P.~N. and {He}, X. and {Houck}, J.~C. and {Karovska}, M. and {Kashyap}, V.~L. and {Lauer}, J. and {McCollough}, M.~L. and {McDowell}, J.~C. and {Miller}, J.~B. and {Mitschang}, A.~W. and {Morgan}, D.~L. and {Mossman}, A.~E. and {Nichols}, J.~S. and {Nowak}, M.~A. and {Plummer}, D.~A. and {Refsdal}, B.~L. and {Rots}, A.~H. and {Siemiginowska}, A. and {Sundheim}, B.~A. and {Tibbetts}, M.~S. and {van}, Stone D.~W. and {Winkelman}, S.~L. and {Zografou}, P.},
        title = "{VizieR Online Data Catalog: The Chandra Source Catalog (CSC), Release 2.0 (Evans+, 2019)}",
 howpublished = {VizieR On-line Data Catalog: IX/57.  Originally published in: 2010ApJS..189...37E},
         year = 2019,
        month = nov,
          eid = {IX/57},
       adsurl = {https://ui.adsabs.harvard.edu/abs/2019yCat.9057....0E}
}

@ARTICLE{Euclid2025b,
       author = {{Euclid Collaboration} and {Huertas-Company}, M. and {Walmsley}, M. and {Siudek}, M. and {Iglesias-Navarro}, P. and {Knapen}, J.~H. and {Serjeant}, S. and {Dickinson}, H.~J. and {Fortson}, L. and {Garland}, I. and et al.},
        title = "{Euclid Quick Data Release (Q1): XI. A first look at the fraction of bars in massive galaxies at z < 1}",
      journal = {\aap},
         year = 2026,
        month = jun,
       volume = {711},
          eid = {A11},
        pages = {A11},
          doi = {10.1051/0004-6361/202554606},
archivePrefix = {arXiv},
       eprint = {2503.15311},
 primaryClass = {astro-ph.GA},
       adsurl = {https://ui.adsabs.harvard.edu/abs/2026A&A...711A..11E}
}

@ARTICLE{Ellison2008,
    author  = {{Ellison}, S.~L. and {Patton}, D.~R. and {Simard}, L. and
               {McConnachie}, A.~W.},
    title   = "{Galaxy Pairs in the Sloan Digital Sky Survey. I. Star
               Formation, Active Galactic Nucleus Fraction, and the
               Luminosity/Mass-Metallicity Relation}",
    journal = {AJ}, year = 2008, volume = {135}, pages = {1877}
 }

@ARTICLE{Goh2025,
       author = {{Goh}, Kam Meng and {Lim}, Derrick Hiang Yaol and {Sham}, Zhen Dong and {Prakash}, Kolla Bhanu},
        title = "{An Interpretable Galaxy Morphology Classification Approach Using Modified SqueezeNet and Local Interpretable Model-agnostic Explanation}",
      journal = {Research in Astron. and Astrophys.},
         year = 2025,
        month = jun,
       volume = {25},
       number = {6},
          eid = {065018},
        pages = {065018},
          doi = {10.1088/1674-4527/adce8f},
       adsurl = {https://ui.adsabs.harvard.edu/abs/2025RAA....25f5018G}
}

@ARTICLE{Heestars2025,
       author = {{Heesters}, Nick and {Chemaly}, David and {M{\"u}ller}, Oliver and {Sola}, Elisabeth and {Fabbro}, S{\'e}bastien and {Ferreira}, Ashley and {McConnachie}, Alan W. and {Magnier}, Eugene and {Hudson}, Michael J. and {Chambers}, Kenneth and {Hammer}, Fran{\c{c}}ois and {Sanchez-Janssen}, Ruben},
        title = "{Galaxies OBserved as Low-luminosity Identified Nebulae (GOBLIN): Catalog of 43 000 high-probability dwarf galaxy candidates in the UNIONS survey}",
      journal = {\aap},
         year = 2025,
        month = jul,
       volume = {699},
          eid = {A232},
        pages = {A232},
          doi = {10.1051/0004-6361/202554501},
archivePrefix = {arXiv},
       eprint = {2505.18307},
 primaryClass = {astro-ph.GA},
       adsurl = {https://ui.adsabs.harvard.edu/abs/2025A&A...699A.232H}
}

@article{Hart2016,
  author  = {Hart, Ross E. and Bamford, Steven P. and Willett, Kyle W.
             and Masters, Karen L. and Cardamone, Carolin
             and Lintott, Chris J. and Mackay, Robert J.
             and Nichol, Robert C. and Rosslowe, Christopher K.
             and Simmons, Brooke D. and Smethurst, Rebecca J.},
  title   = {Galaxy Zoo: Comparing the Demographics of Spiral Arm Number
             and a New Method for Correcting Redshift Bias},
  journal = {\mnras},
  year    = {2016},
  volume  = {461},
  number  = {4},
  pages   = {3663--3682},
  doi     = {10.1093/mnras/stw1588},
  eprint  = {1607.01019},
  archivePrefix = {arXiv},
  primaryClass  = {astro-ph.GA}
}

@ARTICLE{Huertas2023,
       author = {{Huertas-Company}, M. and {Lanusse}, F.},
        title = "{The Dawes Review 10: The impact of deep learning for the analysis of galaxy surveys}",
      journal = {\pasa},
         year = 2023,
        month = jan,
       volume = {40},
          eid = {e001},
        pages = {e001},
          doi = {10.1017/pasa.2022.55},
archivePrefix = {arXiv},
       eprint = {2210.01813},
 primaryClass = {astro-ph.IM},
       adsurl = {https://ui.adsabs.harvard.edu/abs/2023PASA...40....1H}
}

@ARTICLE{HummerStorey1987,
  author = {{Hummer}, D.~G. and {Storey}, P.~J.},
  title = "{Recombination-line intensities for hydrogenic ions. I.
            Case B calculations for H I and He II}",
  journal = {MNRAS}, year = 1987, volume = {224}, pages = {801}}

@ARTICLE{Izvuekova2026,
       author = {{Izviekova}, I.~O. and {Kompaniiets}, O.~V. and {Vavilova}, I.~B.},
        title = "{Revealing a multi-zone circumnuclear dust response in UGC\raisebox{-0.5ex}\textasciitilde11487 through phase-resolved WISE W1--W2 diagnostics}",
      journal = {arXiv e-prints},
         year = 2026,
        month = jul,
          eid = {arXiv:2607.17815},
        pages = {arXiv:2607.17815},
          doi = {10.48550/arXiv.2607.17815},
archivePrefix = {arXiv},
       eprint = {2607.17815},
 primaryClass = {astro-ph.GA},
       adsurl = {https://ui.adsabs.harvard.edu/abs/2026arXiv260717815I}
}

@ARTICLE{Holwerda2019,
       author = {{Holwerda}, Benne W. and {Kelvin}, Lee and {Baldry}, Ivan and {Lintott}, Chris and {Alpaslan}, Mehmet and {Pimbblet}, Kevin A. and {Liske}, Jochen and {Kitching}, Thomas and {Bamford}, Steven and {de Jong}, Jelte and {Bilicki}, Maciej and {Hopkins}, Andrew and {Bridge}, Joanna and {Steele}, R. and {Jacques}, A. and {Goswami}, S. and {Kusmic}, S. and {Roemer}, W. and {Kruk}, S. and {Popescu}, C.~C. and {Kuijken}, K. and {Wang}, L. and {Wright}, A. and {Kitching}, T.},
        title = "{The Frequency of Dust Lanes in Edge-on Spiral Galaxies Identified by Galaxy Zoo in KiDS Imaging of GAMA Targets}",
      journal = {\aj},
         year = 2019,
        month = sep,
       volume = {158},
       number = {3},
          eid = {103},
        pages = {103},
          doi = {10.3847/1538-3881/ab2886},
archivePrefix = {arXiv},
       eprint = {1909.07461},
 primaryClass = {astro-ph.GA},
       adsurl = {https://ui.adsabs.harvard.edu/abs/2019AJ....158..103H}
}

@article{Kewley2001,
  author = {Kewley, L. J. and Dopita, M. A. and Sutherland, R. S. and Heisler, C. A. and Trevena, J.},
  title = {Theoretical Modeling of Starburst Galaxies},
  journal = {\apj},
  volume = {556},
  number = {1},
  pages = {121--140},
  year = {2001},
  doi = {10.1086/321545}
}

@article{Kauffmann2003,
  author = {Kauffmann, G. and Heckman, T. M. and Tremonti, C. and Brinchmann, J. and Charlot, S. and White, S. D. M. and Ridgway, S. E. and Brinkmann, J. and Fukugita, M. and Hall, P. B. and Ivezić, Ž. and Richards, G. T. and Schneider, D. P.},
  title = {The host galaxies of active galactic nuclei},
  journal = {\mnras},
  volume = {346},
  number = {4},
  pages = {1055--1077},
  year = {2003},
  doi = {10.1046/j.1365-2966.2003.07154.x}
}

@ARTICLE{Kompaniiets2026,
       author = {{Kompaniiets}, O.~V. and {Vavilova}, I.~B. and {Vasylkivskyi}, Y.~V. and {Konovalenko}, O.~O. and {Pastoven}, O.~S. and {Izviekova}, I.~O. and {Junais} and {Dmytrenko}, A.~M. and {Dobrycheva}, D.~V. and {Fedorov}, P.~N. and {Khramtsov}, V.~P. and {Sergijenko}, O. and {Vasylenko}, A.~A.},
        title = "{NGC 3521 as the Milky Way near-twin: UV─radio-decameter spectral energy distribution}",
      journal = {\aap},
         year = 2026,
        month = may,
       volume = {710},
          eid = {A11},
        pages = {A11},
          doi = {10.1051/0004-6361/202558597},
archivePrefix = {arXiv},
       eprint = {2512.14664},
 primaryClass = {astro-ph.GA},
       adsurl = {https://ui.adsabs.harvard.edu/abs/2026A&A...710A..11K}
}

@ARTICLE{Kompan2026,
       author = {{Kompaniiets}, O.~V. and {Vasylenko}, A.~A. and {Vavilova}, I.~B.},
        title = "{The 2MIG isolated AGNs ─ 2. X-ray general properties and peculiarities}",
      journal = {\mnras},
         year = 2026,
        month = jul,
       volume = {549},
       number = {3},
          eid = {stag1023},
        pages = {stag1023},
          doi = {10.1093/mnras/stag1023},
archivePrefix = {arXiv},
       eprint = {2506.14348},
 primaryClass = {astro-ph.HE},
       adsurl = {https://ui.adsabs.harvard.edu/abs/2026MNRAS.549g1023K}
}

@ARTICLE{Khramtsov2022,
       author = {{Khramtsov}, V. and {Vavilova}, I.~B. and {Dobrycheva}, D.~V. and {Vasylenko}, M. Yu. and {Melnyk}, O.~V. and {Elyiv}, A.~A. and {Akhmetov}, V.~S. and {Dmytrenko}, A.~M.},
        title = "{Machine learning technique for morphological classification of galaxies from the SDSS. III. Image-based inference of detailed features}",
      journal = {Space Sci. \& Technol.},
         year = 2022,
        month = sep,
       volume = {28},
       number = {5},
        pages = {27-55},
          doi = {10.15407/knit2022.05.027},
archivePrefix = {arXiv},
       eprint = {2209.12194},
 primaryClass = {astro-ph.GA},
       adsurl = {https://ui.adsabs.harvard.edu/abs/2022KosNT..28e..27K}
}

@ARTICLE{Lambas2003,
author  = {{Lambas}, D.~G. and {Tissera}, P.~B. and {Alonso}, M.~S. and
{Coldwell}, G.},
title   = "{Galaxy pairs in the 2dF survey - I. Effects of interactions on
star formation in the field}",
journal = {MNRAS}, year = 2003, volume = {346},
    pages = {1189}
}

@ARTICLE{Li2024,
       author = {{Li}, Fei and {Rahman}, Mubdi and {Murray}, Norman and {Kere{\v{s}}}, Du{\v{s}}an and {Wetzel}, Andrew and {Faucher-Gigu{\`e}re}, Claude-Andr{\'e} and {Hopkins}, Philip F. and {Moreno}, Jorge},
        title = "{The Effect of Galaxy Interactions on Starbursts in Milky Way-mass Galaxies in FIRE Simulations}",
      journal = {\apj},
         year = 2025,
        month = jan,
       volume = {979},
       number = {1},
          eid = {7},
        pages = {7},
          doi = {10.3847/1538-4357/ad94ef},
archivePrefix = {arXiv},
       eprint = {2411.10373},
 primaryClass = {astro-ph.GA},
       adsurl = {https://ui.adsabs.harvard.edu/abs/2025ApJ...979....7L}
}

@ARTICLE{Lintott2008,
       author = {{Lintott}, Chris J. and {Schawinski}, Kevin and {Slosar}, An{\v{z}}e and {Land}, Kate and {Bamford}, Steven and {Thomas}, Daniel and {Raddick}, M. Jordan and {Nichol}, Robert C. and {Szalay}, Alex and {Andreescu}, Dan and {Murray}, Phil and {Vandenberg}, Jan},
        title = "{Galaxy Zoo: morphologies derived from visual inspection of galaxies from the Sloan Digital Sky Survey}",
      journal = {\mnras},
         year = 2008,
        month = sep,
       volume = {389},
       number = {3},
        pages = {1179-1189},
          doi = {10.1111/j.1365-2966.2008.13689.x},
archivePrefix = {arXiv},
       eprint = {0804.4483},
 primaryClass = {astro-ph},
       adsurl = {https://ui.adsabs.harvard.edu/abs/2008MNRAS.389.1179L}
}

@ARTICLE{Martin2020,
       author = {{Martin}, G. and {Kaviraj}, S. and {Hocking}, A. and {Read}, S.~C. and {Geach}, J.~E.},
        title = "{Galaxy morphological classification in deep-wide surveys via unsupervised machine learning}",
      journal = {\mnras},
         year = 2020,
        month = jan,
       volume = {491},
       number = {1},
        pages = {1408-1426},
          doi = {10.1093/mnras/stz3006},
archivePrefix = {arXiv},
       eprint = {1909.10537},
 primaryClass = {astro-ph.GA},
       adsurl = {https://ui.adsabs.harvard.edu/abs/2020MNRAS.491.1408M}
}

@ARTICLE{Masters2025,
       author = {{Masters}, Karen},
        title = "{Morphological Classification of Galaxies}",
      journal = {arXiv e-prints},
         year = 2025,
        month = feb,
          eid = {arXiv:2502.09610},
        pages = {arXiv:2502.09610},
          doi = {10.48550/arXiv.2502.09610},
archivePrefix = {arXiv},
       eprint = {2502.09610},
 primaryClass = {astro-ph.GA},
       adsurl = {https://ui.adsabs.harvard.edu/abs/2025arXiv250209610M}
}

@ARTICLE{Merrow2024,
       author = {{Merrow}, Alex and {Grand}, Robert J.~J. and {Fragkoudi}, Francesca and {Martig}, Marie},
        title = "{Did the Gaia Enceladus/Sausage merger form the Milky Way's bar?}",
      journal = {\mnras},
         year = 2024,
        month = jun,
       volume = {531},
       number = {1},
        pages = {1520-1533},
          doi = {10.1093/mnras/stae1250},
archivePrefix = {arXiv},
       eprint = {2312.02318},
 primaryClass = {astro-ph.GA},
       adsurl = {https://ui.adsabs.harvard.edu/abs/2024MNRAS.531.1520M}
}

@ARTICLE{Merloni2024,
       author = {{Merloni}, A. and {Lamer}, G. and {Liu}, T. and {Ramos-Ceja}, M.~E. and {Brunner}, H. and {Bulbul}, E. and {Dennerl}, K. and {Doroshenko}, V. and {Freyberg}, M.~J. and {Friedrich}, S. and {Gatuzz}, E. and {Georgakakis}, A. and {Haberl}, F. and {Igo}, Z. and {Kreykenbohm}, I. and {Liu}, A. and {Maitra}, C. and {Malyali}, A. and {Mayer}, M.~G.~F. and {Nandra}, K. and {Predehl}, P. and {Robrade}, J. and {Salvato}, M. and {Sanders}, J.~S. and {Stewart}, I. and {Tub{\'\i}n-Arenas}, D. and {Weber}, P. and {Wilms}, J. and {Arcodia}, R. and {Artis}, E. and {Aschersleben}, J. and {Avakyan}, A. and {Aydar}, C. and {Bahar}, Y.~E. and {Balzer}, F. and {Becker}, W. and {Berger}, K. and {Boller}, T. and {Bornemann}, W. and {Br{\"u}ggen}, M. and {Brusa}, M. and {Buchner}, J. and {Burwitz}, V. and {Camilloni}, F. and {Clerc}, N. and {Comparat}, J. and {Coutinho}, D. and {Czesla}, S. and {Dannhauer}, S.~M. and {Dauner}, L. and {Dauser}, T. and {Dietl}, J. and {Dolag}, K. and {Dwelly}, T. and {Egg}, K. and {Ehl}, E. and {Freund}, S. and {Friedrich}, P. and {Gaida}, R. and {Garrel}, C. and {Ghirardini}, V. and {Gokus}, A. and {Gr{\"u}nwald}, G. and {Grandis}, S. and {Grotova}, I. and {Gruen}, D. and {Gueguen}, A. and {H{\"a}mmerich}, S. and {Hamaus}, N. and {Hasinger}, G. and {Haubner}, K. and {Homan}, D. and {Ider Chitham}, J. and {Joseph}, W.~M. and {Joyce}, A. and {K{\"o}nig}, O. and {Kaltenbrunner}, D.~M. and {Khokhriakova}, A. and {Kink}, W. and {Kirsch}, C. and {Kluge}, M. and {Knies}, J. and {Krippendorf}, S. and {Krumpe}, M. and {Kurpas}, J. and {Li}, P. and {Liu}, Z. and {Locatelli}, N. and {Lorenz}, M. and {M{\"u}ller}, S. and {Magaudda}, E. and {Mannes}, C. and {McCall}, H. and {Meidinger}, N. and {Michailidis}, M. and {Migkas}, K. and {Mu{\~n}oz-Giraldo}, D. and {Musiimenta}, B. and {Nguyen-Dang}, N.~T. and {Ni}, Q. and {Olechowska}, A. and {Ota}, N. and {Pacaud}, F. and {Pasini}, T. and {Perinati}, E. and {Pires}, A.~M. and {Pommranz}, C. and {Ponti}, G. and {Poppenhaeger}, K. and {P{\"u}hlhofer}, G. and {Rau}, A. and {Reh}, M. and {Reiprich}, T.~H. and {Roster}, W. and {Saeedi}, S. and {Santangelo}, A. and {Sasaki}, M. and {Schmitt}, J. and {Schneider}, P.~C. and {Schrabback}, T. and {Schuster}, N. and {Schwope}, A. and {Seppi}, R. and {Serim}, M.~M. and {Shreeram}, S. and {Sokolova-Lapa}, E. and {Starck}, H. and {Stelzer}, B. and {Stierhof}, J. and {Suleimanov}, V. and {Tenzer}, C. and {Traulsen}, I. and {Tr{\"u}mper}, J. and {Tsuge}, K. and {Urrutia}, T. and {Veronica}, A. and {Waddell}, S.~G.~H. and {Willer}, R. and {Wolf}, J. and {Yeung}, M.~C.~H. and {Zainab}, A. and {Zangrandi}, F. and {Zhang}, X. and {Zhang}, Y. and {Zheng}, X.},
        title = "{The SRG/eROSITA all-sky survey. First X-ray catalogues and data release of the western Galactic hemisphere}",
      journal = {\aap},
         year = 2024,
        month = feb,
       volume = {682},
          eid = {A34},
        pages = {A34},
          doi = {10.1051/0004-6361/202347165},
archivePrefix = {arXiv},
       eprint = {2401.17274},
 primaryClass = {astro-ph.HE},
       adsurl = {https://ui.adsabs.harvard.edu/abs/2024A&A...682A..34M}
}

@ARTICLE{Moiseev2011,
       author = {{Moiseev}, Alexei V. and {Smirnova}, Ksenia I. and {Smirnova}, Aleksandrina A. and {Reshetnikov}, Vladimir P.},
        title = "{A new catalogue of polar-ring galaxies selected from the Sloan Digital Sky Survey}",
      journal = {\mnras},
         year = 2011,
        month = nov,
       volume = {418},
       number = {1},
        pages = {244-257},
          doi = {10.1111/j.1365-2966.2011.19479.x},
archivePrefix = {arXiv},
       eprint = {1107.1966},
 primaryClass = {astro-ph.CO},
       adsurl = {https://ui.adsabs.harvard.edu/abs/2011MNRAS.418..244M}
}

@ARTICLE{Mao2025,
       author = {{Mao}, Yu and {Tu}, Liangping and {Xu}, Zhenyang and {Jiang}, Yue and {Zheng}, Mingyu},
        title = "{Galaxy Morphology Classification Based on DenseNet-SE4 Algorithm}",
      journal = {Research in Astron. Astrophys,},
         year = 2025,
        month = aug,
       volume = {25},
       number = {8},
          eid = {085010},
        pages = {085010},
          doi = {10.1088/1674-4527/ade22c},
       adsurl = {https://ui.adsabs.harvard.edu/abs/2025RAA....25h5010M}
}

@ARTICLE{Mukundan2024,
       author = {{Mukundan}, Kavya and {Nair}, Preethi and {Bailin}, Jeremy and {Li}, Wenhao},
        title = "{Automating galaxy morphology classification using k-nearest neighbours and non-parametric statistics}",
      journal = {\mnras},
         year = 2024,
        month = sep,
       volume = {533},
       number = {1},
        pages = {292-312},
          doi = {10.1093/mnras/stae1684},
       adsurl = {https://ui.adsabs.harvard.edu/abs/2024MNRAS.533..292M}
}

@ARTICLE{Nair2010,
       author = {{Nair}, Preethi B. and {Abraham}, Roberto G.},
        title = "{A Catalog of Detailed Visual Morphological Classifications for 14,034 Galaxies in the Sloan Digital Sky Survey}",
      journal = {\apjs},
         year = 2010,
        month = feb,
       volume = {186},
       number = {2},
        pages = {427-456},
          doi = {10.1088/0067-0049/186/2/427},
archivePrefix = {arXiv},
       eprint = {1001.2401},
 primaryClass = {astro-ph.CO},
       adsurl = {https://ui.adsabs.harvard.edu/abs/2010ApJS..186..427N}
}

@ARTICLE{Oh2018,
       author = {{Oh}, Kyuseok and {Koss}, Michael and {Markwardt}, Craig B. and {Schawinski}, Kevin and {Baumgartner}, Wayne H. and {Barthelmy}, Scott D. and {Cenko}, S. Bradley and {Gehrels}, Neil and {Mushotzky}, Richard and {Petulante}, Abigail and {Ricci}, Claudio and {Lien}, Amy and {Trakhtenbrot}, Benny},
        title = "{The 105-Month Swift-BAT All-sky Hard X-Ray Survey}",
      journal = {\apjs},
         year = 2018,
        month = mar,
       volume = {235},
       number = {1},
          eid = {4},
        pages = {4},
          doi = {10.3847/1538-4365/aaa7fd},
archivePrefix = {arXiv},
       eprint = {1801.01882},
 primaryClass = {astro-ph.HE},
       adsurl = {https://ui.adsabs.harvard.edu/abs/2018ApJS..235....4O}
}

@ARTICLE{Patton2000,
    author  = {{Patton}, D.~R. and {Carlberg}, R.~G. and {Marzke}, R.~O. and
               {Pritchet}, C.~J. and {da Costa}, L.~N. and {Pellegrini}, P.~S.},
    title   = "{New Techniques for Relating Dynamically Close Galaxy Pairs to
               Merger and Accretion Rates: Application to the Second Southern
               Sky Redshift Survey}",
    journal = {ApJ}, year = 2000, volume = {536}, pages = {153}
 }

@ARTICLE{Ren2023,
       author = {{Ren}, Jian and {Li}, Nan and {Liu}, F.~S. and {Cui}, Qifan and {Fu}, Mingxiang and {Zheng}, Xian Zhong},
        title = "{Revisiting Galaxy Evolution in Morphology in the Cosmic Evolution Survey Field (COSMOS-ReGEM). I. Merging Galaxies}",
      journal = {\apj},
         year = 2023,
        month = nov,
       volume = {958},
       number = {1},
          eid = {96},
        pages = {96},
          doi = {10.3847/1538-4357/acfeee},
archivePrefix = {arXiv},
       eprint = {2309.16531},
 primaryClass = {astro-ph.GA},
       adsurl = {https://ui.adsabs.harvard.edu/abs/2023ApJ...958...96R}
}

@ARTICLE{Sormani2019,
       author = {{Sormani}, Mattia C. and {Barnes}, Ashley T.},
        title = "{Mass inflow rate into the Central Molecular Zone: observational determination and evidence of episodic accretion}",
      journal = {\mnras},
         year = 2019,
        month = mar,
       volume = {484},
       number = {1},
        pages = {1213-1219},
          doi = {10.1093/mnras/stz046},
archivePrefix = {arXiv},
       eprint = {1901.00867},
 primaryClass = {astro-ph.GA},
       adsurl = {https://ui.adsabs.harvard.edu/abs/2019MNRAS.484.1213S}
}

@ARTICLE{Sormani2019b,
       author = {{Sormani}, Mattia C. and {Tre{\ss}}, Robin G. and {Glover}, Simon C.~O. and {Klessen}, Ralf S. and {Barnes}, Ashley T. and {Battersby}, Cara D. and {Clark}, Paul C. and {Hatchfield}, H. Perry and {Smith}, Rowan J.},
        title = "{The geometry of the gas surrounding the Central Molecular Zone: on the origin of localized molecular clouds with extreme velocity dispersions}",
      journal = {\mnras},
         year = 2019,
        month = oct,
       volume = {488},
       number = {4},
        pages = {4663-4673},
          doi = {10.1093/mnras/stz2054},
archivePrefix = {arXiv},
       eprint = {1906.10129},
 primaryClass = {astro-ph.GA},
       adsurl = {https://ui.adsabs.harvard.edu/abs/2019MNRAS.488.4663S}
}

@ARTICLE{Schawinski2007,
       author = {{Schawinski}, Kevin and {Thomas}, Daniel and {Sarzi}, Marc and {Maraston}, Claudia and {Kaviraj}, Sugata and {Joo}, Seok-Joo and {Yi}, Sukyoung K. and {Silk}, Joseph},
        title = "{Observational evidence for AGN feedback in early-type galaxies}",
      journal = {\mnras},
         year = 2007,
        month = dec,
       volume = {382},
       number = {4},
        pages = {1415-1431},
          doi = {10.1111/j.1365-2966.2007.12487.x},
archivePrefix = {arXiv},
       eprint = {0709.3015},
 primaryClass = {astro-ph},
       adsurl = {https://ui.adsabs.harvard.edu/abs/2007MNRAS.382.1415S}
}

@ARTICLE{Somawanshi2024,
       author = {{Somawanshi}, Devang and {Bhattacharya}, Souradeep and {Kataria}, Manish and {Kobayashi}, Chiaki},
        title = "{Understanding stellar populations in thin and thick discs of edge-on galaxies with MUSE - I. The case of the reignited S0 galaxy ESO 544-27}",
      journal = {\mnras},
         year = 2024,
        month = jul,
       volume = {531},
       number = {4},
        pages = {4336-4348},
          doi = {10.1093/mnras/stae1392},
archivePrefix = {arXiv},
       eprint = {2405.17999},
 primaryClass = {astro-ph.GA},
       adsurl = {https://ui.adsabs.harvard.edu/abs/2024MNRAS.531.4336S}
}

@ARTICLE{Su2024,
       author = {{Su}, Yang and {Zhang}, Shiyu and {Sun}, Yan and {Yang}, Ji and {Yan}, Qing-Zeng and {Zhang}, Shaobo and {Chen}, Zhiwei and {Chen}, Xuepeng and {Zhou}, Xin and {Yuan}, Lixia},
        title = "{Revealing Gas Inflows Toward the Galactic Central Molecular Zone}",
      journal = {\apjl},
         year = 2024,
        month = aug,
       volume = {971},
       number = {1},
          eid = {L6},
        pages = {L6},
          doi = {10.3847/2041-8213/ad656d},
archivePrefix = {arXiv},
       eprint = {2407.10857},
 primaryClass = {astro-ph.GA},
       adsurl = {https://ui.adsabs.harvard.edu/abs/2024ApJ...971L...6S}
}

@ARTICLE{Thuruthipilly2025,
       author = {{Thuruthipilly}, H. and {Junais} and {Koda}, J. and {Pollo}, A. and {Yagi}, M. and {Yamanoi}, H. and {Komiyama}, Y. and {Romano}, M. and {Ma{\l}ek}, K. and {Donevski}, D.},
        title = "{DES to HSC: Detecting low-surface-brightness galaxies in the Abell 194 cluster using transfer learning}",
      journal = {\aap},
         year = 2025,
        month = mar,
       volume = {695},
          eid = {A106},
        pages = {A106},
          doi = {10.1051/0004-6361/202452934},
archivePrefix = {arXiv},
       eprint = {2502.03142},
 primaryClass = {astro-ph.GA},
       adsurl = {https://ui.adsabs.harvard.edu/abs/2025A&A...695A.106T}
}

@ARTICLE{Timmis2017,
       author = {{Timmis}, Ian and {Shamir}, Lior},
        title = "{A Catalog of Automatically Detected Ring Galaxy Candidates in PanSTARSS}",
      journal = {\apjs},
         year = 2017,
        month = jul,
       volume = {231},
       number = {1},
          eid = {2},
        pages = {2},
          doi = {10.3847/1538-4365/aa78a3},
       adsurl = {https://ui.adsabs.harvard.edu/abs/2017ApJS..231....2T}
}

@ARTICLE{Tremonti2004,
       author = {{Tremonti}, Christy A. and {Heckman}, Timothy M. and {Kauffmann}, Guinevere and {Brinchmann}, Jarle and {Charlot}, St{\'e}phane and {White}, Simon D.~M. and {Seibert}, Mark and {Peng}, Eric W. and {Schlegel}, David J. and {Uomoto}, Alan and {Fukugita}, Masataka and {Brinkmann}, Jon},
        title = "{The Origin of the Mass-Metallicity Relation: Insights from 53,000 Star-forming Galaxies in the Sloan Digital Sky Survey}",
      journal = {\apj},
         year = 2004,
        month = oct,
       volume = {613},
       number = {2},
        pages = {898-913},
          doi = {10.1086/423264},
archivePrefix = {arXiv},
       eprint = {astro-ph/0405537},
 primaryClass = {astro-ph},
       adsurl = {https://ui.adsabs.harvard.edu/abs/2004ApJ...613..898T}
}

@ARTICLE{Tsukui2025,
       author = {{Tsukui}, Takafumi and {Wisnioski}, Emily and {Bland-Hawthorn}, Joss and {Freeman}, Ken},
        title = "{The emergence of galactic thin and thick discs across cosmic history}",
      journal = {\mnras},
         year = 2025,
        month = jul,
       volume = {540},
       number = {4},
        pages = {3493-3522},
          doi = {10.1093/mnras/staf604},
archivePrefix = {arXiv},
       eprint = {2409.15909},
 primaryClass = {astro-ph.GA},
       adsurl = {https://ui.adsabs.harvard.edu/abs/2025MNRAS.540.3493T}
}

@ARTICLE{Vavilova2021,
       author = {{Vavilova}, I.~B. and {Dobrycheva}, D.~V. and {Vasylenko}, M. Yu. and {Elyiv}, A.~A. and {Melnyk}, O.~V. and {Khramtsov}, V.},
        title = "{Machine learning technique for morphological classification of galaxies from the SDSS. I. Photometry-based approach}",
      journal = {\aap},
         year = 2021,
        month = apr,
       volume = {648},
          eid = {A122},
        pages = {A122},
          doi = {10.1051/0004-6361/202038981},
archivePrefix = {arXiv},
       eprint = {1712.08955},
 primaryClass = {astro-ph.GA},
       adsurl = {https://ui.adsabs.harvard.edu/abs/2021A&A...648A.122V}
}

@ARTICLE{Vavilova2022,
       author = {{Vavilova}, I.~B. and {Khramtsov}, V. and {Dobrycheva}, D.~V. and {Vasylenko}, M. Yu. and {Elyiv}, A.~A. and {Melnyk}, O.~V.},
        title = "{Machine learning technique for morphological classification of galaxies from SDSS. II. The image-based morphological catalogs of galaxies at 0.02<z<0.1}",
      journal = {Space Sci. \& Technol.},
         year = 2022,
        month = apr,
       volume = {28},
       number = {1},
        pages = {03-22},
          doi = {10.15407/knit2022.01.003},
archivePrefix = {arXiv},
       eprint = {2203.06373},
 primaryClass = {astro-ph.GA},
       adsurl = {https://ui.adsabs.harvard.edu/abs/2022KosNT..28a...3V}
}

@dataset{Vavilova2021Cat,
       author = {{Vavilova}, I.~B. and {Dobrycheva}, D.~V. and {Vasylenko}, M. Yu. and {Elyiv}, A.~A. and {Melnyk}, O.~V. and {Khramtsov}, V.},
        title = "{VizieR Online Data Catalog: SDSS galaxies morphological classification (Vavilova+, 2021)}",
 howpublished = {VizieR On-line Data Catalog: J/A+A/648/A122. Originally published in: 2021A\&A...648A.122V},
         year = 2021,
        month = mar,
          eid = {J/A+A/648/A122},
          doi = {10.26093/cds/vizier.36480122},
       adsurl = {https://ui.adsabs.harvard.edu/abs/2021yCat..36480122V}
}

@ARTICLE{Vavilova2022Cat,
       author = {{Vavilova}, I.~B. and {Khramtsov}, V. and {Dobrycheva}, D.~V. and {Vasylenko}, M. Yu. and {Elyiv}, A.~A. and {Melnyk}, O.~V.},
        title = "{VizieR Online Data Catalog: Galaxies at 0.02<z<0.1 morphological catalog (Vavilova+, 2022)}",
      journal = {VizieR Online Data Catalog (other)},
         year = 2022,
        month = sep,
       volume = {0710},
          eid = {J/other/KNIT/28},
        pages = {J/other/KNIT/28},
       adsurl = {https://ui.adsabs.harvard.edu/abs/2022yCatp071002801V}
}

@ARTICLE{Vavilova2023Cat,
       author = {{Vavilova}, I.~B. and {Khramtsov}, V. and {Dobrycheva}, D.~V. and {Vasylenko}, M. Yu. and {Elyiv}, A.~A. and {Melnyk}, O.~V.},
        title = "{VizieR Online Data Catalog: Galaxies at 0.02<z<0.1 morphological catalog (Vavilova+, 2022)}",
      journal = {VizieR Online Data Catalog (other)},
         year = 2023,
        month = feb,
       volume = {0710},
          eid = {J/other/KNIT/28},
        pages = {J/other/KNIT/28},
       adsurl = {https://ui.adsabs.harvard.edu/abs/2023yCatp071002801V}
}

@INPROCEEDINGS{Vavilova2024,
       author = {{Vavilova}, I.~B. and {Dobrycheva}, D.~V. and {Khramtsov}, V. and {Vasylenko}, M.~Y. and {Elyiv}, A.~A.},
        title = "{Machine Learning of Galaxy Classification by their Images and Photometry}",
    booktitle = {Astromical Data Analysis Software and Systems XXXI},
         year = 2024,
       editor = {{Hugo}, B.~V. and {Van Rooyen}, R. and {Smirnov}, O.~M.},
       series = {Astronomical Society of the Pacific Conference Series},
       volume = {535},
        month = may,
        pages = {103},
       adsurl = {https://ui.adsabs.harvard.edu/abs/2024ASPC..535..103V}
}

@ARTICLE{Vavilova2024b,
       author = {{Vavilova}, I.~B. and {Fedorov}, P.~M. and {Dobrycheva}, D.~V. and {Sergijenko}, O. and {Vasylenko}, A.~A. and {Dmytrenko}, A.~M. and {Khramtsov}, V.~P. and {Kompaniiets}, O.~V.},
        title = "{An advanced approach to the definition of the ``Milky Way galaxies-analogues''}",
      journal = {Space Sci. \& Technol.},
         year = 2024,
        month = jan,
       volume = {30},
       number = {4},
        pages = {81},
          doi = {10.15407/knit2024.04.081},
archivePrefix = {arXiv},
       eprint = {2410.04411},
 primaryClass = {astro-ph.GA},
       adsurl = {https://ui.adsabs.harvard.edu/abs/2024KosNT..30d..81V}
}

@ARTICLE{Vehlova2022,
       author = {{Venhola}, Aku and {Peletier}, Reynier F. and {Salo}, Heikki and {Laurikainen}, Eija and {Janz}, Joachim and {Haigh}, Caroline and {Wilkinson}, Michael H.~F. and {Iodice}, Enrichetta and {Hilker}, Michael and {Mieske}, Steffen and {Cantiello}, Michele and {Spavone}, Marilena},
        title = "{The Fornax Deep Survey with the VST. XII. Low surface brightness dwarf galaxies in the Fornax cluster}",
      journal = {\aap},
         year = 2022,
        month = jun,
       volume = {662},
          eid = {A43},
        pages = {A43},
          doi = {10.1051/0004-6361/202141756},
       adsurl = {https://ui.adsabs.harvard.edu/abs/2022A&A...662A..43V}
}

@ARTICLE{Vavilova2026,
       author = {{Vavilova}, I.~B. and {Dmytrenko}, A.~M. and {Dobrycheva}, D.~V. and {Fedorov}, P.~N. and {Izviekova}, I.~O. and {Khramtsov}, V.~P. and {Kompaniiets}, O.~V. and {Kukhar}, O.~N. and {Pastoven}, O.~S. and {Sergijenko}, O. and {Vasylenko}, A.~A.},
        title = "{Milky Way Near Twins (MWNeTs). I. A Hierarchical Framework for Identifying the Evolutionary Counterparts of the Milky Way}",
      journal = {arXiv e-prints},
         year = 2026,
        month = jul,
          eid = {arXiv:2607.08336},
        pages = {arXiv:2607.08336},
          doi = {10.48550/arXiv.2607.08336},
archivePrefix = {arXiv},
       eprint = {2607.08336},
 primaryClass = {astro-ph.GA},
       adsurl = {https://ui.adsabs.harvard.edu/abs/2026arXiv260708336V}
}

@ARTICLE{Vav2026,
       author = {{Vavilova}, I.~B. and {Izviekova}, I.~O. and {Dmytrenko}, A.~M. and {Kompaniiets}, O.~V. and {Kukhar}, O.~N. and {Pastoven}, O.~S. and {Denyschenko}, S.~I. and {Dobrycheva}, D.~V. and {Fedorov}, P.~N. and {Khramtsov}, V.~P. and {Sergijenko}, O. and {Vasylenko}, A.~A.},
        title = "{Milky Way Near Twins (MWNeTs). II. Applications of a Hierarchical Framework to NGC 3521, NGC 2903, UGC 11487, and CGCG 047-026}",
      journal = {arXiv e-prints},
         year = 2026,
        month = oct,
          eid = {arXiv:2610.23327},
        pages = {arXiv:2610.23327},
          doi = {10.48550/arXiv.2610.23327},
archivePrefix = {arXiv},
       eprint = {2610.23327},
 primaryClass = {astro-ph.GA},
       adsurl = {https://ui.adsabs.harvard.edu/abs/2026arXiv261023327V}
}

@ARTICLE{Walmsley2025,
       author = {{Euclid Collaboration} and {Walmsley}, M. and {Huertas-Company}, M. and {Quilley}, L. and {Masters}, K.~L. and {Kruk}, S. and {Remmelgas}, K.~A. and {Popp}, J.~J. and {Romelli}, E. and {O'Ryan}, D. and et al.},
        title = "{Euclid Quick Data Release (Q1): IX. First visual morphology catalogue}",
      journal = {\aap},
         year = 2026,
        month = jun,
       volume = {711},
          eid = {A9},
        pages = {A9},
          doi = {10.1051/0004-6361/202555140},
archivePrefix = {arXiv},
       eprint = {2503.15310},
 primaryClass = {astro-ph.GA},
       adsurl = {https://ui.adsabs.harvard.edu/abs/2026A&A...711A...9E}
}

@ARTICLE{Walmsley2022,
       author = {{Walmsley}, Mike and {Lintott}, Chris and {G{\'e}ron}, Tobias and {Kruk}, Sandor and {Krawczyk}, Coleman and {Willett}, Kyle W. and {Bamford}, Steven and {Kelvin}, Lee S. and {Fortson}, Lucy and {Gal}, Yarin and {Keel}, William and {Masters}, Karen L. and {Mehta}, Vihang and {Simmons}, Brooke D. and {Smethurst}, Rebecca and {Smith}, Lewis and {Baeten}, Elisabeth M. and {Macmillan}, Christine},
        title = "{Galaxy Zoo DECaLS: Detailed visual morphology measurements from volunteers and deep learning for 314 000 galaxies}",
      journal = {\mnras},
         year = 2022,
        month = jan,
       volume = {509},
       number = {3},
        pages = {3966-3988},
          doi = {10.1093/mnras/stab2093},
archivePrefix = {arXiv},
       eprint = {2102.08414},
 primaryClass = {astro-ph.GA},
       adsurl = {https://ui.adsabs.harvard.edu/abs/2022MNRAS.509.3966W}
}

@ARTICLE{Wei2022,
       author = {{Wei}, Shoulin and {Li}, Yadi and {Lu}, Wei and {Li}, Nan and {Liang}, Bo and {Dai}, Wei and {Zhang}, Zhijian},
        title = "{Unsupervised Galaxy Morphological Visual Representation with Deep Contrastive Learning}",
      journal = {\pasp},
         year = 2022,
        month = nov,
       volume = {134},
       number = {1041},
          eid = {114508},
        pages = {114508},
          doi = {10.1088/1538-3873/aca04e},
archivePrefix = {arXiv},
       eprint = {2211.07168},
 primaryClass = {astro-ph.GA},
       adsurl = {https://ui.adsabs.harvard.edu/abs/2022PASP..134k4508W}
}

@ARTICLE{Willett2013,
       author = {{Willett}, Kyle W. and {Lintott}, Chris J. and {Bamford}, Steven P. and {Masters}, Karen L. and {Simmons}, Brooke D. and {Casteels}, Kevin R.~V. and {Edmondson}, Edward M. and {Fortson}, Lucy F. and {Kaviraj}, Sugata and {Keel}, William C. and {Melvin}, Thomas and {Nichol}, Robert C. and {Raddick}, M. Jordan and {Schawinski}, Kevin and {Simpson}, Robert J. and {Skibba}, Ramin A. and {Smith}, Arfon M. and {Thomas}, Daniel},
        title = "{Galaxy Zoo 2: detailed morphological classifications for 304 122 galaxies from the Sloan Digital Sky Survey}",
      journal = {\mnras},
         year = 2013,
        month = nov,
       volume = {435},
       number = {4},
        pages = {2835-2860},
          doi = {10.1093/mnras/stt1458},
archivePrefix = {arXiv},
       eprint = {1308.3496},
 primaryClass = {astro-ph.CO},
       adsurl = {https://ui.adsabs.harvard.edu/abs/2013MNRAS.435.2835W}
}

@ARTICLE{Xu2023,
       author = {{Xu}, Quanfeng and {Shen}, Shiyin and {de Souza}, Rafael S. and {Chen}, Mi and {Ye}, Renhao and {She}, Yumei and {Chen}, Zhu and {Ishida}, Emille E.~O. and {Krone-Martins}, Alberto and {Durgesh}, Rupesh},
        title = "{From images to features: unbiased morphology classification via variational auto-encoders and domain adaptation}",
      journal = {\mnras},
         year = 2023,
        month = dec,
       volume = {526},
       number = {4},
        pages = {6391-6400},
          doi = {10.1093/mnras/stad3181},
archivePrefix = {arXiv},
       eprint = {2303.08627},
 primaryClass = {astro-ph.GA},
       adsurl = {https://ui.adsabs.harvard.edu/abs/2023MNRAS.526.6391X}
}

@ARTICLE{Yoachim2005,
       author = {{Yoachim}, Peter and {Dalcanton}, Julianne J.},
        title = "{The Kinematics of Thick Disks in External Galaxies}",
      journal = {\apj},
         year = 2005,
        month = may,
       volume = {624},
       number = {2},
        pages = {701-713},
          doi = {10.1086/428922},
archivePrefix = {arXiv},
       eprint = {astro-ph/0501394},
 primaryClass = {astro-ph},
       adsurl = {https://ui.adsabs.harvard.edu/abs/2005ApJ...624..701Y}
}

@ARTICLE{Zaw2019,
       author = {{Zaw}, Ingyin and {Chen}, Yan-Ping and {Farrar}, Glennys R.},
        title = "{A Uniformly Selected, All-sky, Optical AGN Catalog}",
      journal = {\apj},
         year = 2019,
        month = feb,
       volume = {872},
       number = {2},
          eid = {134},
        pages = {134},
          doi = {10.3847/1538-4357/aaffaf},
archivePrefix = {arXiv},
       eprint = {1902.03799},
 primaryClass = {astro-ph.GA},
       adsurl = {https://ui.adsabs.harvard.edu/abs/2019ApJ...872..134Z}
}

@ARTICLE{Zeraatgari2024b,
       author = {{Zeraatgari}, Fatemeh Zahra and {Hafezianzadeh}, Fatemeh and {Zhang}, Yanxia and {Mei}, Liquan and {Ayubinia}, Ashraf and {Mosallanezhad}, Amin and {Zhang}, Jingyi},
        title = "{Machine learning-based photometric classification of galaxies, quasars, emission-line galaxies, and stars}",
      journal = {\mnras},
         year = 2024,
        month = jan,
       volume = {527},
       number = {3},
        pages = {4677-4689},
          doi = {10.1093/mnras/stad3436},
archivePrefix = {arXiv},
       eprint = {2311.02951},
 primaryClass = {astro-ph.GA},
       adsurl = {https://ui.adsabs.harvard.edu/abs/2024MNRAS.527.4677Z}
}

\clearpage

\begin{appendix}

\section{Emission-line measurements, classification, and completeness}
\label{app:spectra}

\subsection{Spectroscopic matching and quality criteria}
\label{app:matching}

Spectroscopic counterparts were retrieved from SDSS DR17 \citep{Abdurrouf2022} by matching the catalogue \texttt{SDSSobjID} to \texttt{SpecObjAll.bestObjID}. Emission-line fluxes and uncertainties were taken from the MPA--JHU \texttt{galSpecLine} products \citep{Brinchmann2004,Tremonti2004}, joined through \texttt{specObjID} with the reliability information of \texttt{galSpecInfo}. These are legacy products derived from SDSS DR8-era spectroscopy \citep{Aihara2011} and do not cover all spectra released in DR17; the MPA--JHU pipeline subtracts the stellar continuum, including underlying absorption, and corrects for Galactic foreground extinction.

We required a positional separation of at most 2~arcsec, $|z_{\rm spec}-z_{\rm catalogue}|\leq 0.003$, and $0.003<z_{\rm spec}<0.1$. Accepted spectra additionally satisfied \texttt{sciencePrimary}=1, \texttt{class}=\texttt{GALAXY}, \texttt{zWarning}=0, and \texttt{RELIABLE}=1. Where several candidate spectra were available, the preferred match was chosen by spectral quality, angular separation, and redshift agreement. Companions were matched independently through their \texttt{CompID}, so that their spectral diagnostics refer to the companion itself.

For each diagnostic, we required finite positive fluxes and valid positive uncertainties in the relevant lines, defining the signal-to-noise ratio of a line as its flux divided by its reported uncertainty. Objects failing these criteria are retained in the catalogues with their quality flags but without a classification.

\subsection{BPT classification}
\label{app:bptmethod}

A reliable BPT classification required S/N$\geq3$ independently in all four lines. Emission-line fluxes were corrected for internal attenuation using the observed Balmer decrement, adopting an intrinsic H$\alpha$/H$\beta$ ratio of 2.86 for Case~B recombination under typical H\,\textsc{ii}-region conditions \citep{HummerStorey1987} and the extinction law of \citet{Cardelli1989} with $R_V=3.1$, which gives $A_\lambda/A_V$ of 0.818, 1.164, 0.815, and 1.120 for these four lines, respectively. We constrained the attenuation to $A_V\geq0$, setting $A_V=0$ where the observed ratio fell below the intrinsic value. Both the observed and the corrected ratios are retained in the released catalogues; the classifications quoted here use the corrected ones. Following \citet{Kauffmann2003} and \citet{Kewley2001}, we distinguished star-forming, composite, and AGN-like activity, separating the AGN branch into Seyfert and LINER-like classes with the demarcation of \citet{Schawinski2007}.

Classification uncertainties were quantified with 3000 Monte Carlo realisations of the measured fluxes, drawn from Gaussian distributions defined by their uncertainties; for each positive-flux realisation the Balmer decrement, attenuation correction, and BPT class were recomputed. We retained the resulting class probabilities, the modal class, and its 95\% Wilson interval. The formal $1\sigma$ uncertainties on the corrected BPT coordinates were obtained by propagating the flux errors through both the Balmer correction and the logarithmic ratios, assuming independent input errors. Neither these uncertainties nor the Monte Carlo probabilities include stellar-continuum systematics or covariance between the original line measurements.

\subsection{WHAN classification}
\label{app:whanmethod}

The WHAN diagram \citep{Fernandes2011} combines $\log([\mathrm{N\,II}]\lambda6584/\mathrm{H}\alpha)$ with the rest-frame H$\alpha$ equivalent width, EW(H$\alpha$), computed as the ratio of the continuum-subtracted H$\alpha$ flux from the simultaneous Gaussian fit of MPA--JHU to the local continuum flux density. Following the MPA--JHU rest-frame convention \citep[Appendix~C]{Cybulski2014}, no additional $(1+z)$ correction was applied. The EW uncertainty was propagated from the flux and continuum errors assuming zero covariance, and the selection required S/N$\geq3$ in both H$\alpha$ and [N\,\textsc{ii}]$\lambda6584$.

No internal attenuation correction was applied for WHAN: [N\,\textsc{ii}]$\lambda6584$ and H$\alpha\lambda6563$ are separated by only 21~\AA, so differential reddening affects neither their ratio nor the measured equivalent width appreciably. We distinguished star-forming-like, strong-AGN-like, weak-AGN-like, retired-like, and very-low-equivalent-width classes following \citet{Fernandes2011}. Because WHAN was applied independently of the BPT selection, galaxies with weak H$\beta$ or [O\,\textsc{iii}] retain a two-line classification; since significant H$\alpha$ and [N\,\textsc{ii}] detections are still required, the very-low-EW category is not a complete census of passive galaxies.

\subsection{Spectroscopic completeness and sensitivity to the quality threshold}
\label{app:completeness}

Of the 30\,918 visually confirmed morphological sample entries, 78.4\% satisfy the BPT selection criteria. Among the rest, 85.8\% fail the S/N$\geq3$ requirement in at least one diagnostic line, 12.8\% have missing or non-positive fluxes, and 1.4\% lack MPA--JHU measurements or fail the matching and quality criteria; insufficient signal-to-noise is therefore the dominant cause of BPT incompleteness.

Spectroscopic selection acts unevenly across the samples. At S/N$\geq3$ the BPT coverage ranges from 56.5\% for close pairs to 89.7\% for irregular galaxies and the WHAN coverage from 72.8\% to 98.4\% (Table~\ref{tab:bpt_whan_coverage}); it also depends on brightness and distance, the BPT retention of ring galaxies falling from 81.6\% at $r<16$ to 63.7\% at $16\leq r<17$, and from 89.6\% at $z<0.05$ to 66.6\% at $z\geq0.05$.

Raising the threshold to S/N$\geq5$ reduces the BPT-selected sample from 24\,229 to 18\,155 entries and lowers the LINER-like fraction from 52.5\% to 40.2\% for ringed and from 43.6\% to 33.3\% for barred galaxies, which is why the fractions in Sect.~\ref{sec:nuclear_spectral} are described as properties of the spectroscopically selected subsamples.

\clearpage
\onecolumn

\section{BPT and WHAN diagnostic diagrams for galaxies with visually verified morphologies selected from the CNN-based catalogues}
\label{app:bpt}


\begingroup
\setlength{\parindent}{0pt}
\centering

\noindent\makebox[\textwidth][c]{%
\includegraphics[width=0.325\textwidth]{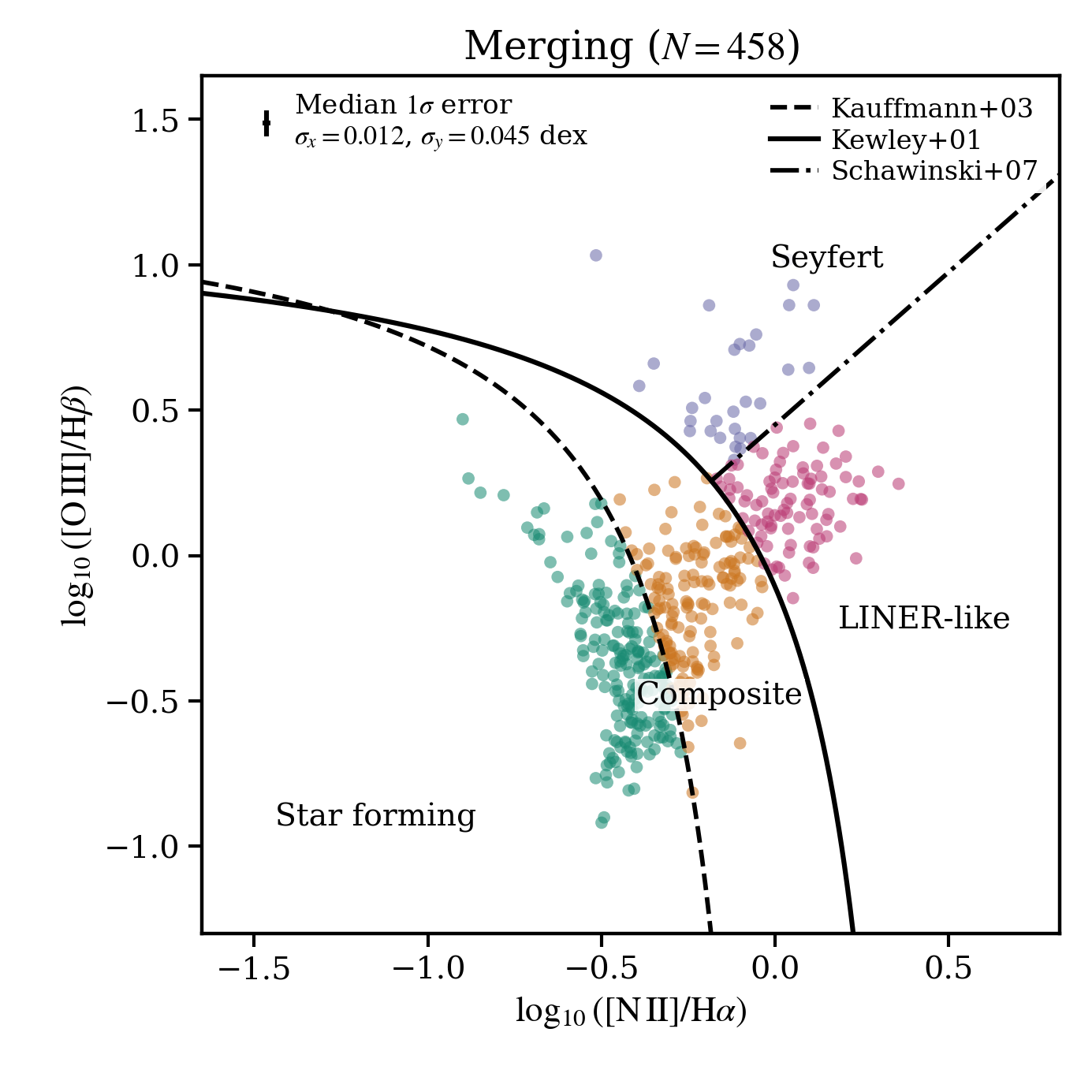}%
\hspace{1mm}%
\includegraphics[width=0.325\textwidth]{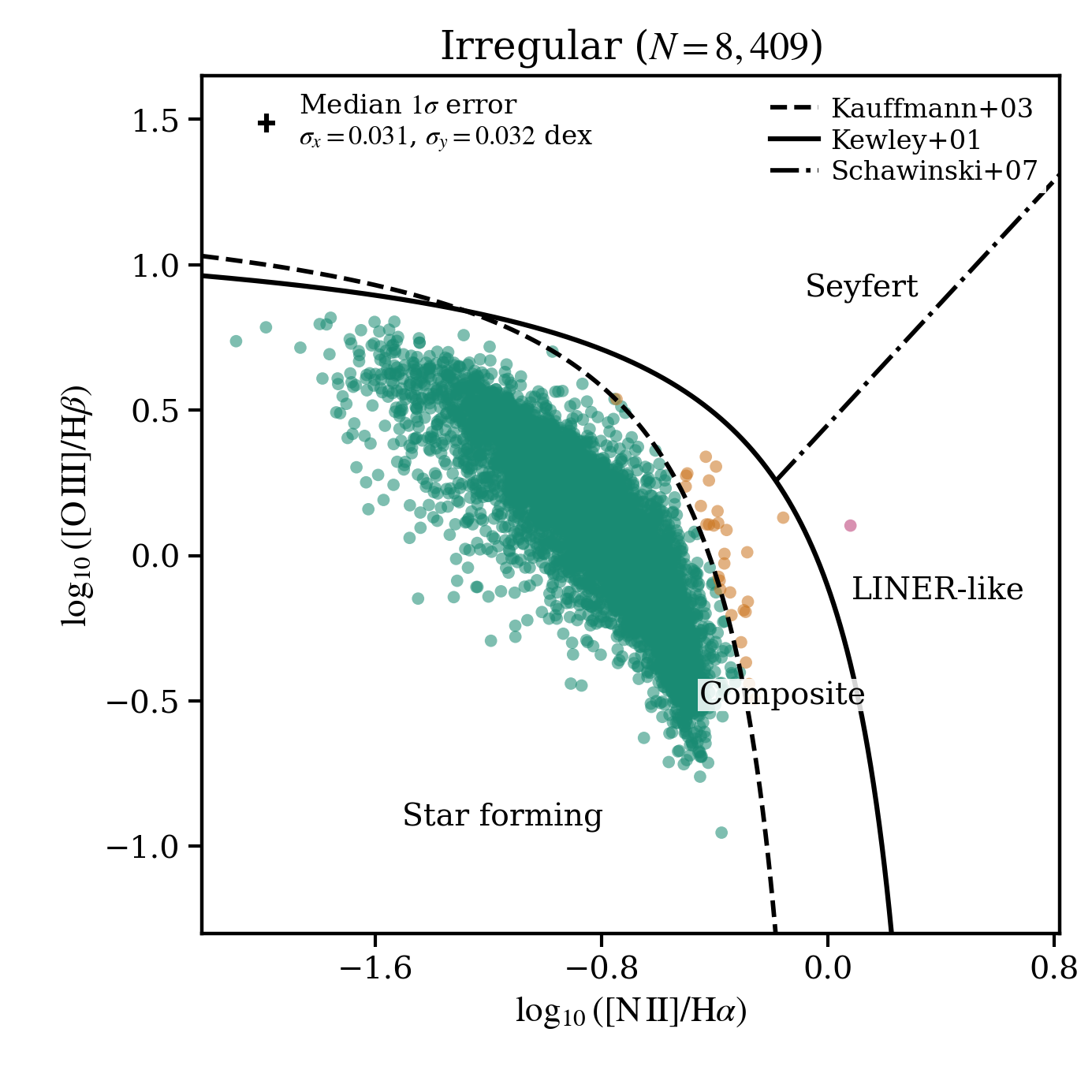}%
\hspace{1mm}%
\includegraphics[width=0.325\textwidth]{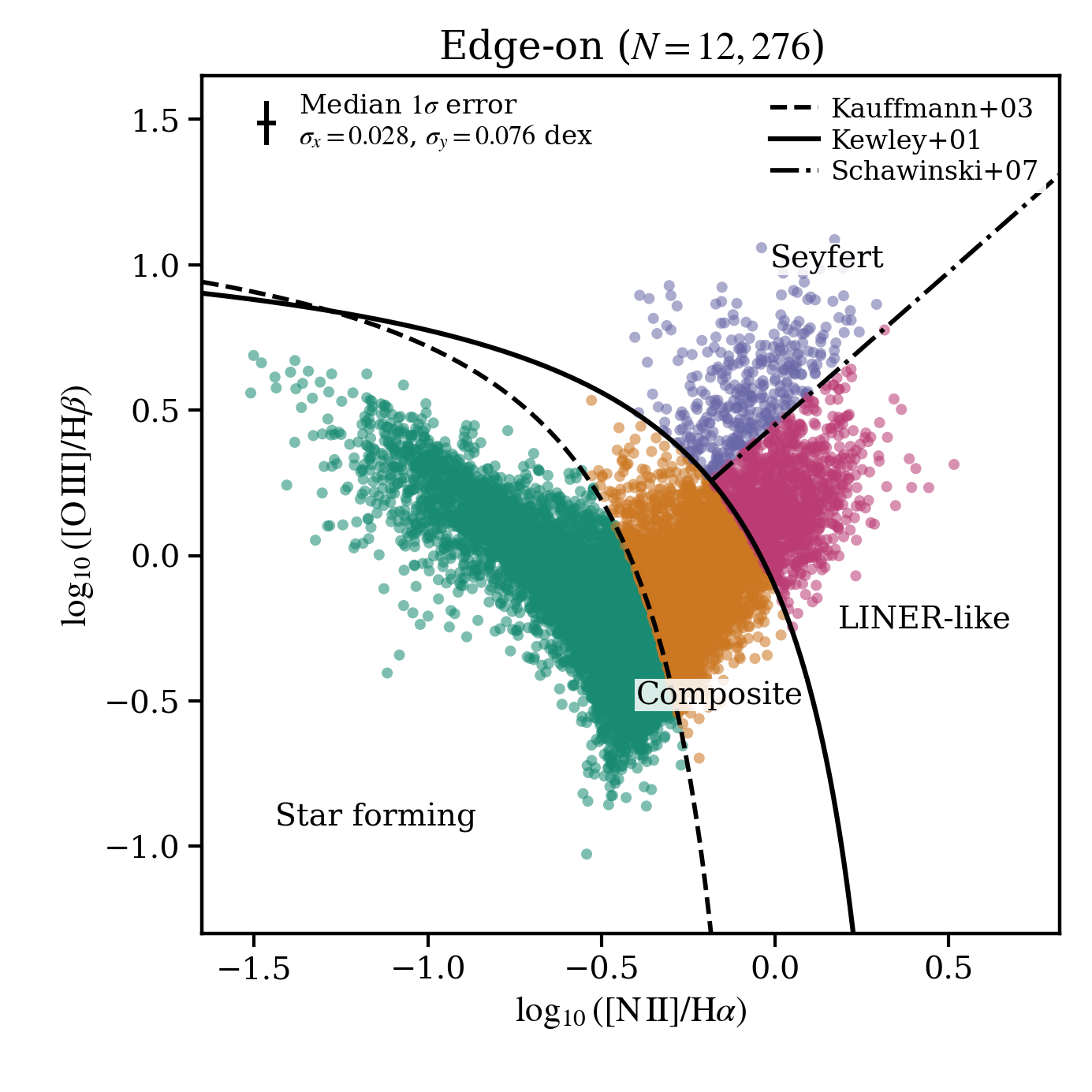}%
}

\par\vspace{2mm}

\noindent\makebox[\textwidth][c]{%
\includegraphics[width=0.325\textwidth]{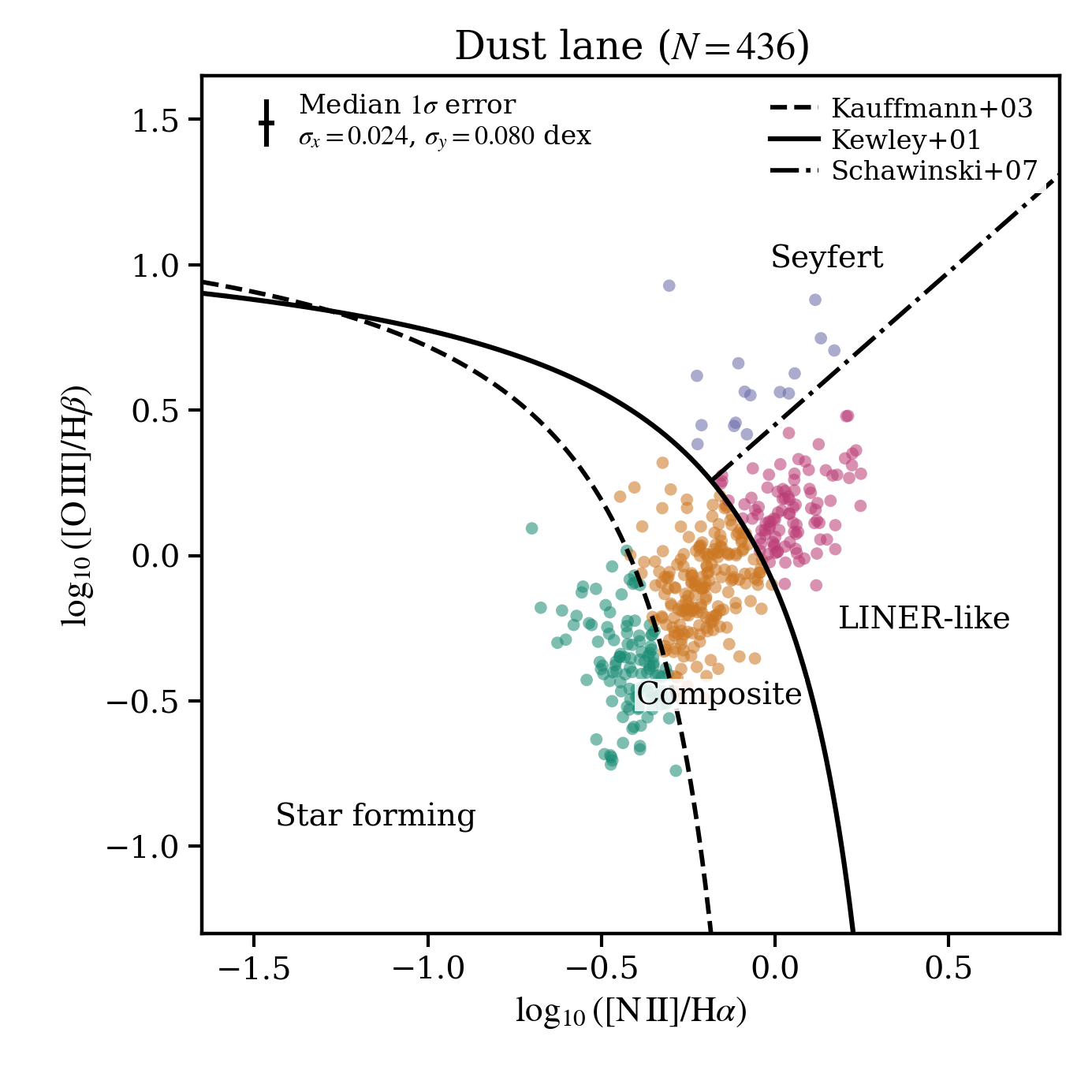}%
\hspace{1mm}%
\includegraphics[width=0.325\textwidth]{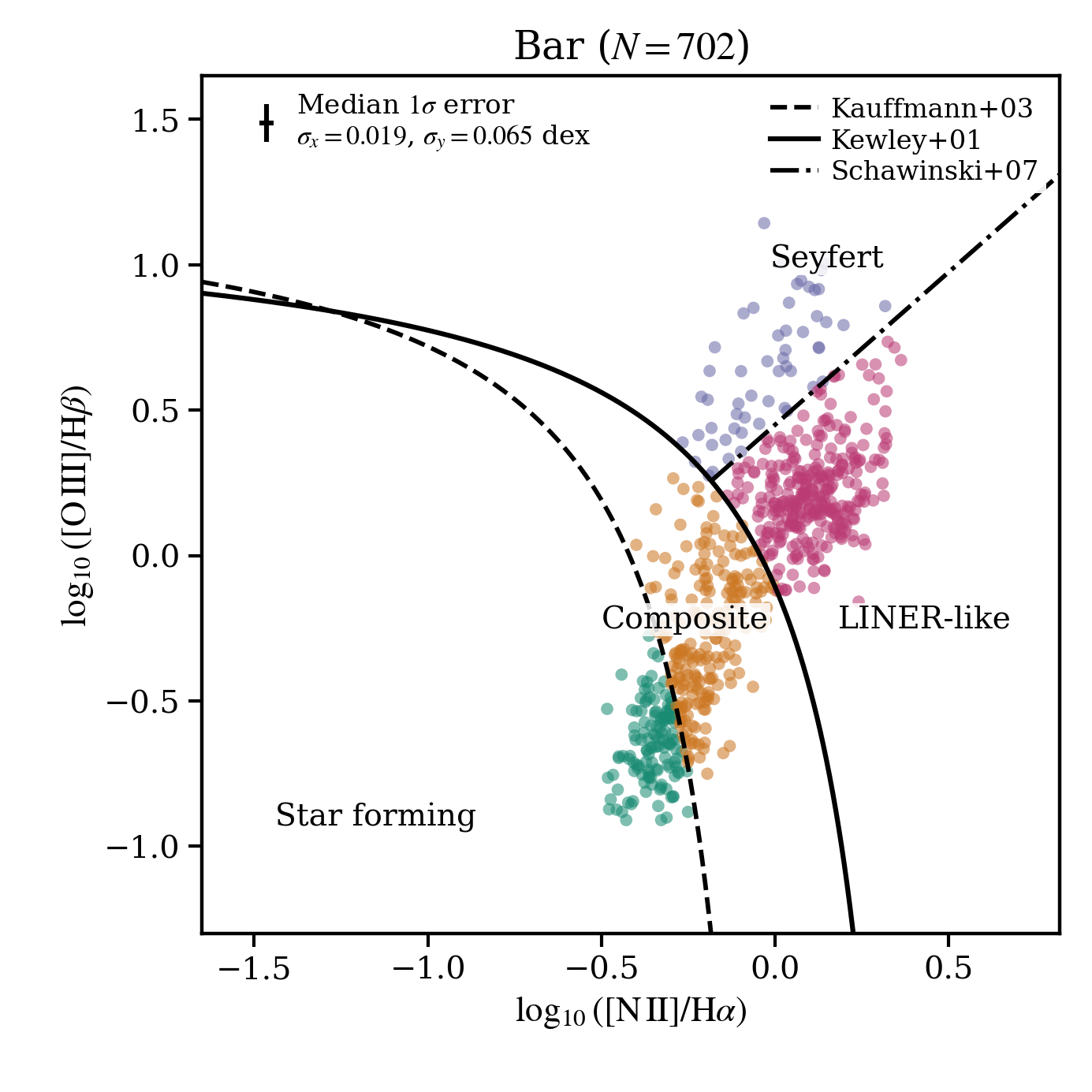}%
\hspace{1mm}%
\includegraphics[width=0.325\textwidth]{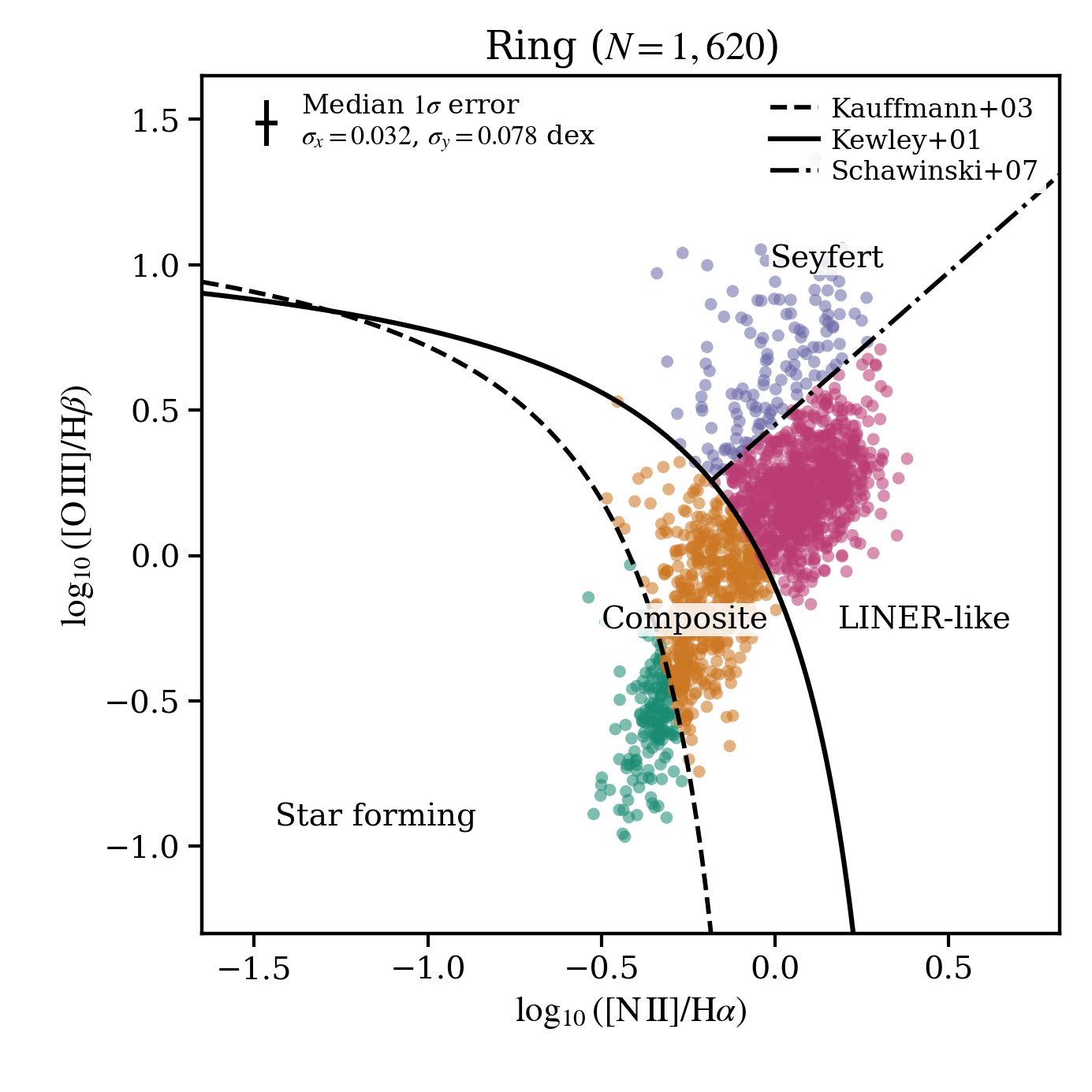}%
}

\par\vspace{2mm}

\noindent\makebox[\textwidth][c]{%
\begin{minipage}[c]{0.325\textwidth}
\centering
\includegraphics[width=\linewidth]{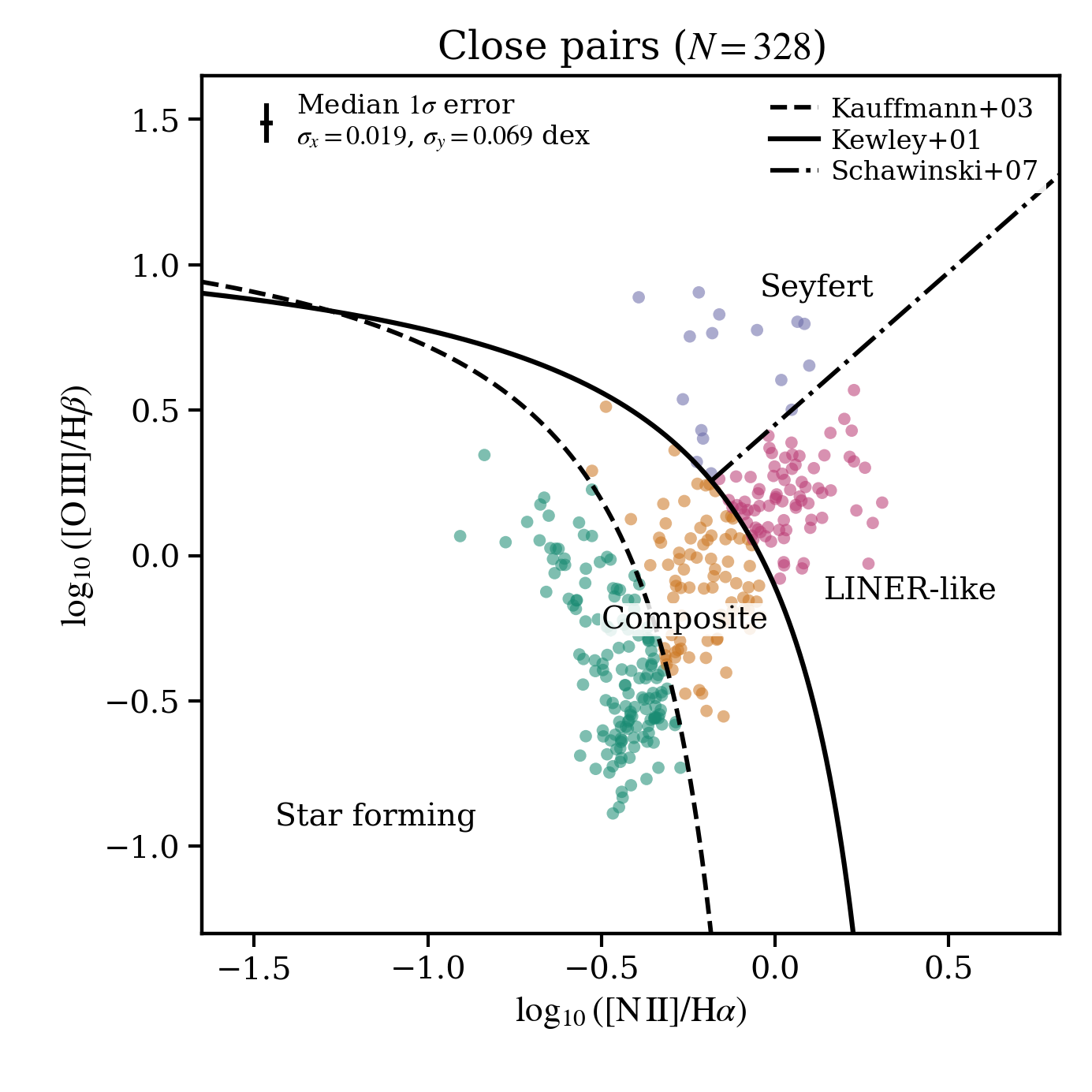}
\end{minipage}%
\hspace{6mm}%
\begin{minipage}[c]{0.59\textwidth}
\captionof{figure}{BPT diagnostic diagrams
for seven visually confirmed morphological samples.
Each panel indicates the sample size and the median
formal $1\sigma$ uncertainties along both diagnostic axes.}
\label{fig:BPT_all}
\end{minipage}%
}

\par
\endgroup

\clearpage

\begingroup
\setlength{\parindent}{0pt}
\centering

\noindent\makebox[\textwidth][c]{%
\includegraphics[width=0.325\textwidth]{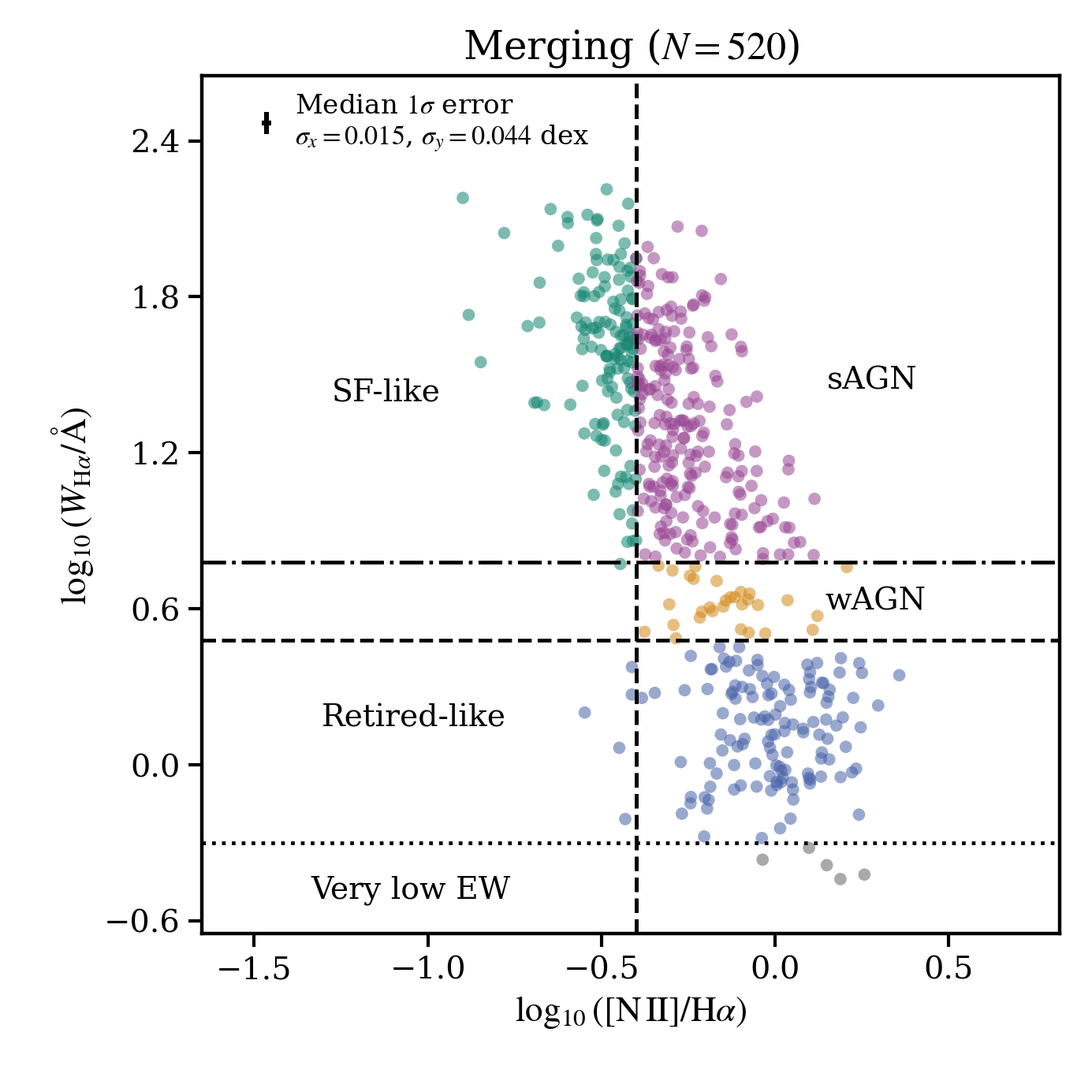}%
\hspace{1mm}%
\includegraphics[width=0.325\textwidth]{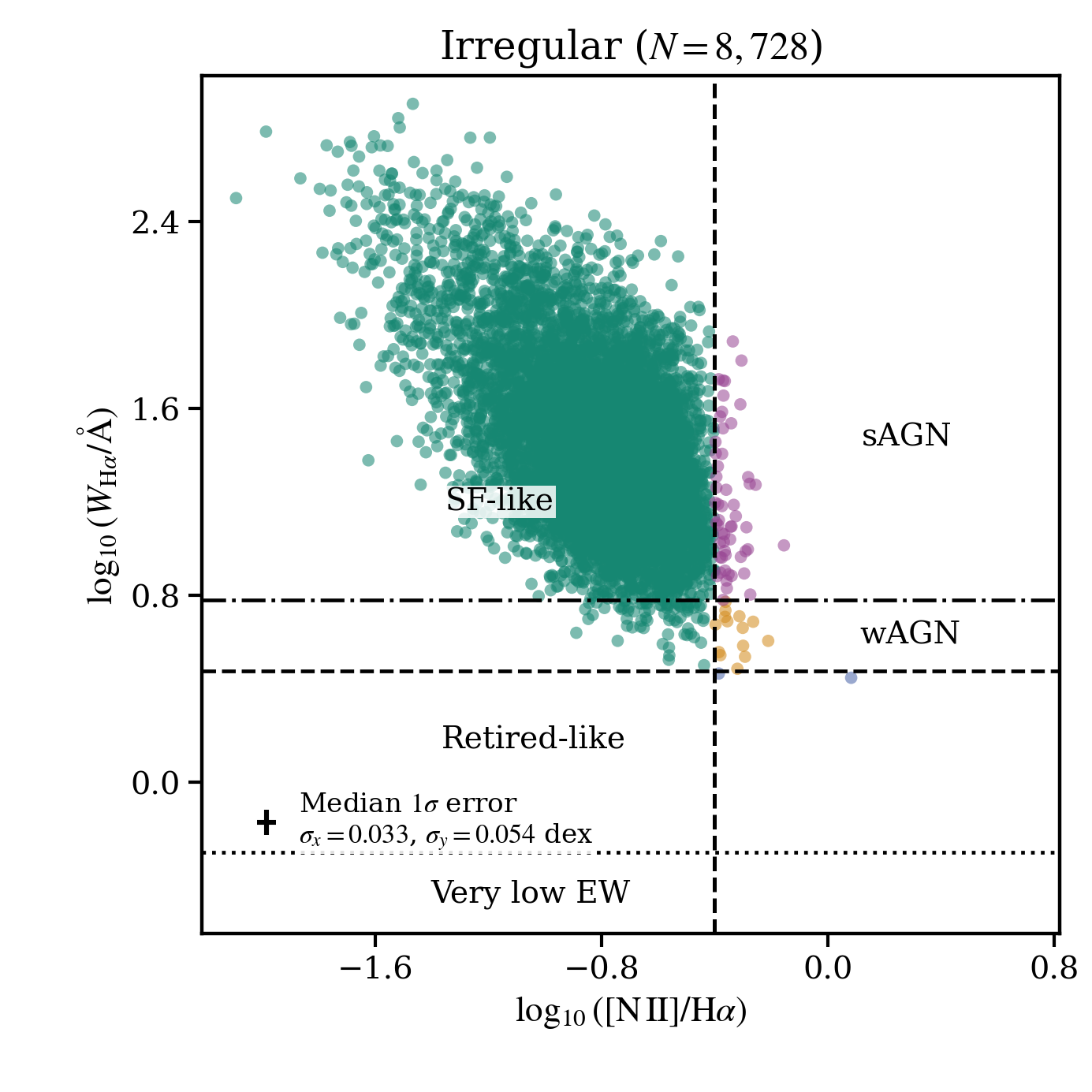}%
\hspace{1mm}%
\includegraphics[width=0.325\textwidth]{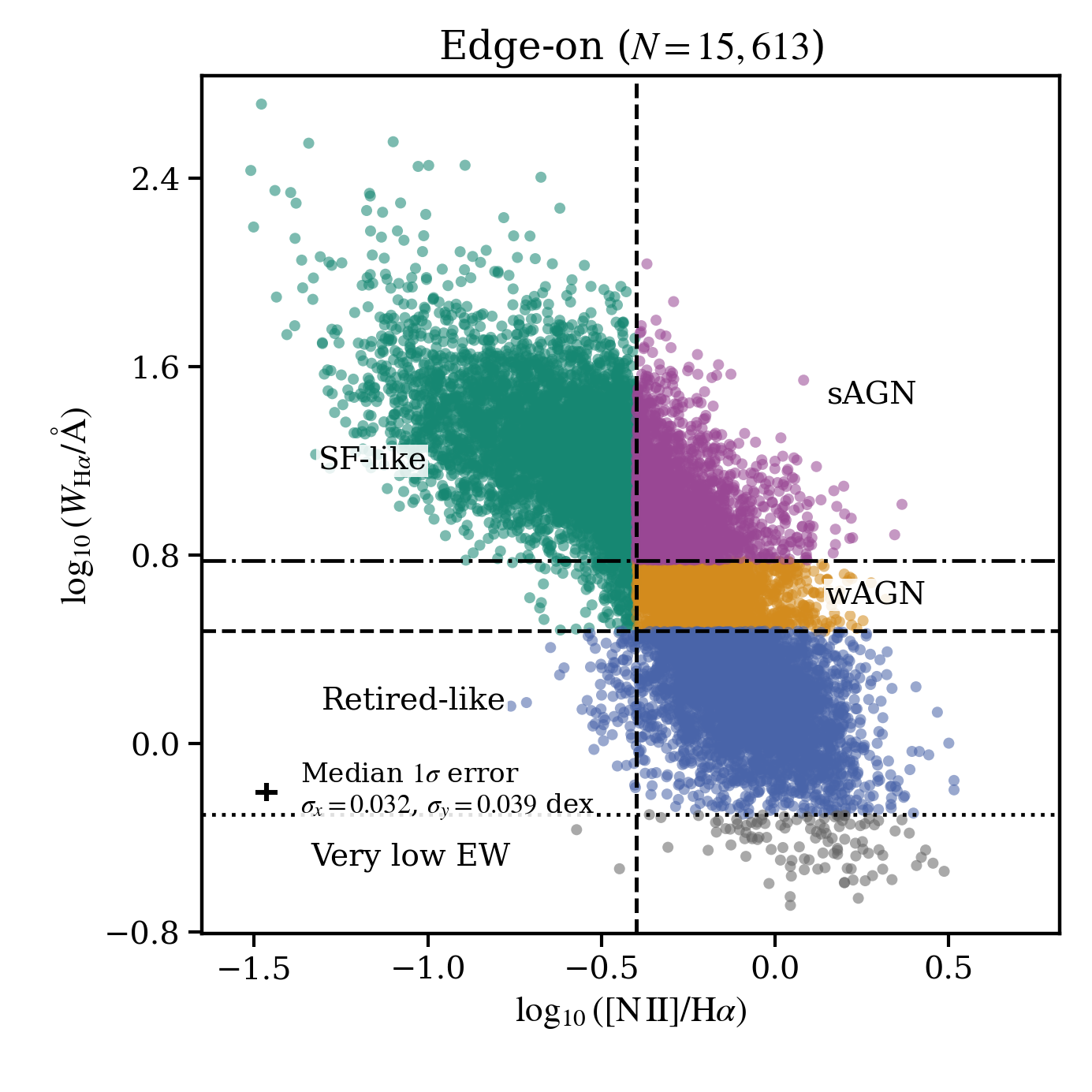}%
}

\par\vspace{2mm}

\noindent\makebox[\textwidth][c]{%
\includegraphics[width=0.325\textwidth]{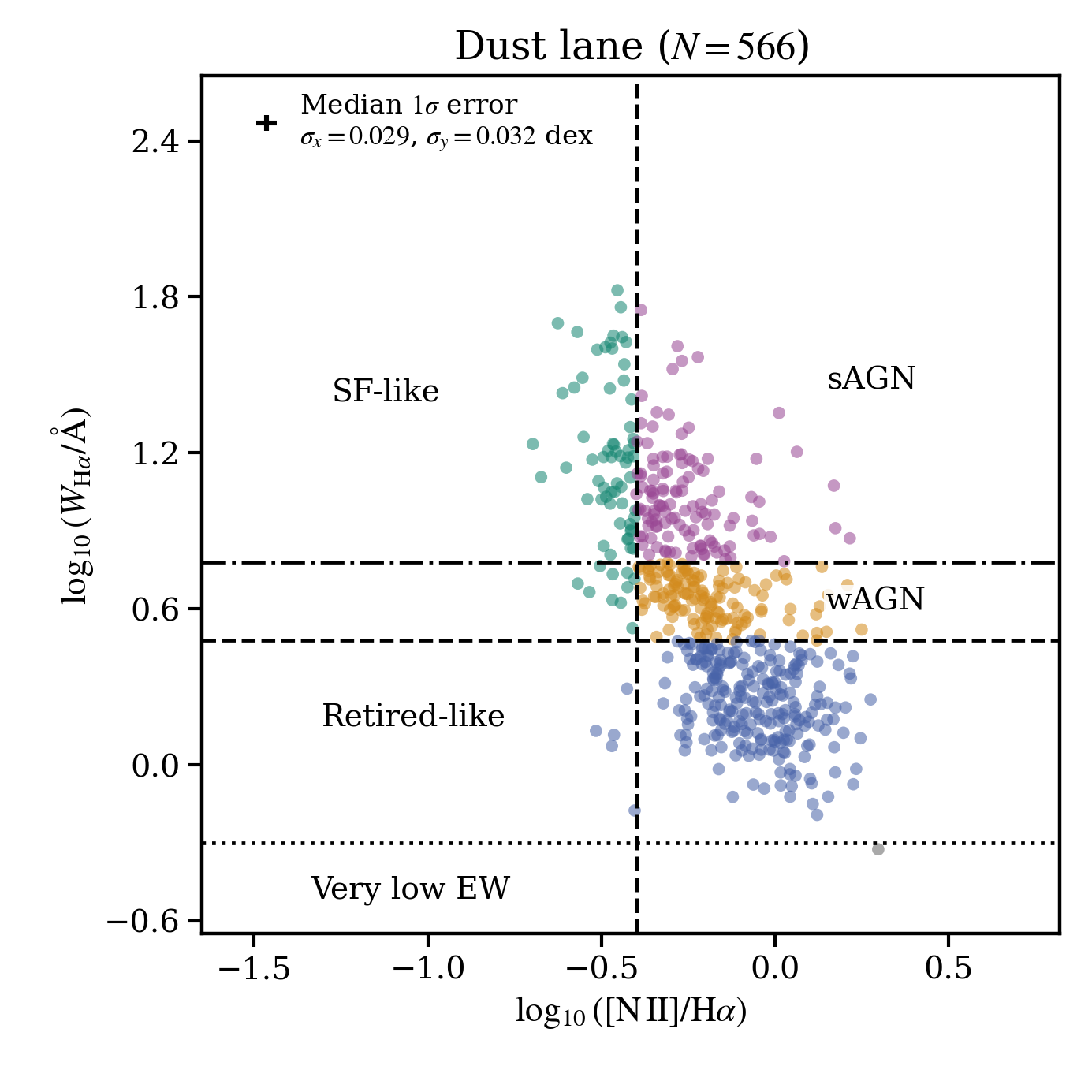}%
\hspace{1mm}%
\includegraphics[width=0.325\textwidth]{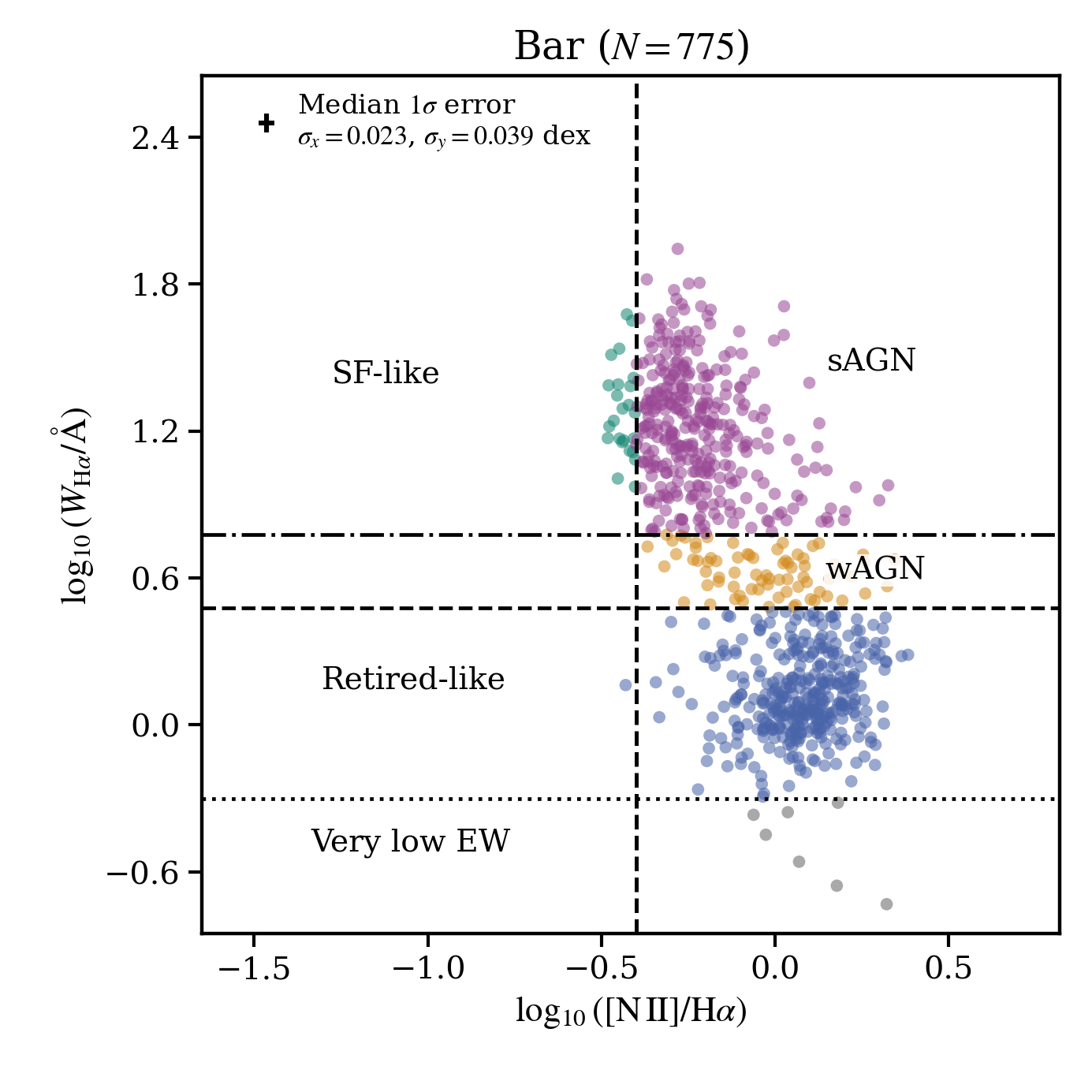}%
\hspace{1mm}%
\includegraphics[width=0.325\textwidth]{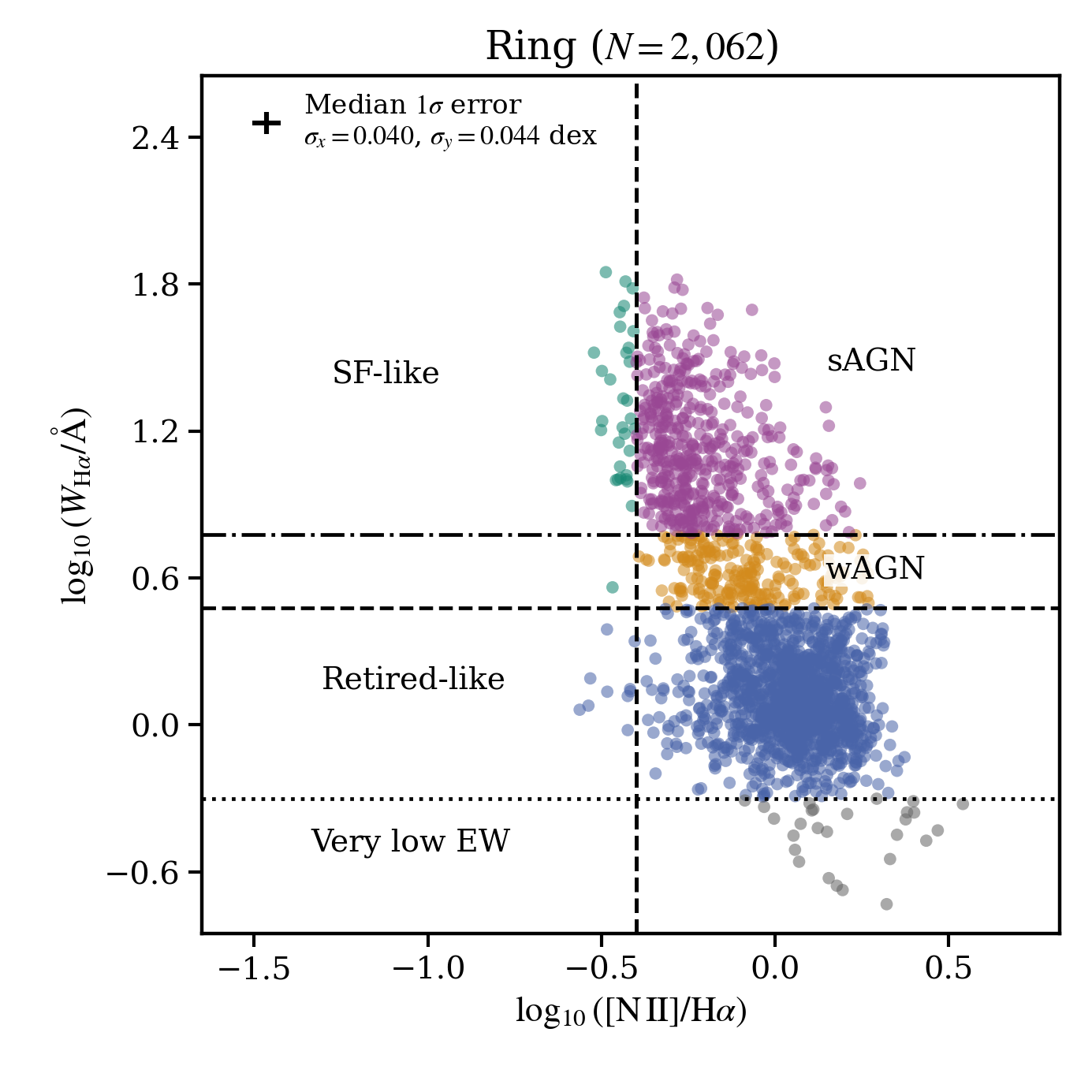}%
}

\par\vspace{2mm}

\noindent\makebox[\textwidth][c]{%
\begin{minipage}[c]{0.325\textwidth}
\centering
\includegraphics[width=\linewidth]{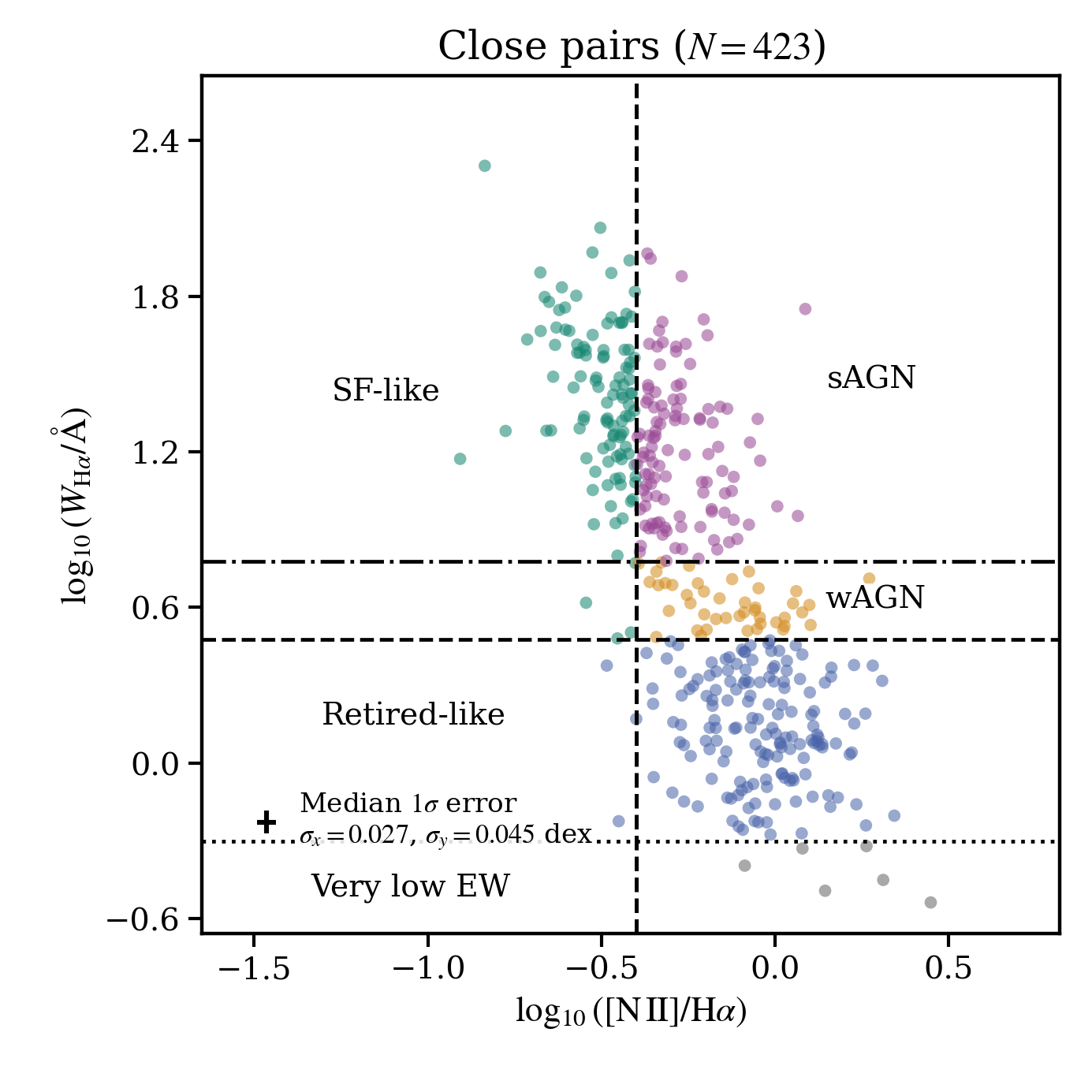}
\end{minipage}%
\hspace{6mm}%
\begin{minipage}[c]{0.59\textwidth}
\captionof{figure}{WHAN diagnostic diagrams
for seven visually confirmed morphological samples.
Each panel indicates the sample size and the median
formal $1\sigma$ uncertainties along both diagnostic axes.}
\label{fig:WHAN_all}
\end{minipage}%
}

\par
\endgroup

\end{appendix}

\end{document}